\documentclass[epj,nopacs,fleqn]{svjour}

\usepackage{listings}
\newcommand{\n}{\nonumber}

\newcommand{\epstld}{\tilde\epsilon}
\newcommand{\tld}{\tilde}
\newcommand{\eps}{\epsilon}
\newcommand{\del}{\partial}
\newcommand{\lam}{\lambda}
\newcommand{\sgn}{{\rm sgn}}
\newcommand{\vect}{\overrightarrow}
\newcommand{\h}{\lambda}
\newcommand{\helas}{{\tt HELAS}\ }
\newcommand{\HELAS}{{\tt HELAS}}
\newcommand{\qvec}{\overrightarrow q}
\newcommand{\0}{\varepsilon}
\newcommand{\dgr}{\dagger}

\usepackage{graphics}
\usepackage{amsmath,amssymb,bbm,bm}
\usepackage{slashed}
\usepackage{hyperref}
\usepackage{color}
\usepackage{tikz-feynhand}
\usepackage{mcite,cite}
\usepackage{multirow}
\usepackage{comment}

\newcommand{\vrev}[1]{{\color{black}#1}}

\begin{document}\vrev

\title{Helicity amplitudes without gauge cancellation\\ for electroweak processes}
\author{
	Junmou Chen\inst{1,}\thanks{chenjm@jnu.edu.cn}
	\and
	Kaoru Hagiwara\inst{2,}\thanks{kaoru.hagiwara@kek.jp}
	\and
	Junichi Kanzaki\inst{3,}\thanks{junichi.kanzaki@ipmu.jp}
	\and
	Kentarou Mawatari\inst{4,}\thanks{mawatari@iwate-u.ac.jp}
}                     
%
%
\institute{
	Department of Physics, Jinan University, Guangzhou, Guangdong Province, 510632, China
	\and 
	KEK Theory Center and Sokendai, Tsukuba, Ibaraki 305-0801, Japan
	\and
	Kavli IPMU (WPI), UTIAS, The University of Tokyo, Kashiwa, Chiba 277-8583, Japan
	\and
	Faculty of Education, Iwate University, Morioka, Iwate 020-8550, Japan
}
%
\date{}

\abstract{
\vrev{
In the 5-component representation of weak bosons, 
the first four components make a Lorentz four vector, representing the transverse and longitudinal polarizations excluding the scalar component of the weak bosons, whereas 
its fifth component corresponds to the Goldstone boson. 
We obtain the $5\times 5$ component propagators of off-shell weak bosons, 
proposed previously and named after the Goldstone boson equivalence theorem,
by starting from the unitary-gauge representation of the tree-level scattering amplitudes,
and by applying the BRST (Becchi--Rouet--Stora--Tyutin) identities to the two sub-amplitudes connected by each off-shell weak-boson line.
}
By replacing all weak boson vertices with those among the 
off-shell 
5-component wavefunctions, we arrive at the expression of the electroweak scattering amplitudes, where the magnitude of each Feynman amplitude has the correct on-shell limits for all internal propagators, and hence with no artificial gauge cancellation among diagrams.
\vrev{
Although our derivation is limited to the tree-level only,
it allows us to study the properties of each Feynman amplitude separately, 
and then learn how they interfere in the full amplitudes.
}
We implement the 5-component weak boson propagators and their vertices in the numerical helicity amplitude calculation code {\tt HELAS} (Helicity Amplitude Subroutines), so that an automatic amplitude generation program such as {\tt MadGraph} can generate the scattering amplitudes without gauge cancellation.
We present results for several high-energy scattering processes where subtle gauge-theory cancellation among diagrams takes place in all the other known approaches.
%
} 
\maketitle

\vspace*{-14cm}
\noindent KEK-TH-2403, IPMU22-0008 
\vspace*{12cm}

\section{Introduction}\label{sec:intro}

It has been customary that scattering amplitudes in QED, QCD, and those in the standard model (SM) are expressed in terms of the Feynman amplitudes, where the gauge-boson propagators are expressed in a given gauge, such as in the Feynman gauge, Landau gauge, light-cone gauge, axial gauge, unitary gauge or renormalizable covariant $R_\xi$ gauge, etc. 
It has been shown that scattering amplitudes, or $S$-matrix elements of external on-shell particles, are independent of the gauge which we need to fix to quantize the gauge bosons. 
The gauge invariance of the amplitudes has thus been used to test our analytic or numerical computations.

Recently, a new type of gauge-boson propagators is proposed for massless gauge bosons~\cite{Hagiwara:2020tbx}, such as photons and gluons:
\begin{align}
  \tilde D_{\mu\nu}(q) = \frac{-i\,\tilde{g}_{\mu\nu}}{q^2+i\0},
\label{propagator_psg}
\end{align}
which has the form of the light-cone gauge propagator,
\begin{align}
  \tilde{g}_{\mu\nu}
 = g_{\mu\nu}-\frac{n_\mu q_\nu+q_\mu n_\nu}{n\cdot q},
\label{gmntld}
\end{align}
where the light-like vector $n^\mu$ is chosen for each propagator as
\begin{align}
  n^\mu = ( \sgn(q^0) , -\vect{q}/|\vect{q}| ),
\label{nmu}
\end{align}
depending on the three momentum of the off-shell gauge bosons.
Here, $\sgn(q^0)=1$ when $q^0\ge 0$, while $\sgn(q^0)=-1$ otherwise.

It has been shown in Ref.~\cite{Hagiwara:2020tbx} that tree-level scattering amplitudes calculated with the above photon and gluon propagators agree exactly with the known gauge invariant helicity amplitudes.
The agreement has been shown to follow from the BRST (Becchi--Rouet--Stora--Tyutin)
identities~\cite{Becchi:1975nq,*Tyutin:1975qk}
\begin{align}
  \langle {\rm phys}| \{ Q_{\rm BRST}, \bar{c} \} |{\rm phys}\rangle
= \langle {\rm phys}| \del_\mu A^\mu |{\rm phys}\rangle = 0,
\end{align}
where the BRST transformation of the anti-ghost field $\bar{c}$ gives the covariant gauge fixing term both in QED and QCD.
The physical states are defined to be BRST invariant, or
\begin{align}
  Q_{\rm BRST} |{\rm phys}\rangle = 0.
\end{align}
When an arbitrary scattering amplitude is connected by a photon/gluon propagator as
\begin{align}
  {\cal M} = T_1^\mu \frac{P^{R_\xi}_{\mu\nu}}{q^2+i\0} T_2^\nu
\end{align}
with
\begin{align}
  P^{R_\xi}_{\mu\nu}
 = -g_{\mu\nu} + (1-\xi)\frac{q_\mu q_\nu}{q^2}
\end{align}
in the covariant ($R_\xi$) gauge, the theorem can be applied for the two sub-amplitudes $T_1^\mu$ and $T_2^\nu$, giving
\begin{align}
  q_\mu T_k^\mu = 0
\label{brstinv}
\end{align}
for both $k=1$ and $2$.
Because of the BRST identities for the two off-shell currents in Eq.~\eqref{brstinv}, we can drop all terms proportional to $q_\mu$ and $q_\nu$ in the polarization tensor as
\begin{align}
  \tilde P_{\mu\nu}
 &= \sum_{\lam=\pm1}\eps^\mu(q,\lam)^*\eps^\nu(q,\lam)
  +\sgn(q^2)\tilde\eps^\mu(q,0)\tilde\eps^\nu(q,0)  \n\\
 &= -\tld{g}_{\mu\nu},
\label{Ptld}
\end{align}
where
\begin{align}
  \epstld^\mu(q,0)   &= \eps^\mu(q,0) - \frac{q^\mu}{Q}, 
\label{etldL}
\end{align}
with $Q=\sqrt{|q^2|}$, when $q^0>0$.
The essence of the propagator in Eqs.~\eqref{propagator_psg} and \eqref{gmntld} is hence that the component of the longitudinally polarized mode of the virtual photon/gluon which grows with $q^\mu$ at high energies is not allowed to propagate.
Indeed, $\epstld^\mu(q,0)$ vanishes at high energies.
Our notation for the polarization vectors is summarized in Appendix~\ref{sec:polvector}.

The striking property of the QED/QCD scattering amplitudes calculated with the propagators of the form~\eqref{propagator_psg} is that there is no subtle cancellation among Feynman amplitudes and that the individual Feynman amplitude shows all the on-shell singularities at all energies in an arbitrary Lorentz frame.
Accordingly, the absolute value squared of the Feynman amplitudes calculated with the propagator~\eqref{propagator_psg} reproduces the well-known DGLAP splitting functions~\cite{Gribov:1972ri,*Lipatov:1974qm,*Altarelli:1977zs,*Dokshitzer:1977sg} in the collinear limit.
Because of this property, the propagator of Eqs.~\eqref{propagator_psg} and \eqref{gmntld} has been named `parton-shower (PS) gauge' in Ref.~\cite{Hagiwara:2020tbx}.

In this paper, we extend the study of Ref.~\cite{Hagiwara:2020tbx} to massive vector bosons in the SM of particle physics, W and Z bosons.
The BRST identities for the physical states now read
\begin{align}
  &\langle {\rm phys}| \{ Q_{\rm BRST}, \bar{c} \} |{\rm phys}\rangle \n\\
  &=\langle {\rm phys}| (\del_\mu V^\mu - \xi m_V \pi_V) |{\rm phys}\rangle = 0
\label{brst}
\end{align}
in the covariant $R_\xi$ gauge~\cite{Fujikawa:1972fe}, where $\pi_V$ corresponds to the Goldstone boson associated with the weak bosons $V = W^\pm$ and $Z$.
We can apply the BRST identities~\eqref{brst} to a pair of an arbitrary tree-level amplitudes which are connected by a weak boson propagator, e.g. as
\vrev{
\begin{align}
  {\cal M} = T_1^\mu \frac{P^{U}_{V\mu\nu}}{q^2-m_V^2+i\0} T_2^\nu
\label{M_unitary}  
\end{align}
}
with
\begin{align}
  P^{U}_{V\mu\nu}
= -g_{\mu\nu} + \frac{q_\mu q_\nu}{m_V^2}
\label{P_unitary}
\end{align}
in the unitary gauge.
We can express the unitary-gauge polarization tensor~\eqref{P_unitary} as
\begin{align}
  P^{U}_{V\mu\nu}
= -\tilde{g}_{\mu\nu}
+\tilde{\epsilon}_\mu(q,0)^* \frac{q_\nu}{Q}
+\frac{q_\mu}{Q} \tilde{\epsilon}_\nu(q,0)
+\frac{q_\mu q_\nu}{m_V^2},
\label{P_unitary_tld}
\end{align}
where $\tilde{g}_{\mu\nu}$ denotes the polarization tensor~\eqref{gmntld} or \eqref{Ptld} of QED and QCD~\cite{Hagiwara:2020tbx}. 
As we show below in Sect.~\ref{sec:deriv}, the BRST identities~\eqref{brst} for the sub-amplitudes $T_k^\mu$ give
\begin{subequations}
\begin{align}
  -iq_\mu T_1^\mu &= m_V T_1^{\pi_V}, \label{brst_amp1}\\
   iq_\nu T_2^\nu &= m_V T_2^{\pi_V}, \label{brst_amp2}
\end{align}\label{brst_amp}%
\end{subequations}
\vrev{
where $T_k^{\pi_V} (k=1,2)$ are the Goldstone-boson amplitudes.
The weak-boson propagators are then expressed as a summation over the contribution of the QED/QCD-like term, 
the mixing terms between the reduced longitudinal polarization vector~\eqref{etldL}
and the Goldstone boson, and the Goldstone-boson exchange term.
All the unitary-gauge propagators are hence expressed in the $5\times5$ matrix form
in the 5-component representation of the weak bosons.
}

The 5-component representation of the weak boson wavefunctions was introduced by Chanowitz and Gaillard~\cite{Chanowitz:1985hj} in their derivation of the Goldstone boson equivalence theorem.
It was adopted
\vrev{
by the authors of Refs.~\cite{Kunszt:1987tk,Borel:2012by,Wulzer:2013mza,Chen:2016wkt} 
in obtaining the equivalent distributions of the weak bosons in massless quarks and leptons.
}
\vrev{
The $5\times5$ matrix form of the weak-boson propagators, Eq.~\eqref{P_unitary_tld}, was introduced in Ref.~\cite{Chen:2016wkt} in their derivation of the electroweak (EW) splitting functions, 
and derived rigorously in Ref.~\cite{Cuomo:2019siu} as a consistent covariant quantization of the EW theory.
Although our derivation is limited to the tree-level amplitudes only, 
as explained in the following section, 
we keep it in the manuscript because it was inspired by the success of obtaining `PS-gauge' amplitudes in unbroken gauge theories~\cite{Hagiwara:2020tbx}
and also because its simplicity has allowed us to develop an automatic amplitude generation code in the SM.
}

In this paper, we present explicit expressions for all the weak boson propagators and their vertices in the SM, and prepare numerical codes to calculate arbitrary tree-level amplitudes 
by updating {\tt HELAS} ({\tt HEL}icity {\tt A}mplitude {\tt S}ubroutines) \cite{Hagiwara:1990dw,helas}
for automatic amplitude generation programs such as {\tt MadGraph}~\cite{Stelzer:1994ta,Alwall:2007st,Alwall:2011uj,Alwall:2014hca}.

The rest of the paper is organized as follows.
In Sect.~\ref{sec:deriv}, we derive the 5-component weak boson wavefunctions and their $5 \times 5$ propagators, starting from the unitary-gauge amplitudes, 
and discuss the relationships with the amplitudes in the $R_\xi$ gauge.
We also explain how to generate an arbitrary tree-level helicity amplitude with the 5-component representation of weak bosons automatically by using 
{\tt MadGraph5\_aMC@NLO}~\cite{Alwall:2014hca}.
In Sect.~\ref{sec:results}, we present numerical results for weak boson scattering processes $VV \to VV$ for $V=W^\pm$ or $Z$, W-boson pair production processes
$l^- l^+ \to W^- W^+ \to f_1 \bar{f}_2 f_3 \bar{f}_4$,  
and 
$l^- l^+ \to l' \bar{l}' VV$ ($l=e$ or $\mu$, $l' = l$ or $\nu_l$).
Section~\ref{sec:summary} summarizes our findings.
We give three appendices.
Appendix~\ref{sec:codes} presents the new \helas codes which are necessary to obtain helicity amplitudes in the 5-component weak-boson formalism.
Appendix~\ref{sec:vertices} gives all the Goldstone-boson couplings in the SM,
while
Appendix~\ref{sec:polvector} summarizes our notation for polarization vectors.

\section{Derivation}\label{sec:deriv}

\vrev{
As stated above, the derivation of the $5\times5$ component weak-boson propagator given in this section 
is valid only for the tree-level scattering amplitudes.
For a more general derivation, we refer the readers to Ref.~\cite{Cuomo:2019siu}.
}

\subsection{5-component polarization vector of weak bosons}

Let us denote a helicity amplitude with an external weak boson $V$ ($V=W^\pm$ or $Z$) as
\begin{align}
  {\cal M}_\lambda = \epsilon_{V\mu}(q,\h)\, T_V^\mu,
\label{M_V}
\end{align}
where $q$ and $\h$ are the four momentum and the helicity of the vector boson.
When the vector boson is in the final state, the polarization vector should be replaced by its complex conjugate.
The three polarization vectors take a simple form
\begin{subequations}
\begin{align}
  &\epsilon^\mu_V(q,\pm) = \frac{1}{\sqrt{2}} (0, \mp 1, -i, 0), \\
  &\epsilon^\mu_V(q,0)
 = \frac{1}{m_V} (|\qvec|, 0, 0, E) 
 = (\sinh y, 0, 0, \cosh y),
\end{align}\label{polvec4}%
\end{subequations}
in the frame where the three momentum $\qvec$ is along the $z$-axis,
\begin{align}
  q^\mu = (E, 0, 0, |\qvec|) = m_V (\cosh y, 0, 0, \sinh y).
  \label{qmu}
\end{align}
Here, $y$ is the rapidity of the vector boson along its three momentum direction.

Because the polarization vector for longitudinally polarized ($\h=0$) vector boson grows with $E/m_V$ or as $e^{y}$ at high energies, the individual Feynman amplitude for longitudinally polarized vector bosons often grows with energy, violating perturbative unitarity.
In theories where the vector boson masses are obtained by spontaneous breaking of the gauge symmetry, all these unitarity violating terms cancel after summing over all the interfering Feynman amplitudes.
The subtle gauge-theory cancellation among interfering amplitudes can lead to the loss of terms of order $(E/m_V)^n$ in numerical calculation of amplitudes for processes with $n$ external weak bosons.
The factor can be orders of $10^8$ for weak boson scattering processes ($n=4$) at $E=10$~TeV.

The BRST invariance of quantized EW theory allows us to obtain the helicity amplitudes without subtle cancellation.
We first split the $\h=0$ polarization vector as
\begin{align}
  \epstld_V^\mu(q,0) = \eps_V^\mu(q,0) - \frac{q^\mu}{m_V},
\end{align}
where
\begin{align}
  \epstld_V^\mu(q,0)
 &= \frac{m_V}{E+|\qvec|} (-1,0,0,1) = e^{-y} (-1,0,0,1),
\label{epstld} 
\end{align}
in the frame where the three momentum of the vector boson is taken along the $z$-axis as in Eq.~\eqref{qmu}.
The BRST identities~\eqref{brst} for the helicity amplitudes ${\cal M}_\lambda$~\eqref{M_V} give
\begin{align}
  i q_\mu T_V^\mu = m_V T^{\pi_V},
\label{T_piv}  
\end{align}
where $T^{\pi_V}$ is the amplitude of the Goldstone boson $\pi_V$ which gives the mass to the gauge boson $V$ when the gauge invariance is broken spontaneously.%
\footnote{
\vrev{
It should be noted that the BRST identities, Eqs.~\eqref{brst_amp1} and \eqref{brst_amp2}, 
 are valid in an arbitrary $R_\xi$ gauge, including the unitary gauge 
 ($\xi\to\infty$) in the tree level. 
 This is because 
 \vrev{
 the sub-amplitudes $T_V^\mu$ and $T^{\pi_V}$ in Eq.~\eqref{T_piv} 
 are truncated, and
 }
 all the Goldstone-boson couplings are $\xi$ independent.
 The BRST identities, Eq.~\eqref{T_piv}, in the unitary gauge has been used in checking the correctness of numerically evaluated scattering amplitudes~\cite{Hagiwara:1990gk,helas}.
}}
The sign of the left-hand-side of Eq.~\eqref{T_piv} should be reversed when the vector boson is in the final state.

An arbitrary amplitude of the longitudinally polarized massive gauge bosons~\eqref{M_V} can hence be expressed as
\begin{align}
  {\cal M}_0 = \epsilon_{V\mu}(q,0)\, T_V^\mu 
 = \epstld_{VM}(q,0)\, T_V^M, 
\label{M0} 
\end{align}
where the index $M$ runs from 0 to 4, $M=\{\mu,4\}$, with
\begin{subequations}
\begin{align} 
 \epstld_V^M(q,\pm) &= (\epstld_V^\mu(q,\pm),\,0)= (\epsilon_V^\mu(q,\pm),\,0), \\
 \epstld_V^M(q,0) &= (\epstld_V^\mu(q,0),\,i), \label{epstld4}
\end{align}\label{pol5}%
\end{subequations}
and
\begin{align}
  T_V^4 = T^{\pi_V},
  \label{eps_T}
\end{align}
when the weak boson $V$ is in the initial state.
When the weak boson $V$ is in the final state,
\vrev{
all the polarization `vectors', $\epstld_V^M(q,\lam)$, should be replaced by their complex conjugates.
}
Since all the components of $\epstld_V^\mu(q,0)$ in Eq.~\eqref{epstld} vanish at high energies as $m_V/(E+|\qvec|)=e^{-y}$, and since the Goldstone boson amplitudes $T^{\pi_V}$ do not grow with energy, we can tell that the $\h=0$ vector boson scattering amplitudes should reduce to the associate Goldstone boson amplitudes at high energies, leading to the Goldstone boson equivalence theorem~\cite{Chanowitz:1985hj,Gounaris:1986cr,*Bagger:1989fc,*Veltman:1989ud}.

\vrev{
As we show in Subsection~\ref{sec:wwww},
we study all $2\to2$ weak-boson scattering amplitudes, 
and verify numerically that the BRST identity~\eqref{T_piv} holds even when one or more external weak-boson polarization vectors are replaced by the 5-component wavefunctions~\eqref{pol5}.
}

If we keep the finite contribution of $\epstld_V^\mu(q,0)$, which can be significant at low energies, we have an exact expression of the amplitudes without gauge-theory cancellation. 
\vrev{
This has been shown first in the axial gauge in obtaining the equivalent real weak-boson distributions~\cite{Kunszt:1987tk,Borel:2012by}, 
and the weak-boson splitting amplitudes~\cite{Chen:2016wkt}, 
and also in the covariant gauge in Refs.~\cite{Wulzer:2013mza,Cuomo:2019siu}.

At this stage, we have obtained the expression of the weak-boson scattering amplitudes which have no external wavefunction components that grow with energy.
However, because we adopt the unitary gauge for the internal propagators,
individual Feynman amplitude can still have components which grow with energy
when the energy of the exchanged off-shell weak boson is large.
This problem can be solved by replacing all the unitary-gauge propagators by the $5\times5$ component propagators of the 5-component weak bosons~\cite{Chen:2016wkt,Cuomo:2019siu}. 
}

\subsection{Propagators of the 5-component weak bosons}

The BRST identities, Eqs.~\eqref{brst_amp1} and \eqref{brst_amp2}, allow us to replace one
unitary-gauge propagator in an arbitrary tree-level scattering amplitude by the $5 \times 5$ component propagator of the 5-component representation of the weak bosons:
\begin{align}
  {\cal M}
= T_1^\mu\frac{P^{U}_{V\mu\nu}}{q^2-m_V^2+i\0} T_2^\nu 
= T_1^M \frac{\tilde{P}_{VMN}}{q^2-m_V^2+i\0} T_2^N,
\label{M5}
\end{align}
with
\begin{subequations}
\begin{align}
  T_k^M &= T_k^\mu &&{\rm when}\ M=\mu=0,1,2,3, \\
  T_k^M &= T_k^{\pi_V} &&{\rm when}\ M=4, 
\end{align}\label{identity5}
\end{subequations}
for $k=1$ and 2, and
\begin{subequations}
\begin{align}
  \tilde{P}_{V\mu\nu} &= -\tilde{g}_{\mu\nu}, \\
  \tilde{P}_{V\mu 4}  &= -i \tilde{\epsilon}_\mu(q,0)^* \frac{m_V}{Q}, \\
  \tilde{P}_{V4 \nu}  &=  i \tilde{\epsilon}_\nu(q,0) \frac{m_V}{Q}, \\
  \tilde{P}_{V4 4}    &= 1,
\end{align}
\label{PMN}
\end{subequations}
or
\begin{align}
  \tilde{P}_{V}{}_{MN}=
  \begin{pmatrix}
	  -\tilde{g}_{\mu\nu} & -i \tilde{\epsilon}_\mu(q,0)^* \dfrac{m_V}{Q} \\
	  i \tilde{\epsilon}_\nu(q,0) \dfrac{m_V}{Q} & 1 \\
  \end{pmatrix}	
\label{PMN_matrix}  
\end{align}
in the matrix representation.

\vrev{
In the first step, the subamplitudes $T_1^\mu$ and $T_2^\nu$ are both off-shell weak-boson currents connected by all Feynman diagrams to on-shell states in the unitary gauge.
What we verified by using our numerical code is that the identities~\eqref{identity5} hold even if the off-shell weak-boson line is connected by the $5\times5$ propagator~\eqref{PMN_matrix}.
Because of this, all the unitary-gauge propagators can be replaced by the $5\times5$ propagators.
}

Let us 
give a few key properties of the $4 \times 4$ components of the polarization tensor $\tilde{P}_{V\mu\nu}$,
which is exactly the same as the special light-cone gauge propagator introduced in Ref.~\cite{Hagiwara:2020tbx} for the photons and gluons,
Eqs.~\eqref{gmntld} and \eqref{Ptld}.
The light-cone vector $n^\mu$~\eqref{nmu} has its three-vector component along $-\qvec$ and the time component is fixed to give
\begin{align}
  n \cdot q = |q^0| + |\qvec|
\label{ndotq}
\end{align}
for an arbitrary four momentum $q^\mu$.
The reduced polarization vector for the $\h=0$ state~\eqref{etldL} can be expressed as
\begin{align}
  \tilde{\epsilon}^\mu(q,0) =  \sgn(q^2)\, e^{-y} (-1,0,0,1),
\end{align}
when the three momentum $\qvec$ is along the positive $z$-axis 
with $q_z=Q\sinh y$ for the time-like, 
and $q_z=Q\cosh y$ for the space-like four momentum.
The reduced polarization tensor becomes
\begin{align}
 \tilde{P}_{V\mu\nu}=-\tilde{g}_{\mu\nu} =\begin{pmatrix}
 e^{-2y}&& 0 &\ & 0&& -e^{-2y} \\
       0&& 1&\ & 0&& 0        \\
       0&& 0&\ & 1&& 0        \\
 -e^{-2y}&& 0&\ & 0&&  e^{-2y} \\ 
\end{pmatrix}
\end{align}
both for the time-like and space-like momentum.
Therefore, the longitudinal polarization mode is suppressed by the $e^{-2y}$ factor as compared to the transverse polarization modes at high energies.

For massive vector bosons, the remaining part of the longitudinally polarized weak bosons
and also the scalar ($j=0$) component can propagate.
For instance, 
the polarization transfer tensor in the unitary gauge can be expressed as
\begin{align}
  P^{U}_{V\mu\nu}
 = -\tilde{g}_{\mu\nu}
   -\frac{n_\mu q_\nu +q_\mu n_\nu}{n\cdot q} 
   +\frac{q_\mu q_\nu}{m_V^2}.
\end{align}
The first term gives the propagation of the spin-1 ($j=1$) components 
just as in QED and QCD~\cite{Hagiwara:2020tbx},
while the second term gives transitions between the reduced longitudinal polarization state
and the spin-0 ($j=0$) component, 
and the last part gives the propagation of the spin-0 component.
This is manifest in the equivalent expression given in Eq.~\eqref{P_unitary_tld}.
Applying the BRST identities~\eqref{brst_amp}, which are valid at an arbitrary $q^\mu$, 
we arrive at the 5-component weak boson propagator  
\begin{align}
  \tilde{P}_{V}{}_{MN}=
  \begin{pmatrix}
	  -\tilde{g}_{\mu\nu} & i m_V \dfrac{n_\mu}{n\cdot q} \\
	  -i m_V \dfrac{n_\nu}{n\cdot q} & 1 \\
  \end{pmatrix}	
  \label{pro5}
\end{align}
in the matrix representation.
Note that the expression in Eq.~\eqref{pro5} agrees with
Eq.~\eqref{PMN_matrix} when $q^2>0$, 
because Eq.~\eqref{PMN_matrix} is obtained from Eq.~\eqref{brst_amp}
when the weak boson is produced by $T_1^\mu$ with the time-like momentum
and then decays into $T_2^\nu$.
The form of our $5 \times 5$ polarization transfer matrix~\eqref{pro5}
is valid for an arbitrary four momentum $q^\mu$, 
\vrev{
including the space-like momentum with $q^2<0$.
}
This can be confirmed by replacing $q^\mu$ by $-q^\mu$ in the propagator;
the diagonal terms in Eq.~\eqref{pro5} are invariant under the replacement.
In the off-diagonal terms, $n\cdot q$, given in Eq.~\eqref{ndotq}, is invariant,
while $n^\mu$ changes sign.
At the same time the Goldstone-boson flow is reversed, 
resulting in the exchange between $i$ and $-i$ in the BRST identities~\eqref{brst_amp}.
We hence recover the same form~\eqref{pro5}.

\subsection{Implementation in {\tt MadGraph5\_aMC@NLO}}
\label{sec:mg5}

The 5-component representation of the weak bosons and their $5 \times 5$ polarization transfer matrix propagators can be a powerful tool in the study of the SM physics if they can be implemented into an automatic amplitude generation program, such as {\tt MadGraph5\_aMC@NLO} ({\tt MG5})~\cite{Alwall:2014hca}.
In this paper, we develop a working model which can be applied to calculate tree-level helicity amplitudes of an arbitrary processes in the SM.

We first generate the amplitudes in the unitary gauge of the SM weak bosons by using {\tt MG5}~\cite{Alwall:2014hca}. 
{\tt MG5} produces numerical codes to compute helicity amplitudes corresponding to each Feynman diagram. 
The numerical codes calculate the helicity amplitudes by successively calling {\tt HELAS} subroutines~\cite{Hagiwara:1990dw,*Murayama:1992gi}, which calculate off-shell fermion and vector boson wavefunctions starting from the external on-shell particle wavefunctions. 
The amplitude is obtained as a complex number when all the off-shell wavefunctions meet at one vertex.%
\footnote{While {\tt MadGraph/MadEvent v4}~\cite{Alwall:2007st} employs the original \helas library~\cite{Hagiwara:1990dw,*Murayama:1992gi}, {\tt MadGraph5}({\tt \_aMC@NLO})~\cite{Alwall:2011uj}(\!\!\!\cite{Alwall:2014hca}) adopts {\tt ALOHA}~\cite{deAquino:2011ub}, which automatically generates the \helas library by using the {\tt UFO} format~\cite{Degrande:2011ua}.}

We modify the program as follows:
\begin{enumerate}
\item When an external particle is a weak boson ($W^\pm$ or $Z$), we replace the external wavefunction with our 5-component wavefunction in Eq.~\eqref{pol5}.
\item All the vertices with external weak bosons should be extended to contain Goldstone boson contributions.
\item All the unitary-gauge propagator of weak bosons should be replaced by the $5 \times 5$ matrix propagator in Eq.~\eqref{pro5}.
\item All photon and gluon propagators should be given in the PS gauge in Eqs.~\eqref{propagator_psg} and \eqref{gmntld}.
\end{enumerate}

The new \helas subroutines for the 5-component representation of a massive vector boson (step 1), those for all the weak boson vertices in the SM (step 2), and those for the 5-component off-shell vector-boson wavefunctions (step 3) have been created, and they are listed in Appendix~\ref{sec:codes}. 
The off-shell photon and gluon propagators in the PS gauge (step 4) have been given in Ref.~\cite{Hagiwara:2020tbx}.

\begin{figure}
  \center
\includegraphics[width=1\columnwidth]{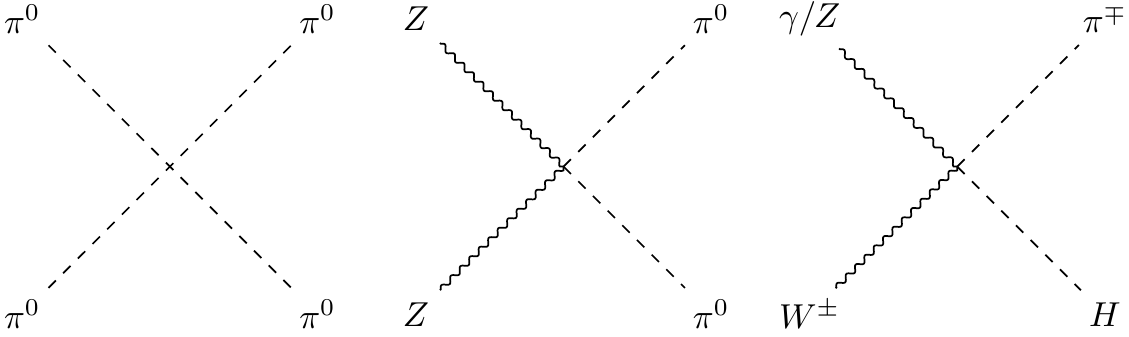}
(a)\hspace*{2.6cm}(b)\hspace*{2.6cm}(c)
\caption{Contact interactions which do not appear in the unitary-gauge weak-boson vertices.}
\label{fig:diagram_GB}
\end{figure}

\vrev{
We note here that all the off-shell \helas subroutines for a $n$-point vertex have $n-1$ inputs and one off-shell output. 
When all the $n-1$ inputs are on-shell states, the subroutine should hence satisfy the BRST identities~\eqref{brst_amp}.
All our new \helas subroutines have been tested explicitly by using the BRST, with the double complex accuracy of 12 to 13 digits.
The BRST identities turn out to be valid when one or all $n-1$ inputs are replaced by off-shell \helas subroutines which are obtained from the external on-shell states.
In this way, we have checked step-by-step the BRST invariance of all off-shell vertices which appear in the tree-level SM amplitudes.
}

It turns out that the above four steps are not sufficient to generate all the SM amplitudes. 
This is because there are four 4-point vertices in the SM which couple only to the Goldstone boson(s)
(the fifth component of our weak-boson wavefunctions) but not to the corresponding weak boson(s) in the unitary gauge. 
They are listed in Fig.~\ref{fig:diagram_GB}: 
(a) The quartic coupling of the Goldstone boson $\pi_Z = \pi^0$. 
(b) The quartic gauge coupling of the Goldstone boson $\pi^0$, which couples to Z-boson pair. 
(c) The quartic coupling among the photon ($\gamma$) or $Z$, $H$, $W^\pm$ and the Goldstone boson of $W^\mp$ ($\pi_{W^\mp}=\pi^\mp$).

Because the only way that we could identify these four missing vertices in the unitary gauge has been to check all Goldstone vertices in the SM, we present our parametrization of the SM Higgs sector in Appendix~\ref{sec:vertices}. 
The quartic coupling of $\pi^0$ (Fig.~\ref{fig:diagram_GB}(a)) appears in the Higgs potential, as a consequence of its custodial SU(2) invariance, under which the three Goldstone bosons $\pi^k$ with $k=1,2,3$ transform as a triplet. 
All the other quartic vertices, Fig.~\ref{fig:diagram_GB}(b,c), appear in the gauge coupling of the Higgs doublet, as so-called seagull terms.

Those four vertices do not appear among the weak boson vertices in the unitary gauge, and hence we should add them as additional vertices in the SM. 
This can easily be achieved in the SM {\tt UFO} model by introducing the following three additional interactions, $ZZZZ$, $WWHZ$ and $WWH\gamma$. 
{\tt MG5} then generates helicity amplitudes with the {\tt HELAS} subroutines with the above vertices. 
It is then straightforward to prepare the {\tt HELAS} subroutines by using the Goldstone boson couplings of the SM. 
These additional codes are also given in Appendix~\ref{sec:codes}.

It should be noted that these quartic vertices are needed in our representation of the scattering amplitudes with the 5-component weak bosons, 
because the BRST identities, 
\vrev{
Eqs.~\eqref{brst_amp1} and \eqref{brst_amp2}, relate the total sum of all the contributing amplitudes rather than individual Feynman amplitudes.
}
For example, the $ZZZZ$ sub-amplitudes with four unitary gauge Z-boson propagators generate the Goldstone vertices like Fig.~\ref{fig:diagram_GB}(a) and (b). 
\vrev{
Therefore, the scalar ($j=0$) polarization component of the physical (unitary-gauge) weak boson interacts with the Goldstone-boson vertices
} 
\vrev{
which are gauge ($\xi$) independent as shown explicitly
}
in the Higgs Lagrangian 
(given in Appendix~\ref{sec:vertices} for the minimal SM).

\subsection{Comments on amplitudes in the $R_\xi$ gauge}

The tree-level amplitudes expressed in the unitary gauge have a special property that when we cut one gauge-boson propagator of a given four momentum in all the Feynman amplitudes, the full helicity amplitude is split into two pieces, each satisfying the BRST invariance.
This is because when the cut vector-boson four momentum is set on-shell, each sub-amplitude can be regarded as an independent scattering amplitude.

In general renormalizable covariant gauge, the $R_\xi$ gauge with finite $\xi$~\cite{Fujikawa:1972fe}, there can appear both a vector-boson and its associate Goldstone-boson propagator which share the same four momentum.
\vrev{
In our 5-component representation of the weak bosons, its fifth component is interacts with the Goldstone boson couplings. 
It might therefore seem that our $5 \times 5$ matrix propagator of the weak bosons contains the associated Goldstone boson propagators in the Feynman gauge ($\xi=1$).
However, this is not the case, as explained below.

This can be seen clearly by denoting the massive weak-boson propagator in the $R_\xi$ gauge
\begin{align}
 D^{R_\xi}_{V\mu\nu}(q)
= \frac{i}{q^2-m_V^2+i\0}
\Big[-g_{\mu\nu} +(1-\xi)\frac{q_\mu q_\nu}{q^2-\xi m_V^2}\Big],
\end{align}
in the following form:
\begin{align}
  D^{R_\xi}_{V\mu\nu}(q)
= D^{U}_{V\mu\nu}(q)
-\frac{q_\mu q_\nu}{m_V^2} \frac{i}{q^2-\xi m_V^2},
\label{Dmn_Rxi}
\end{align}
where $D^{U}_{V\mu\nu}(q)$ is the unitary-gauge propagator.
Upon making use of the BRST identities, the second term in Eq.~\eqref{Dmn_Rxi} gives the negative of the Goldstone-boson contribution in the $R_\xi$ gauge with the propagator
\begin{align}
   D^{R_\xi}_{\pi_V}(q) = \frac{i}{q^2-\xi m_V^2},
   \label{D5_Rxi}
\end{align}
whenever the same four-momentum is shared by the gauge boson and its associated Goldstone boson.
We can hence conclude that their sum should agree with the contribution of the unitary-gauge propagator, and hence that of our $5 \times 5$ propagator.
This statement is valid for an arbitrary $R_\xi$ gauge, including the Feynman gauge ($\xi=1$).
Because of this cancellation, Goldstone-boson propagators cannot appear inside the tree-level scattering amplitudes.

Therefore even if we start from the scattering amplitude expression in the $R_\xi$ gauge, the vector-boson and its associated Goldstone-boson contributions sum up to be replaced by the unitary-gauge propagator contribution. 
\vrev{
We find that in the $R_\xi$ gauge amplitudes, we always have both a gauge-boson and its associated Goldstone-boson lines which share the same four-momentum, 
and this cancellation always takes place.%
}
\footnote{
\vrev{
The exact cancellation of the contributions of the Goldstone boson and the scalar component of the weak boson 
($\del^\mu V_\mu$) in the scattering amplitudes agrees with the quartet decoupling mechanism of the unitarity of covariantly quantized gauge theories~\cite{Kugo:1977zq,*Kugo:1977yx,*Kugo:1977mk,*Kugo:1977mm}, 
where a quartet of the Faddeev--Popov ghost and its anti-ghost~\cite{Faddeev:1967fc}, $\del^\mu V_\mu$ and its associated Goldstone $\pi_V$, which have the common $\xi$-dependent mass,
can appear only as a zero-norm state in the scattering amplitudes. 
In the tree level, it tells that  $\del^\mu V_\mu$ and $\pi_V$ contributions add up to zero,
since ghosts do not contribute.
}
} 
}

\vrev{
Although it may seem that the Goldstone boson propagates inside our amplitudes, 
none of them are the Goldstone bosons of the $R_\xi$ gauge.
We can verify this by keeping the $\xi$ factor in the mass term of the Goldstone-boson and the weak-boson propagators of the $R_\xi$ gauge, Eqs.~\eqref{D5_Rxi} and \eqref{Dmn_Rxi}, respectively.
The two contributions cancel out exactly in the scattering amplitudes,
and 
only the unitary-gauge part of the $R_\xi$-gauge propagators,
the first term in Eq.~\eqref{Dmn_Rxi}, survives.
It is the scalar component of this remaining 
unitary-gauge polarization tensor, Eq.~\eqref{P_unitary_tld}, 
which interacts with the Goldstone-boson vertices by the BRST identities, Eqs.~\eqref{brst_amp1} and \eqref{brst_amp2}.
The propagating degrees of freedom are just that of the unitary-gauge weak bosons, Eq.~\eqref{M_unitary}.
}

\vrev{
We can therefore arrive at the same representation of the amplitudes by starting from the $R_\xi$ gauge with finite $\xi$, and replacing all the pair of the gauge-boson and its associated Goldstone-boson propagators by our $5 \times 5$ component propagators. 
The derivation of our amplitudes given in this section tells that the resulting amplitudes should agree with the one which we obtain from the unitary-gauge expression, with the additional four types of vertices, given in Fig.~\ref{fig:diagram_GB}.
}
\vrev{
In the SM, we identify the four vertices of Fig.~\ref{fig:diagram_GB} which do not have weak-boson counterparts in the unitary gauge.
In models beyond the SM,
we should start from a new set of Feynman rules which treat the Goldstone bosons as the 5th component of the weak bosons.
In the SM effective field theory (SMEFT), the Goldstone bosons can have additional interactions in higher dimensional operators,
whereas in general models with spontaneously broken gauge symmetry,
the Goldstone bosons of the SM and additional weak bosons can have components in 
additional Higgs multiplets.
The Feynman diagrams with the Goldstone-boson couplings are then generated automatically 
as the vertices among the fifth component of the weak bosons. 
}

\section{Sample results}\label{sec:results}

In order to demonstrate the power of the 5-component description of weak bosons
in numerical calculation of the amplitudes,
we study several high-energy EW scattering processes.
In the following, we denote ``New HELAS" as the 5-component weak-boson description introduced in this paper, and ``HELAS" as the original \HELAS~\cite{Hagiwara:1990dw,*Murayama:1992gi}, in which the unitary-gauge and Feynman-gauge propagators are employed for massive and massless gauge bosons, respectively, 
and compare the two results.

\subsection{Weak boson scattering processes}\label{sec:wwww}

\begin{figure}
  \center
\includegraphics[height=0.21\columnwidth]{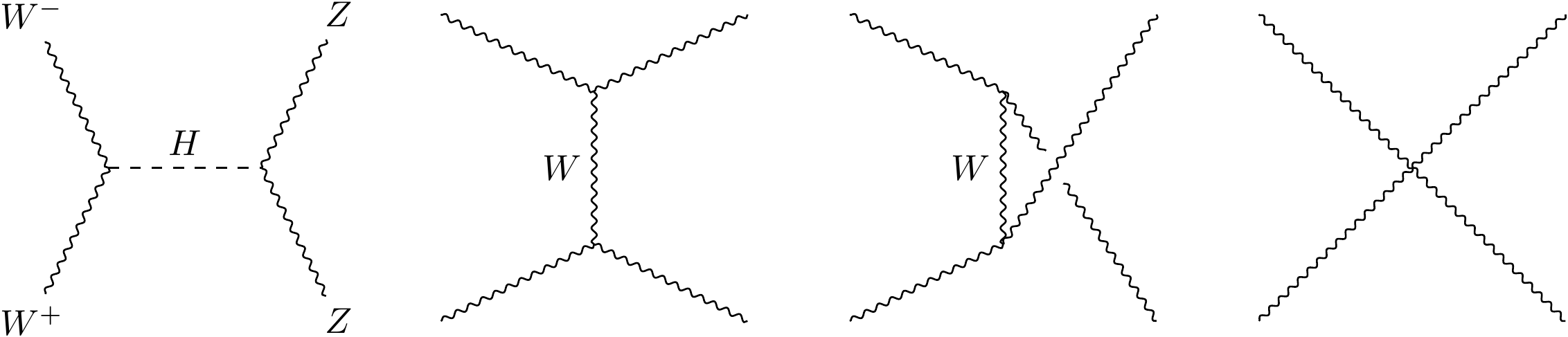}
\hspace*{0.2cm}(a)\hspace*{1.8cm}(b)\hspace*{1.8cm}(c)\hspace*{1.8cm}(d)
\caption{Feynman diagrams for $W^-W^+\to ZZ$.}
\label{fig:diagram_wwzz}
\end{figure}

\begin{figure}
  \center
\includegraphics[height=0.21\columnwidth]{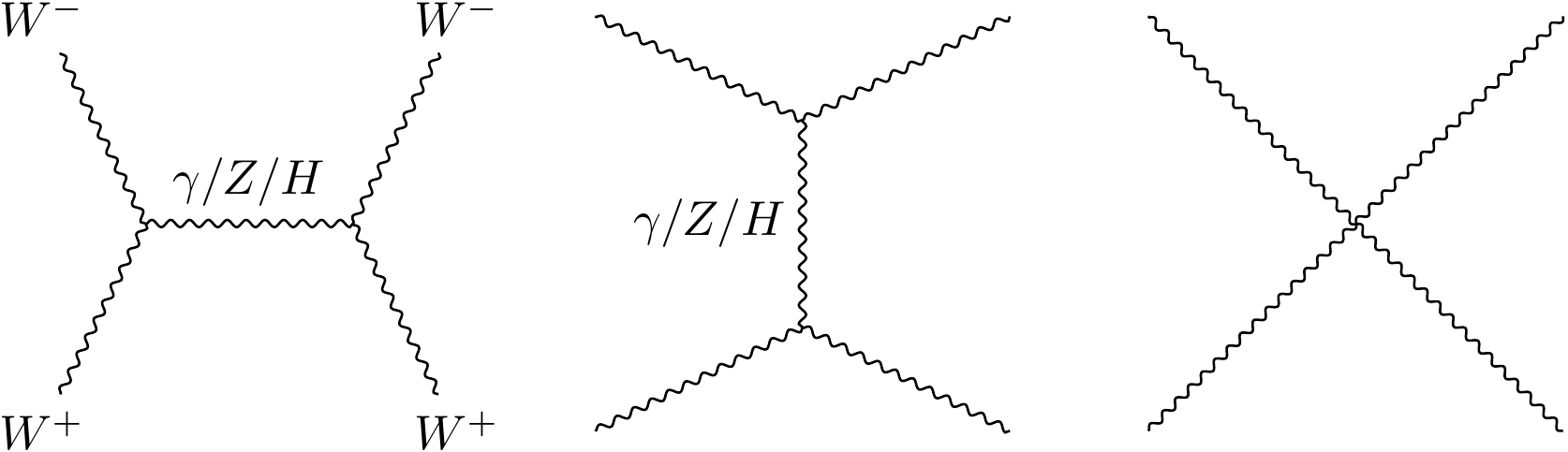}
(a)\hspace*{1.8cm}(b)\hspace*{1.8cm}(c)
\caption{Feynman diagrams for $W^-W^+\to W^-W^+$.}
\label{fig:diagram_wwww}
\end{figure}

\begin{figure}
  \center
\includegraphics[height=0.21\columnwidth]{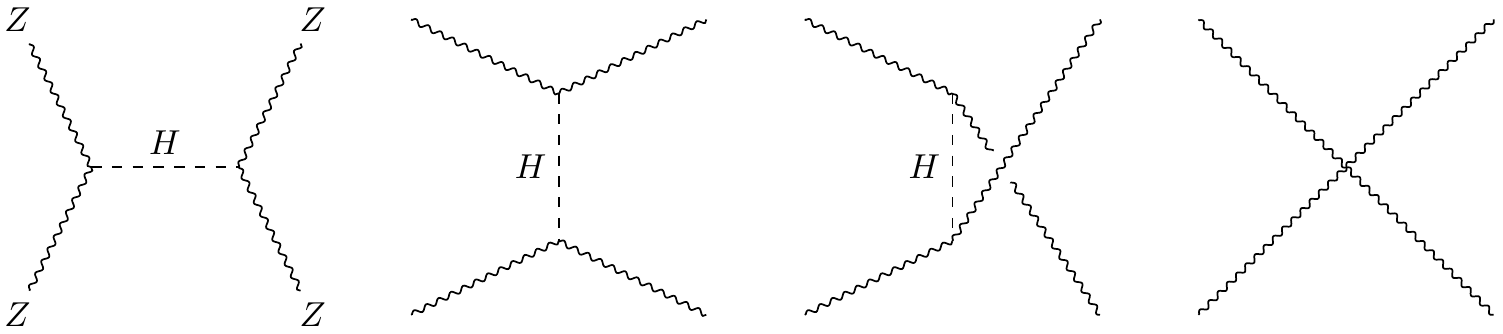}
\hspace*{0.1cm}(a)\hspace*{1.8cm}(b)\hspace*{1.8cm}(c)\hspace*{1.8cm}(d)
\caption{Feynman diagrams for $ZZ\to ZZ$.}
\label{fig:diagram_zzzz}
\end{figure}

First, we study weak boson scattering processes:
\begin{align}
 & W^-W^+\to ZZ, \\
 & W^-W^+\to W^-W^+, \\
 & ZZ\to ZZ,
\end{align}  
whose Feynman diagrams are shown 
in Figs.~\ref{fig:diagram_wwzz}, \ref{fig:diagram_wwww} and \ref{fig:diagram_zzzz},
respectively.
As discussed in Sect.~\ref{sec:mg5}, the 4-point Z-boson diagram, Fig.~\ref{fig:diagram_zzzz}(d), 
has been introduced to account for the Goldstone boson amplitudes, Fig.~\ref{fig:diagram_GB}(a) and \ref{fig:diagram_GB}(b),
via the fifth component of the Z boson wavefunction.

\begin{figure}
\center
\includegraphics[width=1\columnwidth]{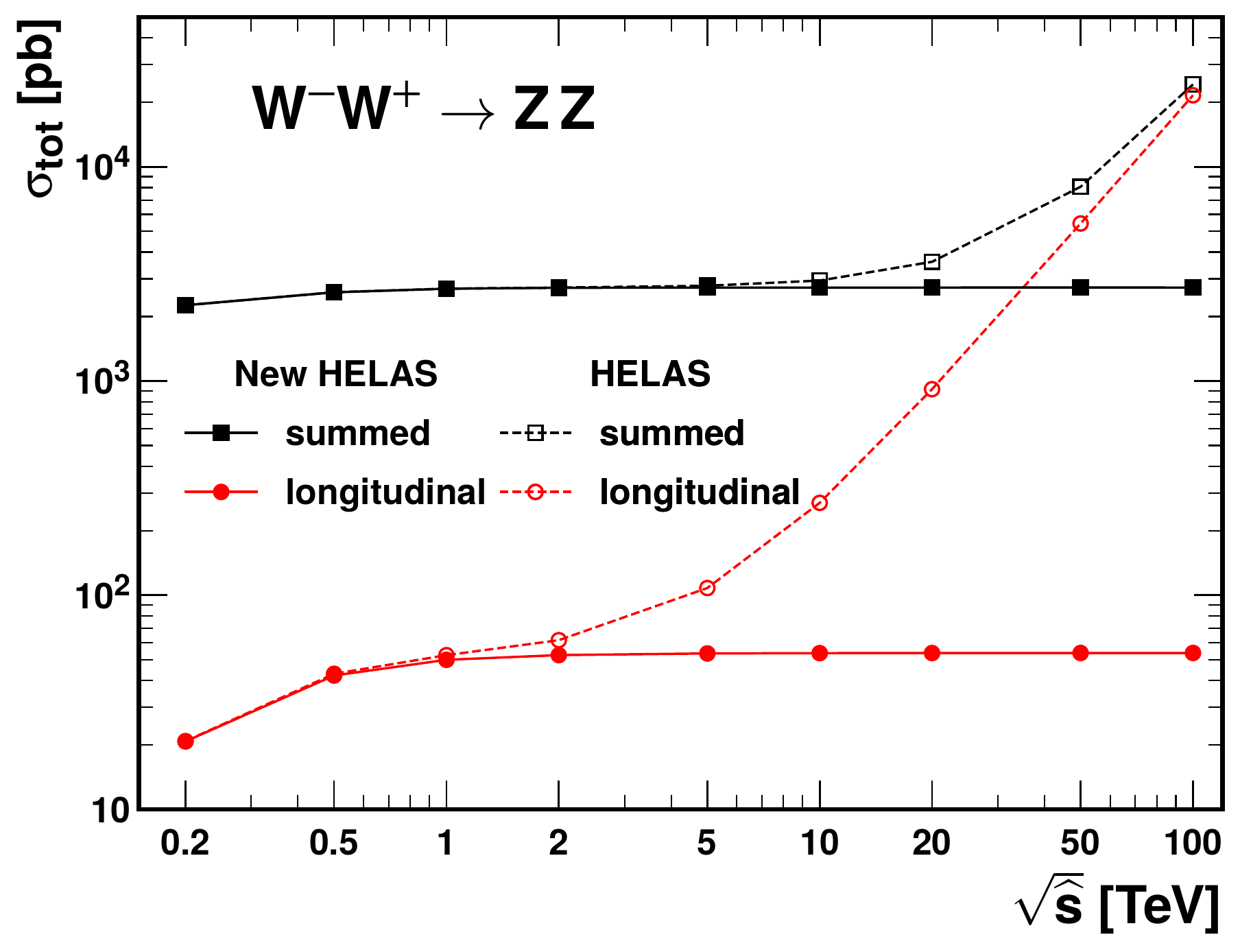}
\caption{
Historical plot of the  
total cross sections for $W^-W^+\to ZZ$ 
as a function of the colliding $WW$ energy,
when W-boson propagators exchanged in the $t/u$-channels are given the physical width.
Solid lines show the cross sections computed by the new \helas subroutines,
while dashed ones are by the original {\tt HELAS}.
Lines with squares (circles) denote helicity-summed (only longitudinally polarized)
cross sections.
}
\label{fig:WWZZ_tot}
\end{figure}

Figure~\ref{fig:WWZZ_tot} shows the 
total cross section of $W^-W^+\to ZZ$ 
as a function of the total energy $\sqrt{\hat s}$ 
from 200~GeV up to 100~TeV.
Solid lines show the cross sections computed by the new \helas subroutines,
while dashed ones are by the original {\tt HELAS}.
Lines with black squares denote helicity-summed cross sections,
while those with red circles represent cross sections among longitudinally polarized weak bosons.
They are both calculated with the W-boson propagator with its physical width as $D_V(q^2)=(q^2-m_V^2+im_V\Gamma_V)^{-1}$, 
since it was a default setting of {\tt MadGraph} until recently.%
\footnote{The weak boson widths in the space-like propagators are automatically set to zero in {\tt MG5} v2.8.0 or later~\cite{Alwall:2014hca}.}

It is well known that the width of the weak bosons spoils gauge cancellation.
Although the uncanceled terms are of higher order in the EW perturbation theory,
when the amplitudes are calculated in the unitary gauge, 
violation of subtle cancellation among individual Feynman amplitudes spoils the accuracy of the cross section at high energies.
On the other hand, the Feynman amplitudes calculated with new {\tt HELAS} 
do not have subtle gauge cancellation at high energies,
and the weak boson width effect, whether it is needed or not, remains small at all energies.

We confirmed that, if we set $\Gamma_V=0$ in the $t/u$-channel propagators of the weak bosons in the original {\tt HELAS} computation, the total cross sections recover the perturbative unitarity and agree with the new \helas results.
In the following studies, we always set the width of weak bosons to zero in the $t/u$-channel propagators,
\vrev{
while we should introduce the width term to unitarize the amplitudes near the weak-boson mass shell,
as we study in Sect.~\ref{sec:eeww}.%
\footnote{
\vrev{
Issues on gauge invariance of the amplitudes around the weak-boson mass-square pole have been studied 
in Ref.~\cite{Cuomo:2019siu}.
}
}
}

\begin{figure*}
\center
 \includegraphics[width=0.825\textwidth]{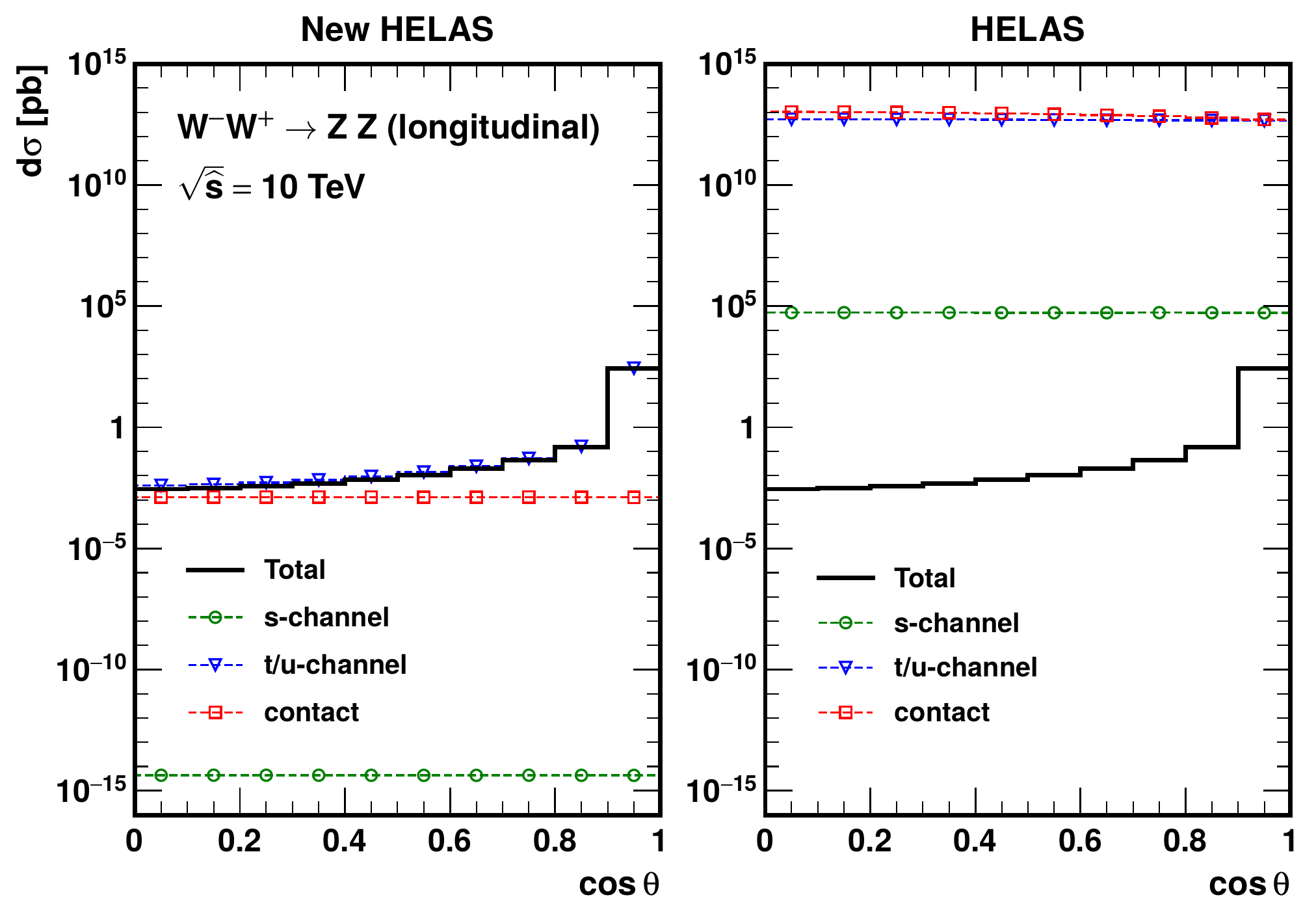}
\caption{
Distributions of the scattering angle for $W^-W^+\to ZZ$ at $\sqrt{\hat s}=10$~TeV
by using the new (original) \helas in left (right) panel, 
where all the helicities for the external bosons are taken to be longitudinal.
A solid line denotes the total distribution, 
while dashed lines with circles, triangles and squares show the distributions of 
the amplitude squared of the $s$-channel, $t/u$-channel and the contact diagrams, respectively.
}
\label{fig:WWZZ_dif}
\end{figure*}

\begin{figure*}
\center
 \includegraphics[width=0.825\textwidth]{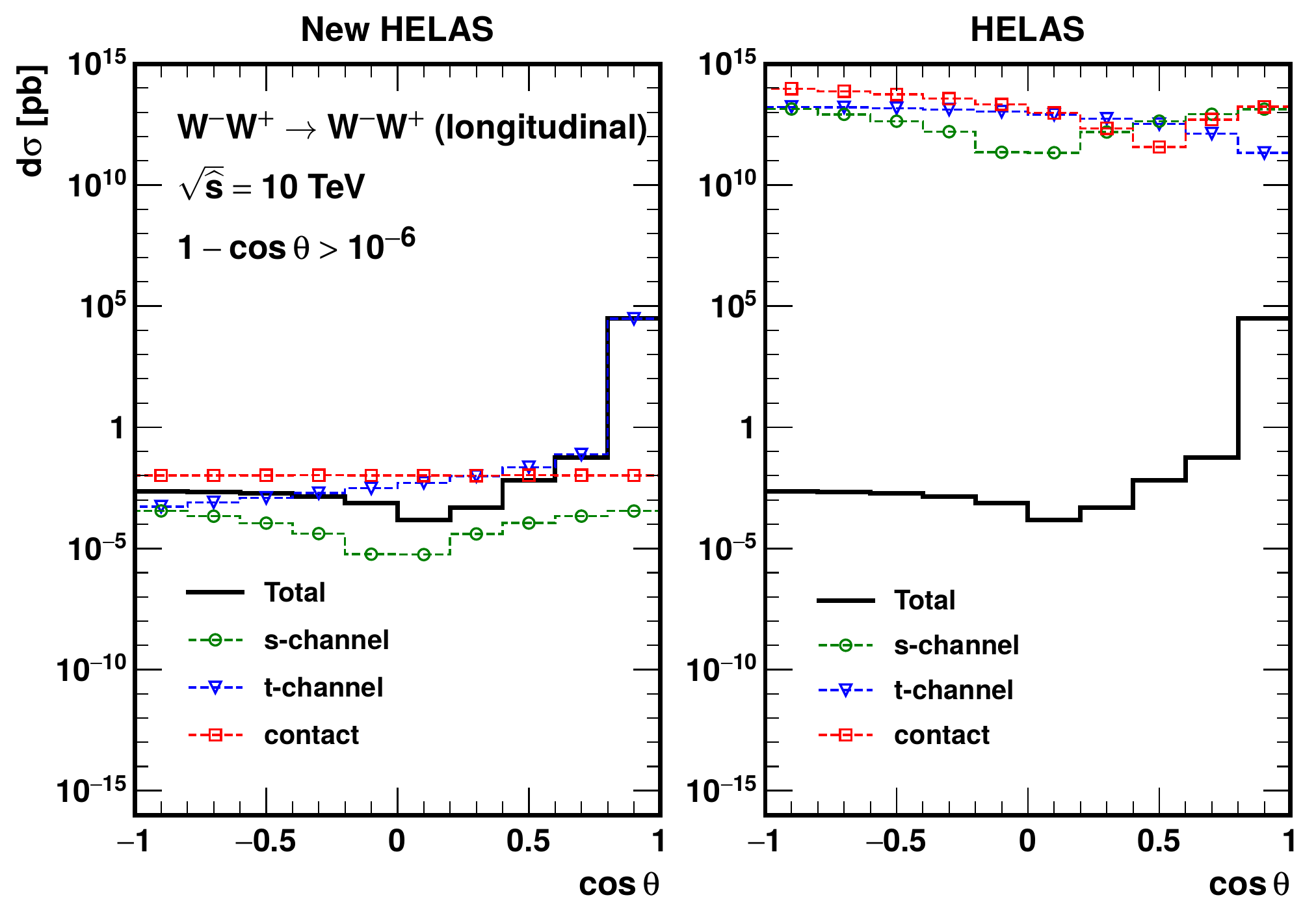}
\caption{
Same as Fig.~\ref{fig:WWZZ_dif}, but for $W^-W^+\to W^-W^+$. 
}
\label{fig:WWWW_dif}
\end{figure*}

\begin{figure*}
\center
 \includegraphics[width=0.825\textwidth]{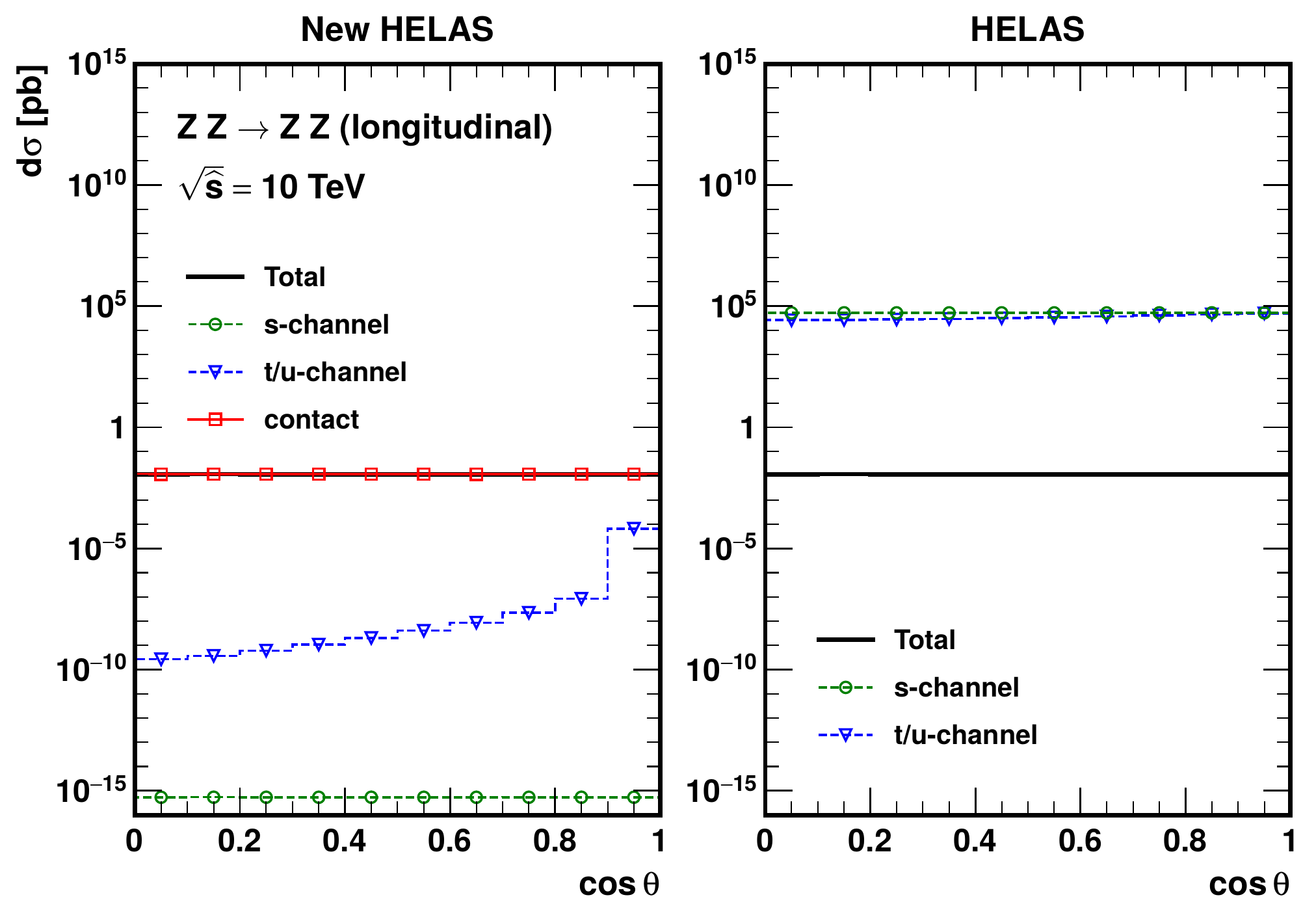}
\caption{
Same as Fig.~\ref{fig:WWZZ_dif}, but for $ZZ\to ZZ$.
}
\label{fig:ZZZZ_dif}
\end{figure*}

In Fig.~\ref{fig:WWZZ_dif} we show 
the scattering angle distribution for $W^-W^+\to ZZ$ at $\sqrt{\hat s}=10$~TeV
by new {\tt HELAS} (left) and by original {\tt HELAS} (right), 
where all the helicities for the external weak bosons are taken to be longitudinal.
A solid line denotes the total differential distribution, 
while dashed lines with circles, triangles and squares show the distributions of 
the amplitude squared of the $s$-channel, $t/u$-channel and the contact diagrams, 
in Fig.~\ref{fig:diagram_wwzz}(a), (b), (c) and (d), respectively.

The physical distributions, calculated from the absolute value square of the total sum of all the Feynman amplitudes,
are identical between the new and original \helas calculations as they should be.
However, the contribution from each amplitude is quite different between the two calculations.
While there is no energy growth of each amplitude at all in the new \HELAS, 
one can observe about ${\cal O}(10^{15})$ cancellation among the amplitudes in the original \HELAS,
which is computationally expensive to obtain physical cross sections accurately at very high energies. 

In the new \helas amplitudes with the 5-component weak-boson description, 
no term which grows with energy ($|q^0|\gg Q$) appears, 
and the absolute value of each Feynman amplitude is expressed as a product of the Lorentz invariant propagator factor
$D_V(q^2)=(q^2-m_V^2)^{-1}$.
As expected, the contributions from the $t/u$-channel amplitudes are dominant
especially for the forward region, while the ones from the contact amplitude are constant and significant only in the central region.
We find that the contribution from the $s$-channel Higgs boson exchange amplitude is completely negligible at this energy.

In contrast, the original \helas amplitudes given in the right panel of Fig.~\ref{fig:WWZZ_dif}
show that the square of $t/u$-channel exchange amplitudes (blue triangles) and the contact term (red squares) are
both almost flat in $\cos\theta$, with 10 to 15 orders of 
magnitude larger than the physical distribution (black solid).

Likewise, in Fig.~\ref{fig:WWWW_dif} we show the angular distributions for $W^-W^+\to W^-W^+$.
In the new \helas results given in the left panel,
the $t$-channel $\gamma$ and $Z$ exchange amplitudes dominate the cross section
at $\cos\theta\sim1$, 
while we can study the interference pattern among the different amplitudes quantitatively
in the central ($\cos\theta\sim0$) and the backward ($\cos\theta<0$) regions.
On the other hand, each amplitude in the unitary gauge,
i.e. in the original \helas calculation,
apparently does not give any useful information 
about physical properties of the scattering process. 
As text books describe, only after summing all the amplitudes,
the energy-growing terms are cancelled by the gauge symmetry, 
and then the physically observable distributions given by black solid curves 
are the same between new \helas (left panel) and the original \helas in the unitary gauge (right panel). 

Figure~\ref{fig:ZZZZ_dif} shows the $\cos\theta$ distribution for $ZZ\to ZZ$.
The physical distribution is flat in both new \helas (left) and original \helas (right).
The origin of the flat distribution is the dominance of the contact interaction term (red squares)
from the Goldstone boson coupling in Fig.~\ref{fig:diagram_GB}(a) or Fig.~\ref{fig:diagram_zzzz}(d) in new \helas results (left panel),
where the $t/u$-channel exchange contribution is subdominant at this energy
($\sqrt{\hat s}=10$~TeV). 

On the other hand, in the unitary gauge with the original \helas (right panel),
the same constant distribution is obtained as a consequence of subtle cancellation 
between the $s$-channel (green circles) and $t/u$-channel (blue triangles) Higgs exchange amplitudes,
each giving three orders of magnitude larger absolute values.

\subsection{W-boson pair production at lepton colliders}\label{sec:eeww}

\begin{figure}
 \center
\includegraphics[height=0.35\columnwidth]{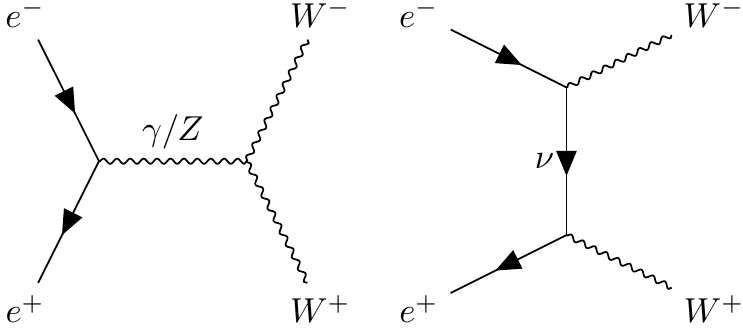}
(a)\hspace*{3.cm}(b)
\caption{Feynman diagrams for $e^-e^+\to W^-W^+$.}
\label{fig:diagram_eeww}
\end{figure}

Next, we study W-boson pair production in lepton collisions,
whose Feynman diagrams are depicted in Fig.~\ref{fig:diagram_eeww}.

\begin{figure*}
\center
\includegraphics[width=0.4\textwidth]{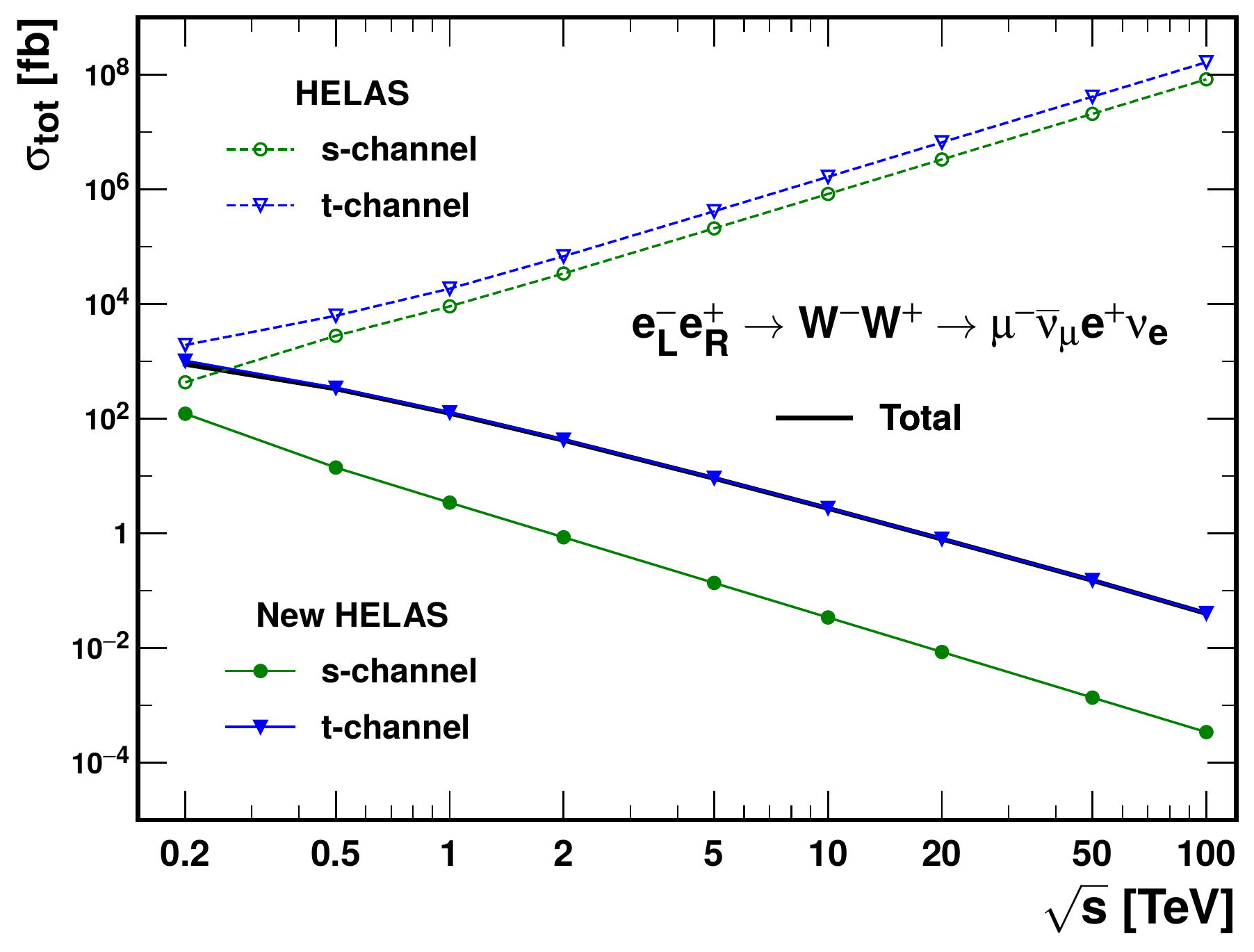}\qquad
\includegraphics[width=0.4\textwidth]{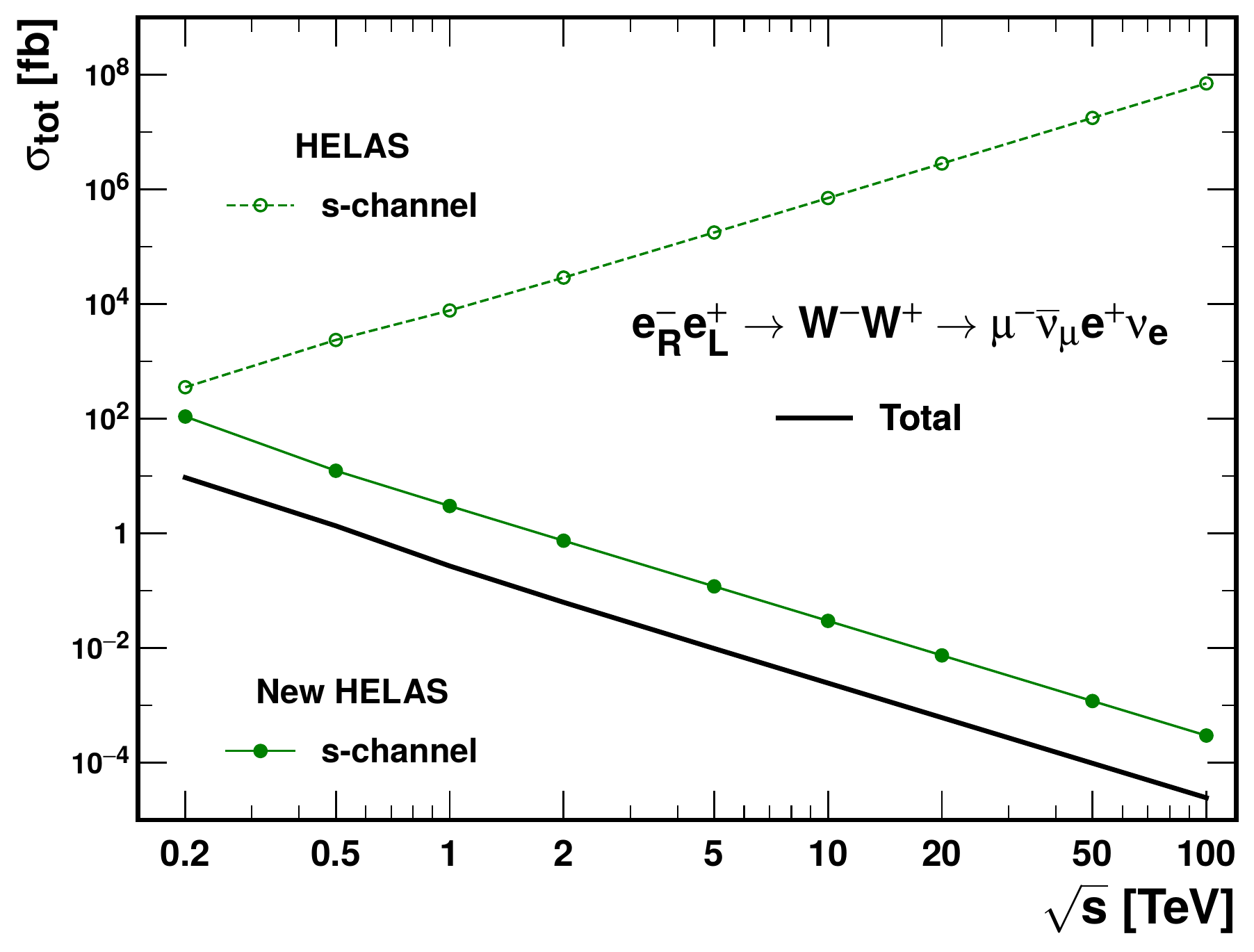}
\caption{
Total cross section for $e^-e^+\to W^-W^+\to\mu^-\bar\nu_\mu e^+\nu_e$ 
with the left-handed (right-handed) electron and the right-handed (left-handed) positron 
as a function of the collision energy in the left (right) panel.
A black solid line shows the total cross section, while lines with symbols denote contributions from the amplitude squared of each diagram. 
Solid lines are the results computed by the new \helas subroutines,
while dashed ones are by the original \HELAS.
}
\label{fig:eeww_tot}
\end{figure*}

Figure~\ref{fig:eeww_tot} shows 
the total cross section for $e^-e^+\to W^-W^+$
including the subsequent leptonic decays as a function of the collision energy. 
In the left (right) panel, the collision with
the left-handed (right-handed) electron and the right-handed (left-handed) positron is given. 

\begin{figure*}
\center
\includegraphics[width=0.485\textwidth]{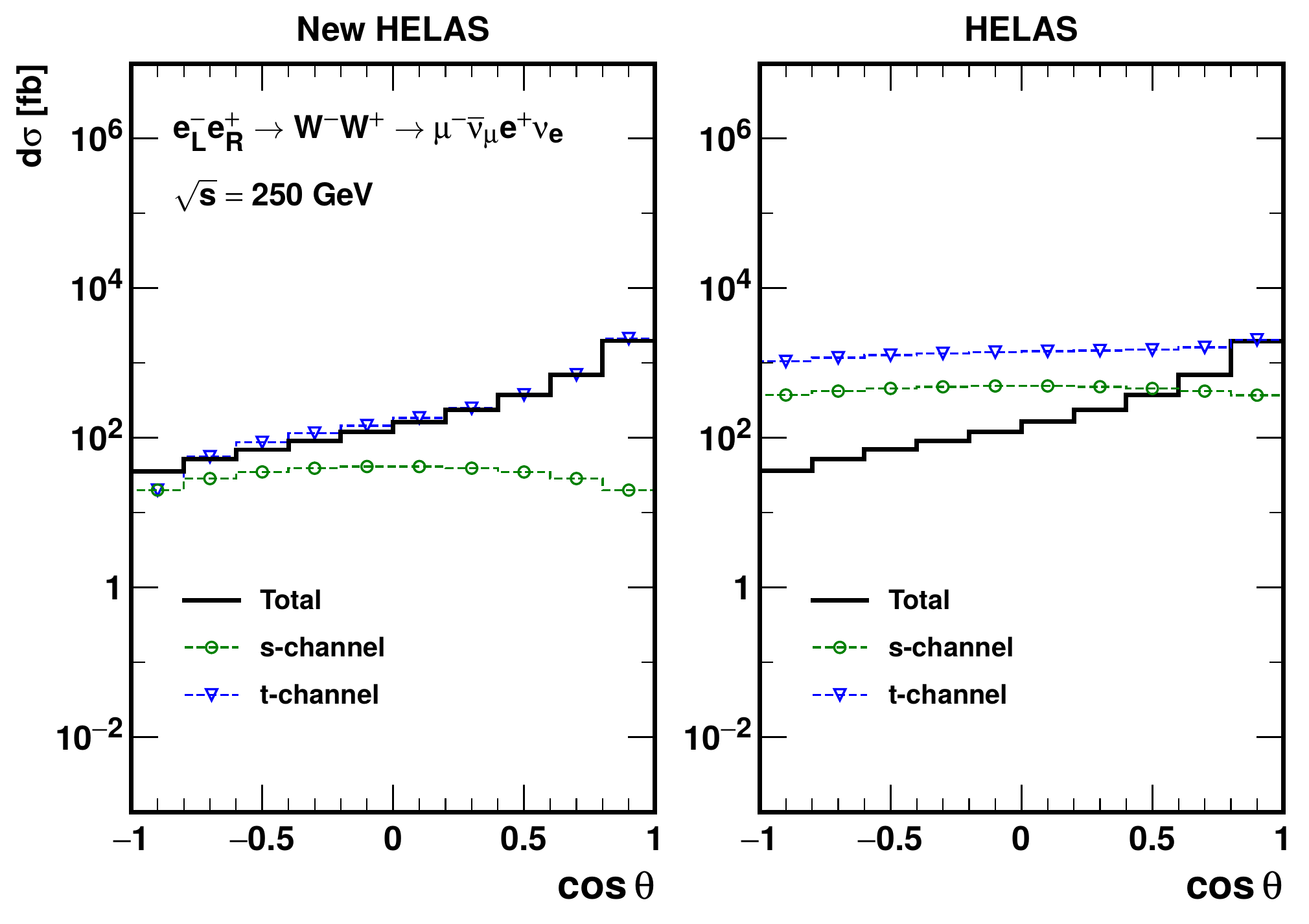}\quad
\includegraphics[width=0.485\textwidth]{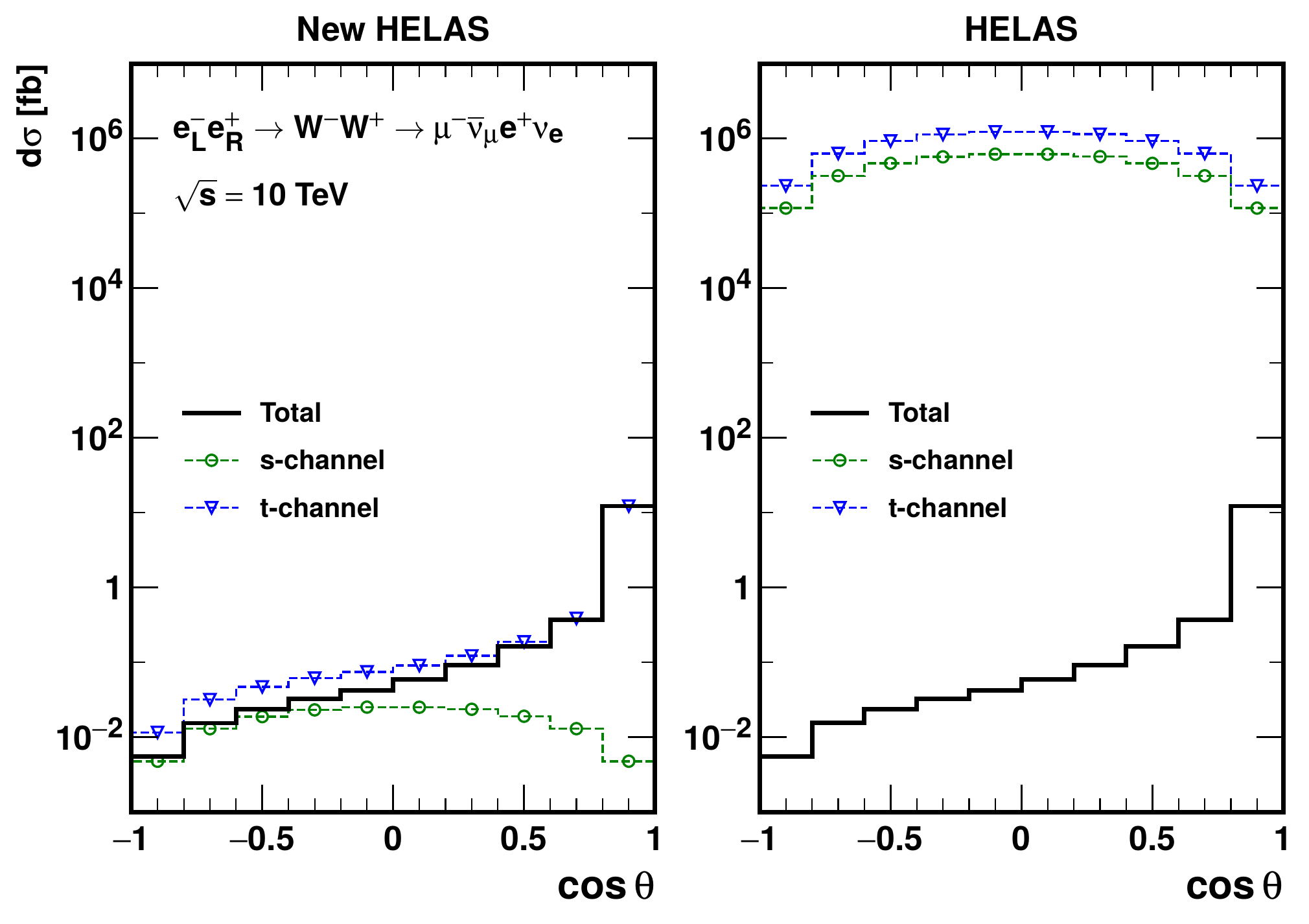}
\hspace*{0.75cm}(a)\hspace*{8.75cm}(b)
\caption{
Distributions of the scattering angle of $W^-$ for $e^-e^+\to W^-W^+\to\mu^-\bar\nu_\mu e^+\nu_e$ 
with the left-handed electron and the right-handed positron
by the new and original \helas at $\sqrt{s}=250$~GeV (a) and 10~TeV (b).
A solid line denotes the total distribution, 
while dashed lines with circles and triangles show the distributions of 
the squared amplitudes of the $s$-channel and $t$-channel diagrams, respectively.
}
\label{fig:eeww_dif}
\end{figure*}

\begin{figure*}
\center
\includegraphics[width=0.485\textwidth]{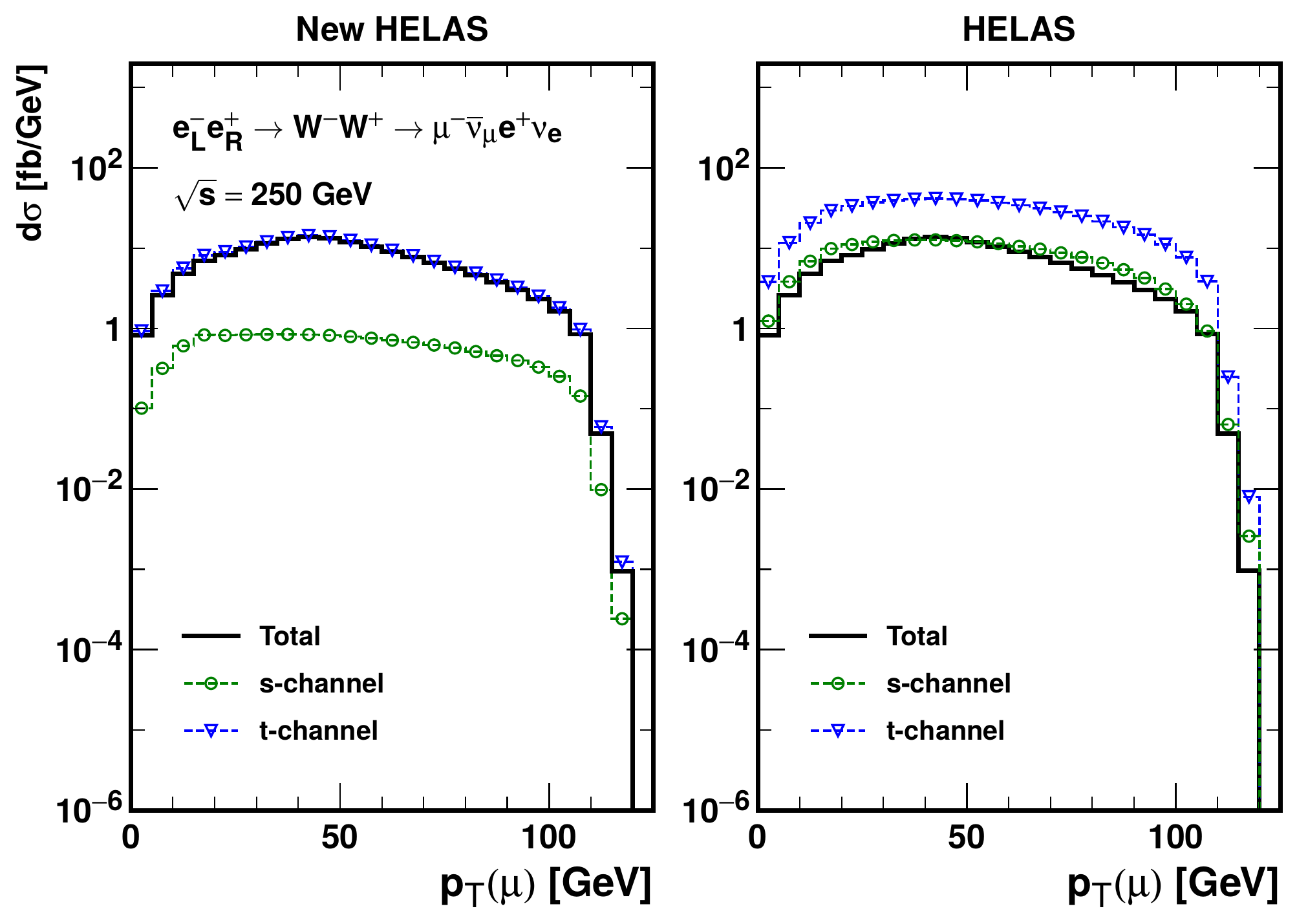}\quad
\includegraphics[width=0.485\textwidth]{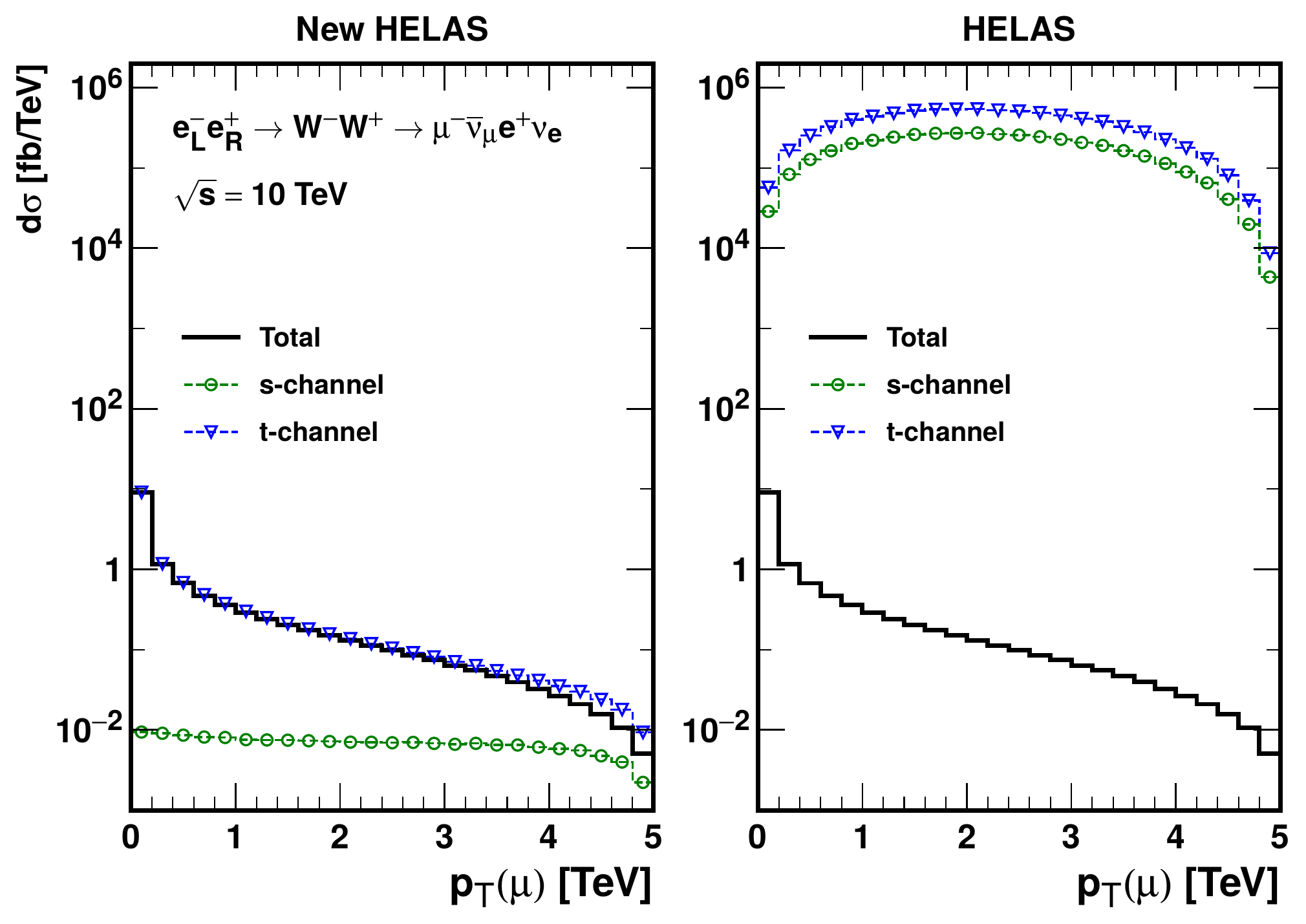}
\hspace*{0.75cm}(a)\hspace*{8.75cm}(b)
\caption{
Same as Fig.~\ref{fig:eeww_dif}, but for $p_T(\mu)$ distributions.
}
\label{fig:eeww_pt}
\end{figure*}

The total cross section, denoted by a black solid line, falls as $1/s$ and 
agree between the two calculations as expected.
However, the contributions from each amplitude are completely different between 
new \helas (solid curves) and the original unitary-gauge \helas (dashed curves). 

In the 5-component description with new \HELAS,
we find that the $t$-channel neutrino-exchange amplitude is dominant for the left-handed electron beam at all the energies. 
For the right-handed electron shown in the right panel, on the other hand, we can identify the interference between the $s$-channel photon and Z-boson amplitudes.
The sum of the square of the photon and Z-boson exchange diagrams, 
as shown by the green solid curves, 
is about ten times larger than the observable cross section, 
given by the black solid curve, at all energies.
This has a clear physics interpretation.
At energies high above the Z-boson mass, only the U(1) gauge boson $B^\mu$ couples to the right-handed electron.
$B^\mu$ deos not couple to the $W$'s because the $W$ hypercharge is zero.
It is only the Goldstone boson component of the $W$, $\pi^\pm$, which couples to the $B^\mu$ boson
since they have hypercharge.
For the individual $\gamma$ and $Z$ exchange amplitudes, both the SU(2)$_L$ triplet $W^\pm$ and the doublet $\pi^\pm$
contributes, 
while in the sum, for the $e_R^-e_L^+$ annihilation, the $W^\pm$ components cancel out at high energies.
The produced $\pi^\pm$ decays via longitudinal $W$ mixing into a lepton pair.

On the other hand, the unitary-gauge amplitudes with original \helas give 
contribution of each Feynman diagram which grows with energy, 
as shown by dashed curves in both left ($e_L^-$) and right ($e_R^-$) plots in Fig.~\ref{fig:eeww_tot}.
At $\sqrt{s}=100$~TeV, the cancellation in the amplitudes give $10^{-9}$ and $10^{-12}$ times the individual contribution for $e_L^-$ and $e_R^-$, respectively.
This `fake' cancellation between the $s$-channel $\gamma$ and $Z$ exchange amplitudes
(with $\gamma WW$ and $ZWW$ couplings) and the $t$-channel $\nu$ exchange amplitudes
gave us a wrong hope that we could have a better sensitivity to the triple gauge boson coupling
at high energies.
In order to overcome such unjustified expectation, 
gauge invariant operator formalism which may be a prototype of SMEFT was introduced
in Ref.~\cite{Hagiwara:1986vm}.
With new \helas amplitudes, 
we can tell that the magnitudes of the $s$-channel $\gamma$ and $Z$ exchange amplitudes
(given by green solid curves in Fig.~\ref{fig:eeww_tot}) are essentially the same 
for $e_L^-$ (left panel) and $e_R^-$ (right panel).
Because of their interference with the dominant $\nu$ exchange amplitude
(whose square is given by a blue solid curve with triangles), 
$e_L^-$ annihilation process is more sensitive to the triple weak boson couplings. 

In Figs.~\ref{fig:eeww_dif} and \ref{fig:eeww_pt}, 
we present differential distributions for $e^-_Le^+_R\to W^-W^+\to\mu^-\bar\nu_\mu e^+\nu_e$ 
at $\sqrt{s}=250$~GeV in the left two panels (a) and at 10~TeV in the right two panels (b).
Fig.~\ref{fig:eeww_dif} gives $\cos\theta$ distribution of $W^-$, 
whereas Fig.~\ref{fig:eeww_pt} gives $p_T(\mu)$ distribution in the lab
(colliding $e^+e^-$ center-of-mass) frame.

For the angular distributions in Fig.~\ref{fig:eeww_dif},
in the unitary gauge by the original \HELAS, 
the cancellation among the amplitudes is still mild 
at $\sqrt{s}=250$~GeV, especially at $\cos\theta\gtrsim 0.5$,
but it becomes very large at $\sqrt{s}=10$~TeV,
where the degree of the cancellation is $10^{-4}$ at $\cos\theta\sim1$, 
and $10^{-7}$ at $\cos\theta\sim-1$.
For the $p_T(\mu)$ distributions shown in Fig.~\ref{fig:eeww_pt}, the unitary-gauge (original \HELAS) amplitudes
suffer from $10^{-1}$ and $10^{-6}$ level cancellation at 250~GeV and 10~TeV, respectively.

In the new \helas calculation, on the other hand, there is no subtle cancellation among the amplitudes.
Moreover, each squared amplitude describes 
\vrev{
can be interpreted as parton distribution of
the corresponding parton-shower history.
For instance in the left-panel of Fig.11(b), we observe that the physical
cross section for $e^-_L e^+_R \to W^- W^+$ is dominated by the $t$-channel
neutrino-exchange contribution in the forward direction, where the collinear
limit is reached at $\cos\theta=1$, whereas at $\cos\theta < 0$, where lower
partial wave contributions dominate, and we observe destructive interference
between the $t$- and $s$-channel amplitudes.
We believe that it is the merit of the new HELAS amplitudes which allows
us to make such qualitative statements on the magnitude and the interference
patterns among interfering Feynman amplitudes.
}

\subsection{Weak boson scattering at lepton colliders}\label{sec:vbf}

\begin{figure*}
\center
 \includegraphics[width=1\columnwidth]{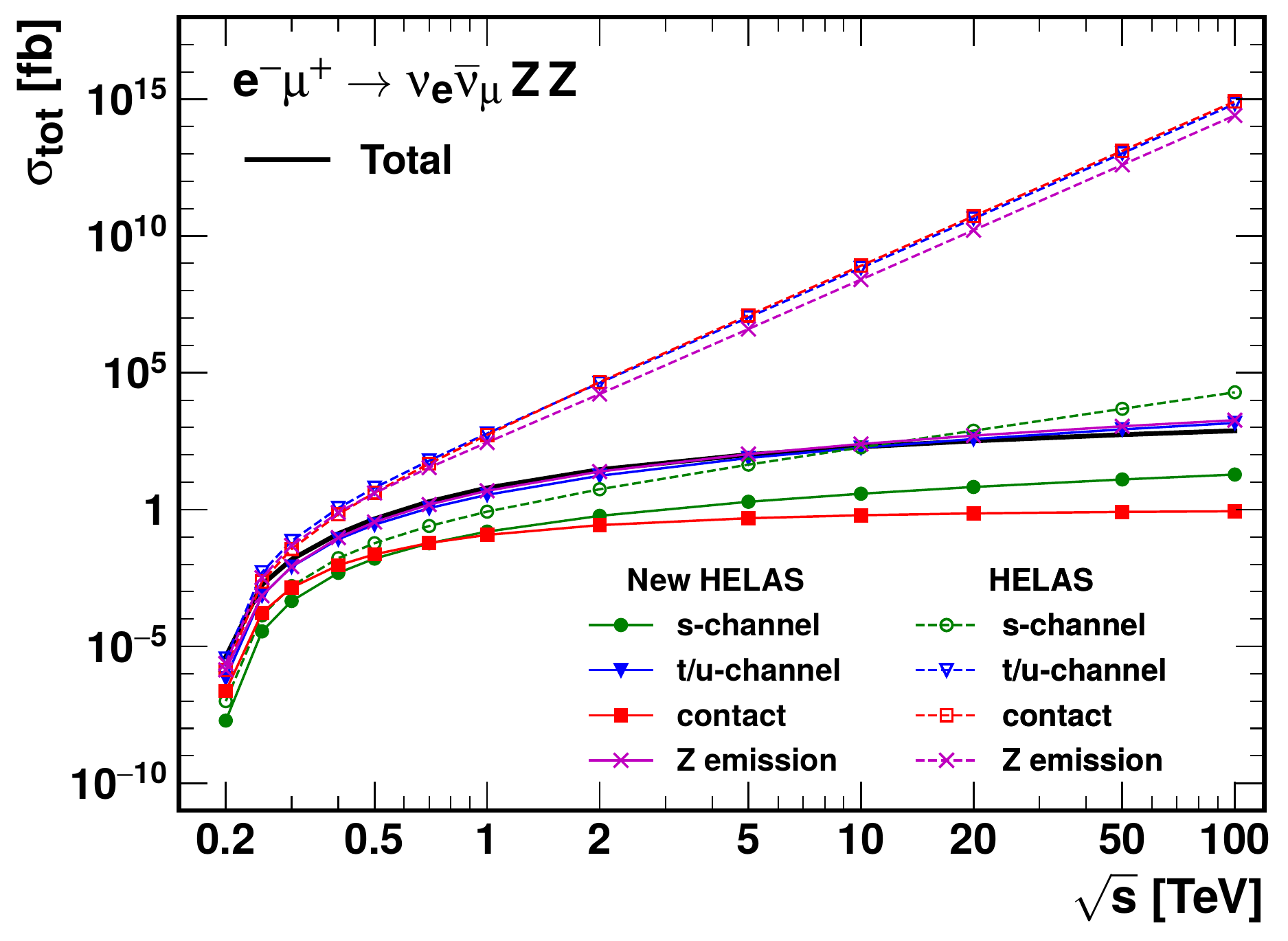}
 \includegraphics[width=1\columnwidth]{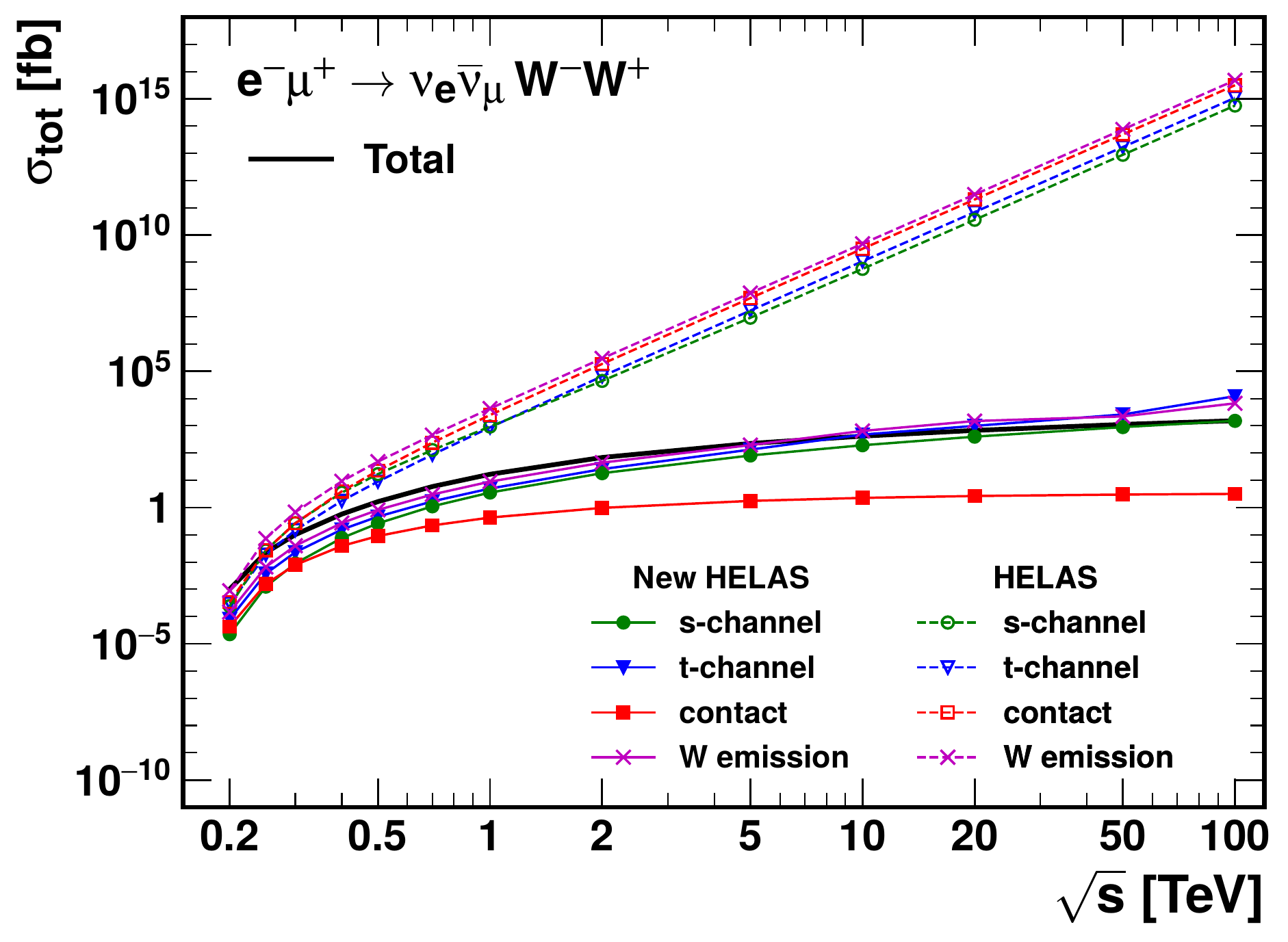}
 \hspace*{1.5cm}(a)\hspace*{8.5cm}(b)\\[0.5mm]
 \includegraphics[width=1\columnwidth]{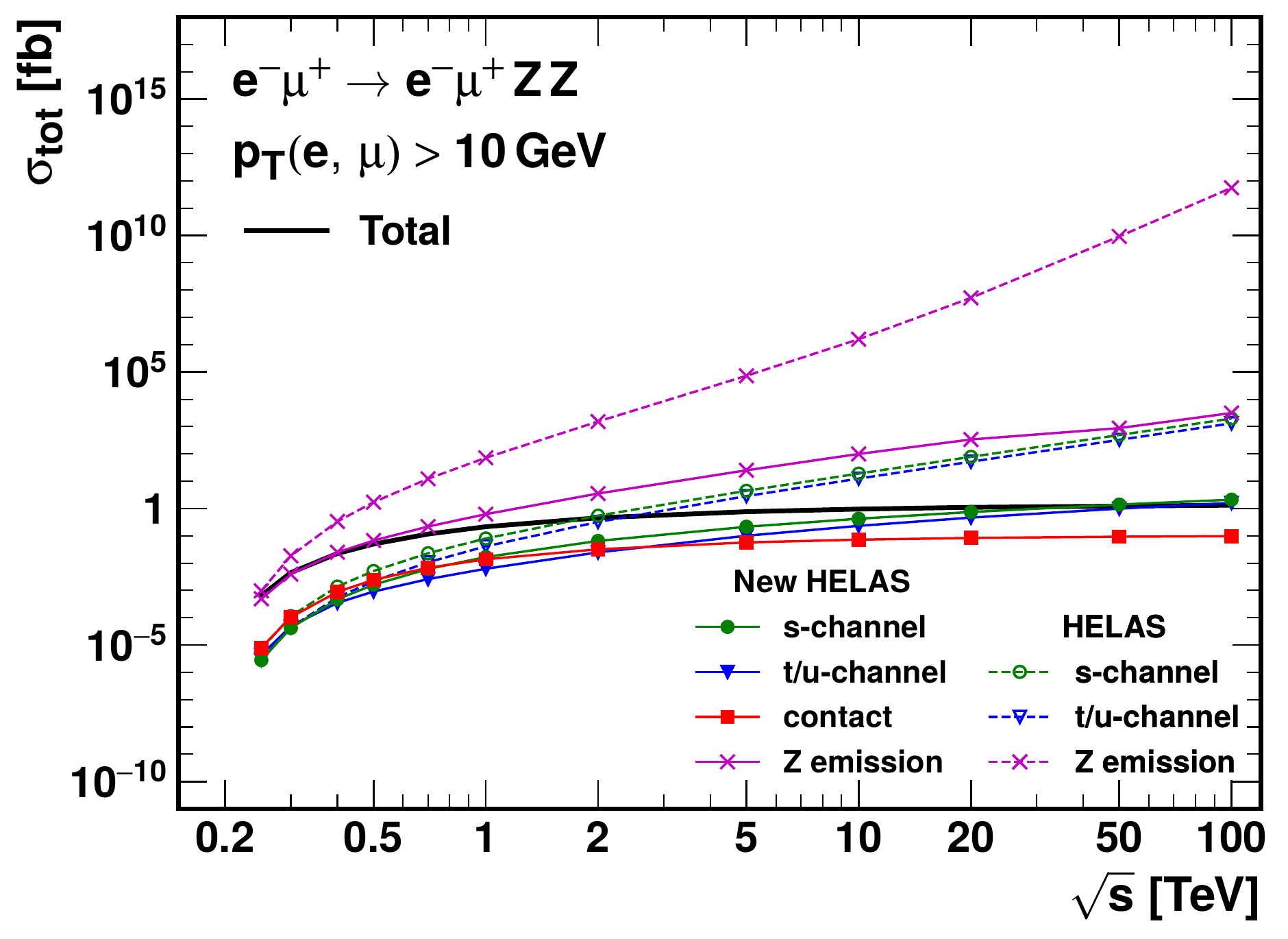}
 \includegraphics[width=1\columnwidth]{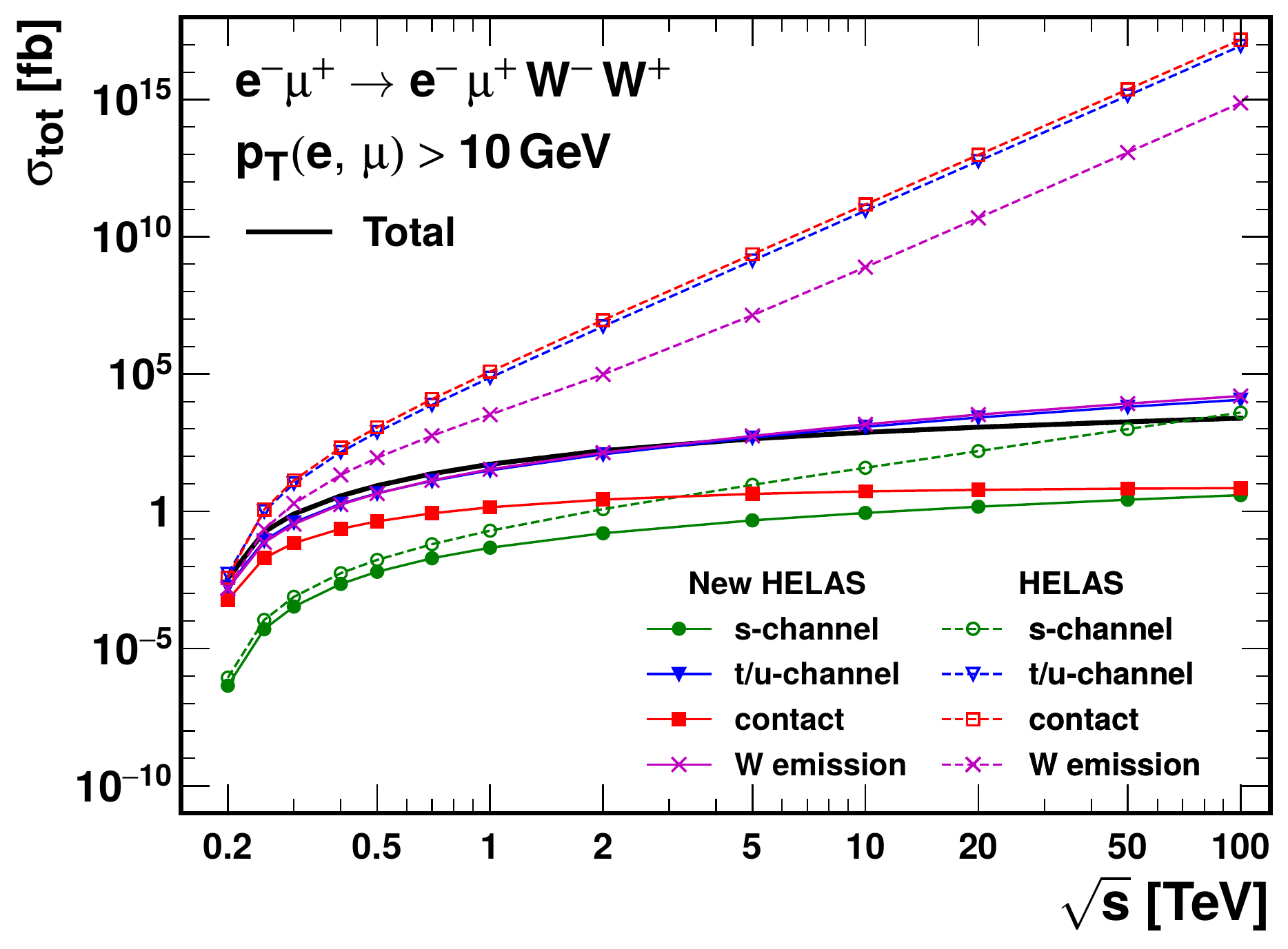} 
  \hspace*{1.5cm}(c)\hspace*{8.5cm}(d)
\caption{
Total cross sections for $l^-l^+\to \nu\bar\nu ZZ$ (a),
$l^-l^+\to \nu\bar\nu W^-W^+$ (b),
$l^-l^+\to l^-l^+ ZZ$ (c), and $l^-l^+\to l^-l^+ W^-W^+$ (d)
as a function of the collision energy.
A Black solid line shows the total cross section, while lines with symbols denote contributions from the amplitude squared of each Feynman diagram in different categories. 
Solid lines show the results computed by the new \HELAS,
while dashed ones are by the original \HELAS.
} 
\label{fig:vbf_tot}
\end{figure*}

Finally, we demonstrate weak boson scattering processes, discussed in Sect.~\ref{sec:wwww},
in a more realistic setup, i.e. in lepton collisions. 
We study
$l^- l^+ \to l' \bar{l}' VV$ ($l=e$ or $\mu$, $l' = l$ or $\nu_l$, $V=W^\pm, Z$),
specifically,
\begin{subequations}
\begin{align}
  e^-\mu^+ &\to \nu_e\bar\nu_\mu ZZ, \label{pa}\\
  e^-\mu^+ &\to \nu_e\bar\nu_\mu W^-W^+, \label{pb} \\
  e^-\mu^+ &\to e^-\mu^+ ZZ, \label{pc}\\
  e^-\mu^+ &\to e^-\mu^+ W^-W^+, \label{pd}  
\end{align}\label{vbfprocesses}%
\end{subequations}
for Z-boson and W-boson pair productions associated with the leptons.
Here, in order to make physics discussions simpler,
we consider $e^-\mu^+$ collisions in order to remove annihilation amplitudes.
While these processes include the following weak boson scatterings: 
\begin{subequations}
\begin{align}
 & W^-W^+\to ZZ, \\
 & W^-W^+\to W^-W^+, \\
 & ZZ\to ZZ,\\
 & ZZ\to W^-W^+,
\end{align} 
\end{subequations} 
respectively,
the majority of the Feynman diagrams consists of 
Z- or W-boson emissions from the external lepton currents.

Figure~\ref{fig:vbf_tot} shows
total cross sections for each process in~\eqref{vbfprocesses}
as a function of the collision energy,
where all the helicities in the initial and final states are summed.
Similar to the previous sub-sections, we separate the contributions 
from each Feynman amplitude and show the sum of each amplitude squared 
by lines with symbols. 
Solid lines show the results computed by the new \HELAS,
while dashed ones are by the original \HELAS.
Since the processes~\eqref{pc} and \eqref{pd} contain the $t$-channel photon exchange diagrams,
we impose $p_T(l)>10$~GeV 
to avoid divergence in the forward region.

The common features for the four processes~\eqref{vbfprocesses} in Fig.~\ref{fig:vbf_tot} are as follows.
In the original \helas computation, each amplitude squared grows as the collision energy is increased,
which results in subtle cancellation among the amplitudes to obtain the physical cross section.
In the new \helas computation, on the other hand, the energy dependence of  
each amplitude squared behaves as the physical cross section, 
and subtle cancellations do not exist.%
\footnote{
In Ref.~\cite{Bailey:2022wqy} the authors report suppression of 
cancellation among the amplitudes for the process $\gamma\gamma\to W^-W^+$ 
in the axial gauge~\cite{Dams:2004vi}.
} 

We note, however, that even in new \helas amplitudes the cancellation can be seen most clearly 
in Fig.~\ref{fig:vbf_tot}(c) for $e^-\mu^+ \to e^-\mu^+ ZZ$~\eqref{pc}. 
Hints of similar cancellation can also be seen in other processes (\ref{vbfprocesses}a,b,d), 
respectively shown in Fig.~\ref{fig:vbf_tot}(a), (b) and (d), at very high energies.
We investigate the origin of the cancellations among new \helas amplitudes below.

\begin{figure}
\center
 \includegraphics[width=1\columnwidth]{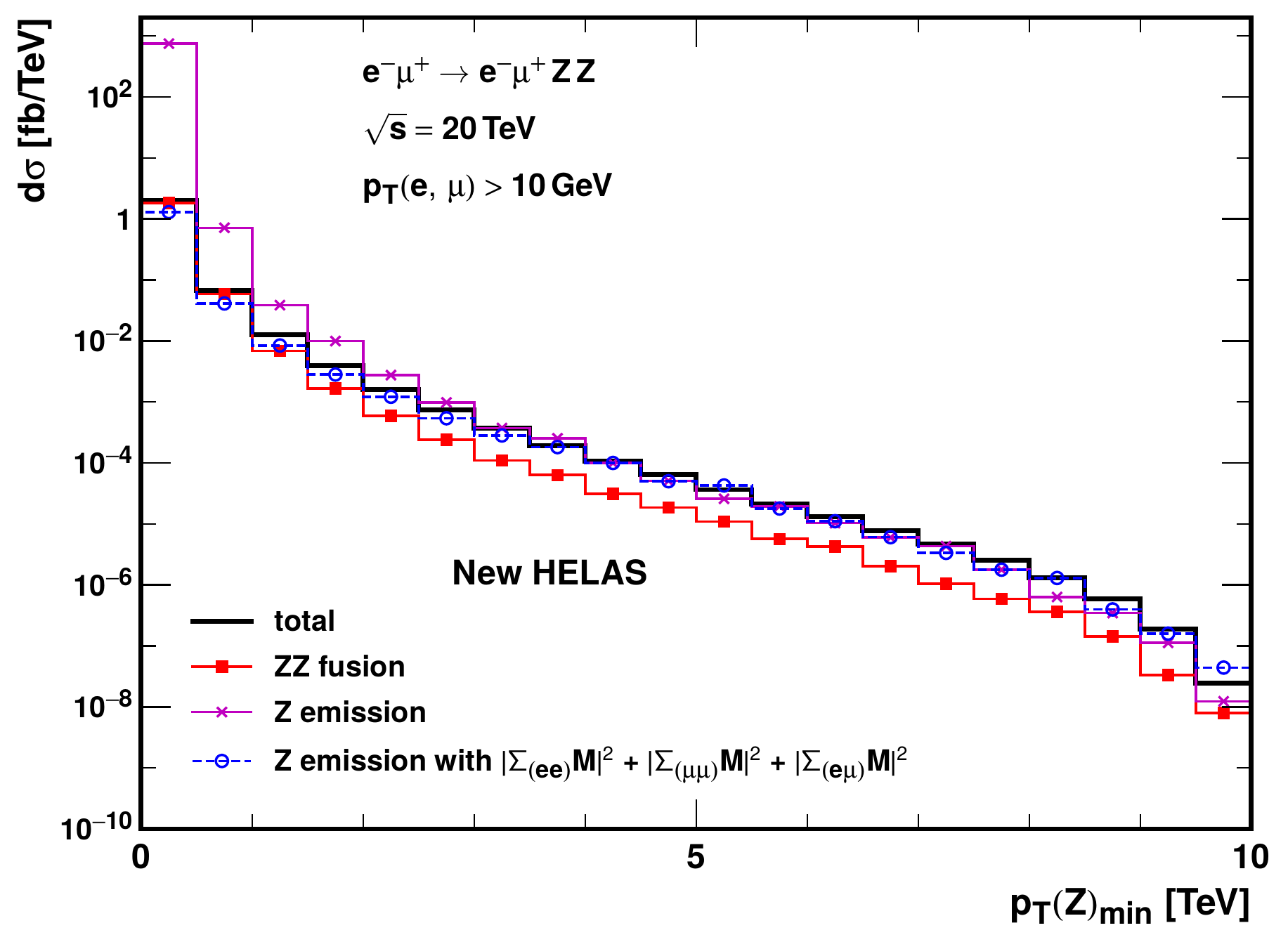}
\caption{
$p_T$ distributions of the Z boson with smaller $p_T$ 
for $e^-\mu^+\to e^-\mu^+ZZ$ at $\sqrt{s}=20$~TeV.
A thick solid line shows the total distribution, while other lines with symbols
are explained in the text.
} 
\label{fig:vbf_dif}
\end{figure}

In Fig.~\ref{fig:vbf_dif} we present $p_T$ distributions of the Z boson with smaller $p_T$ 
for $e^-\mu^+\to e^-\mu^+ZZ$ at $\sqrt{s}=20$~TeV, where $p_T(l)>10$~GeV is imposed 
as in Fig.~\ref{fig:vbf_tot}(c).
The total distribution is shown by a thick solid line. 
A solid line with squares denotes the contribution from the $ZZ\to ZZ$ scattering, 
which is computed by new \helas as 
\begin{align}
 \sum_{k}^{\rm ZZ\, fusion}|{\cal M}_k|^2,
\end{align}
corresponding to the sum of the four amplitude squares, i.e. the $s,t,u$, and
contact amplitudes, shown in Fig.~\ref{fig:diagram_zzzz}. 
A solid line with crosses gives the contributions from the sum of the Z emission amplitude squares:
\begin{align}
 \sum_{k}^{\rm Z\, emission}|{\cal M}_k|^2
\label{amp_z}
\end{align}
as in Fig.\ref{fig:vbf_tot}(c).
We can observe cancellation among the Z emission amplitudes, 
whose square sum in Eq.~\eqref{amp_z} is a factor of $10^3$ larger than the physical cross section
at the smallest $pT(Z)$ bin of $p_T(Z)<0.5$~TeV at $\sqrt{s}=20$~TeV.
From our experiences in QED, we examine the possibility that the cancellation is a consequence of soft Z-boson emission 
from the initial and final charged leptons.

As a test of this postulate,
we separate the Z emission amplitudes into three sub-groups such as Z emissions from the electron current,
from the muon current, and one Z from the electron current and the other from the muon current,
because we know that soft-photon emission amplitudes in QED are factorized into the three groups.
A blue dashed line with circles in Fig.~\ref{fig:vbf_dif} gives the sum of the contributions from the three sub-groups,
where the individual Feynman amplitudes (${\cal M}_k$) are summed within each sub-group 
before taking the absolute value square as
\begin{align}
  \Big|\sum_{k}^{{\rm Z\, emi\,}(ee)}{\cal M}_k\Big|^2
 +\Big|\sum_{k}^{{\rm Z\, emi\,}(\mu\mu)}{\cal M}_k\Big|^2
 +\Big|\sum_{k}^{{\rm Z\, emi\,}(e\mu)}{\cal M}_k\Big|^2.
\end{align}
No more subtle cancellation remains after the above `soft-Z' emission amplitudes are summed up before squaring.
As a further confirmation of the validity of our interpretation, 
we confirm that the majority of the small-$p_T$ Z boson from individual diagrams is transversely polarized. 

As a final remark of our sample studies, the new \helas amplitudes with the 5-component wavefunction
and the $5\times5$ component propagators of the weak bosons are free from unphysical `gauge theory cancellation'
among individual Feynman amplitudes,
and hence we can study physical origin of destructive or constructive interference among contributing amplitudes.

\section{Summary}\label{sec:summary}

\vrev{
In this paper we successfully implement the
$5 \times 5$ representation of the weak boson
propagators~\cite{Chen:2016wkt,Cuomo:2019siu}
in an arbitrary tree-level helicity amplitudes of
the electroweak theory which is free from subtle gauge-theory 
cancellation among interfering amplitudes.  
}

In this representation, the longitudinally polarized 
weak boson states are expressed as a quantum mechanical 
superposition of the vector boson with reduced 
longitudinal polarization vector and its associate 
Goldstone boson with the same four momenta.  
The reduced polarization vector for the longitudinal polarization state 
$\tilde{\epsilon}^\mu(q,0)$ is defined as the difference 
between the longitudinal and the scalar polarization vectors 
as in Eq.~\eqref{etldL}.  
The amplitudes of the scalar polarization component are 
expressed as those of the Goldstone boson amplitudes, by 
making use of the BRST identities, Eq.~\eqref{T_piv}, which are 
satisfied for the physical states in the $S$-matrix elements.
We show that the same BRST identities hold for the two 
sub-amplitudes which are connected by a gauge boson propagator 
of an arbitrary tree-level scattering amplitudes.  
By applying the identities successively to all the gauge boson 
propagators in a scattering amplitude, we arrive at the 
expression of the amplitude in which all the off-shell weak boson
propagators have the $5 \times 5$ polarization transfer 
matrix 
\vrev{
of Refs.~\cite{Chen:2016wkt,Cuomo:2019siu},
}
between the 5-component off-shell wavefunctions. 
The propagator reduces to the `parton-shower gauge' form
for QED and QCD~\cite{Hagiwara:2020tbx}, in which only the transverse and the reduced longitudinal 
polarization states are allowed to propagate.   

In this representation of the amplitudes, all the components 
of individual Feynman amplitudes which grow with four momentum of on-shell 
and off-shell gauge bosons are subtracted systematically, 
while the total sum of all the contributing amplitudes, 
the $S$-matrix element, remains intact by the BRST invariance.
As a consequence of the removal of all components of the 
sub-amplitudes which grow with the weak boson energies, 
all the sub-amplitudes, those 
amplitudes associated with an individual Feynman diagram, 
are expressed as a product of the invariant propagator 
factors, $D_V(q^2) = (q^2-m_V^2)^{-1}$,
with the spin polarization sum at each vertex 
giving the splitting amplitudes~\cite{Hagiwara:2009wt,Chen:2016wkt,Cuomo:2019siu}. 
This allows us to interpret the full scattering amplitude as 
a superposition of all contributing sub-amplitudes (Feynman 
diagrams), and we can study interference among them 
quantitatively.

We believe that the above properties of the amplitudes in 
\vrev{
the $5\times5$ representation of weak-boson propagators~\cite{Chen:2016wkt,Cuomo:2019siu} 
}
are valuable in the study of 
scattering amplitudes at high energies, and prepared a 
set of numerical codes, new \helas codes, which allow us to obtain the 
tree-level amplitudes of an arbitrary SM processes by 
using an automatic Feynman amplitude generation code 
like {\tt MadGraph\_aMC@NLO}~\cite{Alwall:2014hca}. 
\vrev{
Some of the representative codes are presented
}
in Appendix~\ref{sec:codes}, and sample results 
are reported in Sect.~\ref{sec:results}, for the weak boson scattering 
processes ($W^-W^+\to ZZ$, $W^-W^+\to W^-W^+$, $ZZ \to ZZ$), 
and for the processes where the above $2 \to 2$ weak boson 
scattering appears in processes with massless leptons 
in the initial state.  
All our findings look encouraging, allowing us to 
interpret the property of each contributing Feynman amplitude 
and the interference among them. 

The gauge boson propagators are available only after 
gauge fixing, and their explicit forms depend on the 
gauge-fixing condition.  
A particular form of the gauge boson propagators has hence 
been named after the chosen gauge-fixing condition, such 
as the Feynman gauge, Landau gauge, axial or light-cone 
gauge, covariant renormalizable $R_\xi$ gauge, etc. 
The particular form of 
the massless gauge-boson propagators with the reduced longitudinal polarization vector 
has been named as 
`parton shower gauge'~\cite{Hagiwara:2020tbx}, because of the property that 
the absolute value square of each sub-amplitude give the 
lepton, quark, photon and gluon splitting functions~\cite{Gribov:1972ri,*Lipatov:1974qm,*Altarelli:1977zs,*Dokshitzer:1977sg,Hagiwara:2009wt}.  
For the EW gauge bosons,
the 5-component representation of the weak bosons and their $5\times5$ propagators
are
called `equivalent gauge'~\cite{Wulzer:2013mza,Cuomo:2019siu}
or as `Goldstone equivalence gauge'~\cite{Chen:2016wkt}, referring to the 
original 5-component description of the weak bosons adopted 
in deriving the Goldstone boson equivalence theorem~\cite{Chanowitz:1985hj}.  
That this particular form of the propagator is useful in 
obtaining the splitting functions among weak bosons 
and the Higgs boson has been clearly observed in Refs.~\cite{Chen:2016wkt,Cuomo:2019siu}.
Since weak bosons can be treated as partons at high energies, 
the expression like `parton shower' may be a good 
common description of our propagators and amplitudes.  
\vrev{
However, the name `parton shower gauge' was adopted by Nagy and Soper~\cite{Nagy:2007ty,Nagy:2014mqa}
for a specific light-cone gauge in their study of parton-shower properties with quantum interference.
We may call the gauge-boson propagator representations of unbroken~\cite{Hagiwara:2020tbx} and broken~\cite{Chen:2016wkt,Cuomo:2019siu} gauge theories as `Feynman diagram (FD) gauge', 
because of the common property that the individual Feynman amplitude 
\vrev{
can be expressed
} 
as a product of invariant propagator factors for the internal lines and the splitting amplitudes at each vertex.
}

\section*{Acknowledgements}
We would like to thank Kai Ma for collaboration in the early stage,
Yang Ma, Tao Han and Brock Tweedie for discussions, 
and Olivier Mattelaer for technical advice related to {\tt MadGraph5\_aMC@NLO}.
We also wish to thank Hye-Sung Lee at KAIST,
where JMC, KH and JK worked on the project in July--August, 2019.
We also thank Yajuan Zheng for sending us valuable comments 
on the use of preliminary versions of new \helas codes.
All the Feynman diagrams were drawn with {\tt TikZ-FeynHand}~\cite{Ellis:2016jkw,*Dohse:2018vqo}.
JMC is supported by Fundamental Research Funds for the Central Universities of China NO.11620330.
The work was supported in part 
by World Premier International Research Center Initiative (WPI Initiative), MEXT, Japan, and also
by JSPS KAKENHI Grant No. 18K03648,
20H 05239, 21H01077 and 21K03583.

\appendix

\section{HELAS subroutines with 5-component weak bosons}\label{sec:codes}
\newcommand{\sw}{\mbox{$s_{\scriptscriptstyle W}$}}
\newcommand{\sws}{\mbox{$s_{\scriptscriptstyle W}^2$}}
\newcommand{\cw}{\mbox{$c_{\scriptscriptstyle W}$}}
\newcommand{\cws}{\mbox{$c_{\scriptscriptstyle W}^2$}}
\newcommand{\ctwow}{\mbox{$c_{\scriptscriptstyle 2W}$}}
\newcommand{\ctwows}{\mbox{$c_{\scriptscriptstyle 2W}^2$}}
\newcommand{\PL}{\mbox{$P_{\scriptscriptstyle L}$}}
\newcommand{\PR}{\mbox{$P_{\scriptscriptstyle R}$}}
\newcommand{\gL}{\mbox{$g_{\scriptscriptstyle L}$}}
\newcommand{\gR}{\mbox{$g_{\scriptscriptstyle R}$}}
\newcommand{\yL}{\mbox{$y_{\scriptscriptstyle L}$}}
\newcommand{\yR}{\mbox{$y_{\scriptscriptstyle R}$}}

In this appendix we list and present new \HELAS\ subroutines needed to compute amplitudes and off-shell currents with 5-component weak bosons.
We refrain from presenting all of them, but we show the representative subroutines for each type of vertices, i.e. those to compute amplitudes. 
Only for the fermion--fermion--vector boson vertex, we also present the subroutines to compute the off-shell currents as examples. 
They are publicly available on the web site.%
\footnote{HELAS/MadGraph/MadEvent Home Page: \\ https://madgraph.ipmu.jp/IPMU/}\

We already modified the internal structure of \HELAS\ wavefunctions in our previous paper~\cite{Hagiwara:2020tbx} to make it easier to extend them.
In the new \HELAS, which includes the new 5-component weak boson calculation, the size of the complex array of the vector wavefunction becomes seven to have the fifth component of wavefunctions.

The coupling constants of the new \HELAS\ subroutines become the array of double-precision complex numbers to include new 5-component weak-boson interactions.
Tables in each subsection show the coupling constants of each interaction vertex.
The content of the array of coupling constants should be permutated depending on the order of input/output particles because the vertices of the new \HELAS\ library with 5-component weak bosons are not symmetric for some interaction vertices.
Other Tables describe the order of permutations in subsections if necessary.

We note that the original \HELAS\ library~\cite{helas} was coded by {\tt Fortran77}, while the new one was done by {\tt Fortran90}.

\subsection{Parameter file and utility subroutine}

In the module, \texttt{helas\_params}, in \texttt{helas\_modules.f90}, 
a new value of \texttt{helas\_mode=3} is introduced for the 5-component weak boson calculations.
The subroutine, \texttt{hlmode}, can change this internal flag by calling it with a mode number, \texttt{helas\_mode}.
The subroutine call, \texttt{call hlmode(3)},
changes gauge definitions of gauge-boson propagators.

In the module, \texttt{helas\_utils}, some utility subroutines are also included: \texttt{define\_gauge\_dir} calculates $n$-vector (Eq.~\eqref{nmu}) from particle four-momentum.


\subsection{Wavefunction subroutines}

The function, \texttt{vxxxxx} (List~\ref{lst:vxxxxx}), computes the vector particle wavefunction of the 5-component vector boson representation.

\begin{lstlisting}[caption=vxxxxx.f90,label=lst:vxxxxx]
!
! This subroutine computes a VECTOR wavefunction.
!
! input:
!       double precision p(0:3): four-momentum of vector boson
!       double precision vmass : mass          of vector boson
!       integer nhel = -1, 0, 1: helicity      of vector boson
!                                (0 is forbidden if vmass=0.0)
!       integer nsv  = -1 or 1 : +1 for final, -1 for initial
!
! output:
!       double complex vc(7)   : vector wavefunction - epsilon^mu(v)
!     
subroutine vxxxxx(p,vmass,nhel,nsv , vc)

  use helas_params
  use helas_const
  use helas_utils
  
  implicit none

  double precision, intent(in) :: p(0:3),vmass
  integer, intent(in) :: nhel,nsv
  complex(kind(0.d0)), intent(out) :: vc(ndim_wv)

  double precision :: hel,pt,pt2,pp,pzpt,emp

  double precision :: n(0:4)
  double precision :: nk

  double precision, parameter :: sqrhlf = sqrt(0.5d0)


  hel = dble(nhel)

  pt2 = p(1)**2+p(2)**2
  pp = min(p(0),sqrt(pt2+p(3)**2))
  pt = min(pp,sqrt(pt2))

  vc(1) = dcmplx(p(0),p(3))*nsv
  vc(2) = dcmplx(p(1),p(2))*nsv

  if (abs(nhel)==1) then

     ! hel=+-1

     if (vmass/=0.d0) then

        if (pp==0.d0) then

           vc(3) = dcmplx( 0.d0 )
           vc(4) = dcmplx(-hel*sqrhlf )
           vc(5) = dcmplx( 0.d0 , nsv*sqrhlf )
           vc(6) = dcmplx( 0.d0 )

        else

           emp = p(0)/(vmass*pp)
           vc(3) = dcmplx( 0.d0 )
           vc(6) = dcmplx( hel*pt/pp*sqrhlf )
           if (pt/=0.d0) then
              pzpt = p(3)/(pp*pt)*sqrhlf*hel
              vc(4) = dcmplx(-p(1)*pzpt,-nsv*p(2)/pt*sqrhlf)
              vc(5) = dcmplx(-p(2)*pzpt, nsv*p(1)/pt*sqrhlf)
           else
              vc(4) = dcmplx( -hel*sqrhlf )
              vc(5) = dcmplx( 0.d0, nsv*sign(sqrhlf,p(3)) )
           endif

        endif

     else
        
        pp = p(0)
        pt = sqrt(p(1)**2+p(2)**2)
        vc(3) = dcmplx( 0.d0 )
        vc(6) = dcmplx( hel*pt/pp*sqrhlf )
        if (pt/=0.d0) then
           pzpt = p(3)/(pp*pt)*sqrhlf*hel
           vc(4) = dcmplx( -p(1)*pzpt, -nsv*p(2)/pt*sqrhlf )
           vc(5) = dcmplx( -p(2)*pzpt,  nsv*p(1)/pt*sqrhlf )
        else
           vc(4) = dcmplx( -hel*sqrhlf )
           vc(5) = dcmplx( 0.d0 , nsv*sign(sqrhlf,p(3)) )
        endif

     endif

  else

     ! hel=0

     if (pp==0.d0) then

        vc(3:6) = [ 0.d0, 0.d0, 0.d0, 1.d0 ]

     else

        emp = p(0)/(vmass*pp)
        vc(3) = dcmplx( pp/vmass )
        vc(6) = dcmplx( emp*p(3) )
        if (pt/=0.d0) then
           vc(4) = dcmplx( emp*p(1) )
           vc(5) = dcmplx( emp*p(2) )
        else
           vc(4) = dcmplx( 0.d0 )
           vc(5) = dcmplx( 0.d0 )
        endif

     endif

  endif

  if (helas_mode==helas_mode_5cwb) then

     if (pp>0.d0) then

        n(0) = sign(1.d0,dble(p(0)))
        n(1) = -p(1)/pp
        n(2) = -p(2)/pp
        n(3) = -p(3)/pp
        n(4) = 0.d0

     else

        n(0) = 1.d0
        n(1) = 0.d0
        n(2) = 0.d0
        n(3) = -1.d0
        n(4) = 0.d0

     endif

     nk = n(0)*p(0)-n(1)*p(1)-n(2)*p(2)-n(3)*p(3)
     
     if (abs(nhel)==1) then

        vc(7) = dcmplx( 0.d0 )
        
     else

        vc(3:6) = -vmass/nk*n(0:3)
        vc(7) = -nsv*ci
        
     endif
     
  endif
  
  return

end subroutine vxxxxx
\end{lstlisting}

\begin{table*}
\centering
\caption{Vertices for $ff'W^\pm$, $ffZ$ and $ffA$. 
$\PL=\tfrac{1}{2}(1-\gamma^5)$ and $\PR=\tfrac{1}{2}(1+\gamma^5)$. 
$f$ denotes fermions in the initial state; $\bar{f}$ denotes fermions in the final state.
Vector bosons are taken to be outgoing, i.e., in the final state.
$u$ represents up-type quarks ($u, c, t$) and neutrinos, and $y_u$ represents their Yukawa couplings;
$d$ represents down-type quarks ($d, s, b$) and charged leptons ($e, \mu, \tau$), and $y_d$ represents their Yukawa couplings.  The CKM matrix is taken to be ${\bf 1}$. }
\label{tab:ffV}
\renewcommand{\arraystretch}{1.9}
\begin{tabular}{|c|c|c|c|c|c|c|}
\hline 
\hline 
\multicolumn{2}{|c|}{$V^M_{ff'V}$}      
     &     $u\bar{d}W^+$  &   $d\bar{u}W^-$  &  $u\bar{u}Z$  &  $d\bar{d}Z$  &  $f\bar{f}A$  \\
\hline 
\hline
\ \ \ \multirow{2}{2em}{$V^{\mu}$}   &   $-i\gamma^{\mu}(\gL\PL $
&  $\gL=\dfrac{g}{\sqrt{2}}$    & $\gL=\dfrac{g}{\sqrt{2}}$  &  $\gL=\dfrac{g}{\cw}(\dfrac{1}{2}-Q_f\sws)$  &   $\gL=-\dfrac{g}{\cw}(\dfrac{1}{2}+Q_f\sws)$   
&    $\gL=g\sw$           \\
                                              &               \  \  \     \     $+\gR \PR)$
 &   $\gR=0$                          &       $\gR=0$                &          $\gR=-\dfrac{g}{\cw}Q_f\sws$             &          $\gR=-\dfrac{g}{\cw}Q_f\sws$                                  &     $\gR=g\sw$           \\  
\hline
\ \ \  \multirow{2}{2em}{$V^4$ }       &     $\yL \PL$         
  &    $\yL=y_d$                    &       $\yL=y_u$                &          $\yL=\dfrac{y_u}{\sqrt{2}}$              &          $\yL=-\dfrac{y_d}{\sqrt{2}}$
    &    $\yL=0$       \\          
                                              &            $+\yR\PR$      
    &   $\yR=-y_u$                   &          $\yR=-y_d$          &           $\yR=-\dfrac{y_u}{\sqrt{2}}$             &       $\yR=\dfrac{y_d}{\sqrt{2}}$
     &      $\yR=0$       \\
\hline
\end{tabular}
\end{table*}

\subsection{FFV vertex}

Subroutines in this section compute amplitudes (\texttt{iovxxx}) and off-shell currents (\texttt{fvixxxx}, \texttt{fvoxxx}, \texttt{jioxxx}) of the fermion--fermion--vector boson vertex.
The subroutine for the amplitudes is shown explicitly in List~\ref{lst:iovxxx},
while those for one of the off-shell fermion currents and the off-shell vector current
are in List~\ref{lst:fvixxx} and List~\ref{lst:jioxxx}, respectively.

The coupling constants of the fermion--fermion--vector boson vertices are listed in Table~\ref{tab:ffV}.

\begin{lstlisting}[caption=iovxxx.f90,label=lst:iovxxx]
!
! This subroutine computes an amplitude of the 
! fermion-fermion-vector coupling.
!
! input:
!       double complex fi(6) : flow-in  fermion   |fi>
!       double complex fo(6) : flow-out fermion   <fo|
!       double complex vc(7) : input    vector    v
!       double complex gc(2) : coupling constant  gffv 
!
! output:
!       double complex vertex: amplitude          <fo|v|fi>
!     
subroutine iovxxx(fi,fo,vc,gc, vertex)

  use helas_params
  use helas_const
  use helas_utils

  use spinor_operations
  
  implicit none

  complex(kind(0.d0)), intent(in) :: fi(ndim_wf),fo(ndim_wf),vc(ndim_wv)
  complex(kind(0.d0)), intent(in) :: gc(2)
  complex(kind(0.d0)), intent(out) :: vertex

  complex(kind(0.d0)) :: yc(2)

  complex(kind(0.d0)) :: js(0:4)
  
  call spinor_product(fo, fi, gc, js)

  vertex = js(0)*vc(3) -js(1)*vc(4) -js(2)*vc(5) -js(3)*vc(6)

  if (helas_mode==helas_mode_5cwb) then

     vertex = vertex &
           - ci * (yc(1)*(fo(3)*fi(3)+fo(4)*fi(4)) &
               +  yc(2)*(fo(5)*fi(5)+fo(6)*fi(6))) *vc(7)

  endif

  return

end subroutine iovxxx
\end{lstlisting}

\begin{lstlisting}[caption=fvixxx.f90,label=lst:fvixxx]
!
! This subroutine computes an off-shell fermion wavefunction from a
! flowing-IN external fermion and a vector boson.
!
! input:
!       double complex fi(6)   : flow-in  fermion        |fi>
!       double complex vc(6)   : input    vector         v
!       double complex gc(2)   : coupling constants      gvf
!       double precision fmass : mass  of output fermion f'
!       double precision fwidth: width of output fermion f'
!
! output:
!       double complex fvi(6)  : off-shell fermion       |f',v,fi>
!
subroutine fvixxx(fi,vc,gc,fmass,fwidth , fvi)

  use helas_params
  use helas_const
  use helas_utils
  
  implicit none

  complex(kind(0.d0)), intent(in) :: fi(ndim_wf),vc(ndim_wv),gc(2)
  double precision, intent(in) :: fmass,fwidth
  complex(kind(0.d0)), intent(out) :: fvi(ndim_wf)
  
  complex(kind(0.d0)) :: sl1,sl2,sr1,sr2,d
  double precision :: pf(0:3),pf2

  complex(kind(0.d0)) :: f5c(4)

  complex(kind(0.d0)) :: yc(2)


  fvi(1:2) = fi(1:2)-vc(1:2)
  
  pf(0) = dble( fvi(1))
  pf(1) = dble( fvi(2))
  pf(2) = dimag(fvi(2))
  pf(3) = dimag(fvi(1))
  pf2 = pf(0)**2-(pf(1)**2+pf(2)**2+pf(3)**2)

  d = -1.d0/dcmplx( pf2-fmass**2, fmass*fwidth )

  sl1 =   (vc(3)+   vc(6))*fi(3) &
        + (vc(4)-ci*vc(5))*fi(4)
  sl2 =   (vc(4)+ci*vc(5))*fi(3) &
        + (vc(3)-   vc(6))*fi(4)

  sr1 =   (vc(3)-   vc(6))*fi(5) &
        - (vc(4)-ci*vc(5))*fi(6)
  sr2 = - (vc(4)+ci*vc(5))*fi(5) &
        + (vc(3)+   vc(6))*fi(6)
  
  fvi(3) = gc(1)*( (pf(0)   -pf(3))*sl1-(pf(1)-ci*pf(2))*sl2) &
         + gc(2)*fmass*sr1
  fvi(4) = gc(1)*(-(pf(1)+ci*pf(2))*sl1+(pf(0)   +pf(3))*sl2) &
         + gc(2)*fmass*sr2 
  fvi(5) = gc(2)*( (pf(0)   +pf(3))*sr1+(pf(1)-ci*pf(2))*sr2) &
         + gc(1)*fmass*sl1
  fvi(6) = gc(2)*( (pf(1)+ci*pf(2))*sr1+(pf(0)   -pf(3))*sr2) &
         + gc(1)*fmass*sl2

  if (helas_mode==helas_mode_5cwb) then
     
     f5c(1) = yc(2)*( (pf(0)   -pf(3))*fi(5) &
                     -(pf(1)-ci*pf(2))*fi(6)) &
             + yc(1)*fmass*fi(3)
     f5c(2) = yc(2)*(-(pf(1)+ci*pf(2))*fi(5) &
                     +(pf(0)   +pf(3))*fi(6)) &
             + yc(1)*fmass*fi(4)
     f5c(3) = yc(1)*( (pf(0)   +pf(3))*fi(3) &
                     +(pf(1)-ci*pf(2))*fi(4)) &
             + yc(2)*fmass*fi(5)
     f5c(4) = yc(1)*( (pf(1)+ci*pf(2))*fi(3) &
                     +(pf(0)   -pf(3))*fi(4)) &
             + yc(2)*fmass*fi(6)

     fvi(3:6) = fvi(3:6)-ci*f5c(1:4)*vc(7)
     
  endif

  fvi(3:6) = d * fvi(3:6)
  
  return

end subroutine fvixxx
\end{lstlisting}

%
\begin{lstlisting}[caption=jioxxx.f90,label=lst:jioxxx]
!
! This subroutine computes an off-shell vector current from
! an external fermion pair.
!
! input:
!       double complex fi(7)   : flow-in  fermion       |fi>
!       double complex fo(7)   : flow-out fermion       <fo|
!       double complex gc(2)   : coupling constants     gvf
!       double precision vmass : mass  of OUTPUT vector v
!       double precision vwidth: width of OUTPUT vector v
!
! output:
!       double complex jio(7)  : vector current         j^mu(<fo|v|fi>)
!     
subroutine jioxxx(fi,fo,gc,vmass,vwidth , jio)

  use helas_params
  use helas_const
  use helas_utils

  use spinor_operations
  
  implicit none

  complex(kind(0.d0)), intent(in) :: fi(ndim_wf),fo(ndim_wf)
  complex(kind(0.d0)), intent(in) :: gc(2)
  double precision, intent(in) :: vmass, vwidth
  complex(kind(0.d0)), intent(out) :: jio(ndim_wv)

  
  complex(kind(0.d0)) :: jj(0:4)
  complex(kind(0.d0)) :: j12(0:4)
  complex(kind(0.d0)) :: js,d
  complex(kind(0.d0)) :: js1,js2
  complex(kind(0.d0)) :: q(0:4)
  double precision :: qabs2,qabs
  double precision :: q2,vm2

  complex(kind(0.d0)) :: yc(2)

  double precision :: n(0:4)
  double precision :: nq


  jio(1:2) = fo(1:2)-fi(1:2)

  q(0) = -dble( jio(1))
  q(1) = -dble( jio(2))
  q(2) = -dimag(jio(2))
  q(3) = -dimag(jio(1))
  q(4) = -ci*vmass

  qabs2 = dble(q(1))**2+dble(q(2))**2+dble(q(3))**2
  qabs = sqrt(qabs2)
  q2 = dble(q(0))**2-qabs2

  vm2 = vmass**2

  call spinor_product(fo, fi, gc, j12)

  call propagator_factor(q2,vmass,vwidth, d)
  
  if (vmass/=0.d0) then

     js = (q(0)*j12(0)-q(1)*j12(1)-q(2)*j12(2)-q(3)*j12(3))/vm2

     jj(0:3) = j12(0:3) - q(0:3)*js

  else

     jj(0:3) = j12(0:3)
     
  end if

  jio(3:6) = jj(0:3) * d

  if (helas_mode==helas_mode_psg) then

     if (vmass==0.d0) then

        call define_gauge_dir(q, n)

        nq = n(0)*dble(q(0))-n(1)*dble(q(1))-n(2)*dble(q(2))-n(3)*dble(q(3))

        js1 = (n(0)*j12(0)-n(1)*j12(1)-n(2)*j12(2)-n(3)*j12(3)) / nq

        js2 = (q(0)*j12(0)-q(1)*j12(1)-q(2)*j12(2)-q(3)*j12(3)) / nq
        
        jio(3:6) = (j12(0:3)-q(0:3)*js1-n(0:3)*js2) * d
        
     endif
     
  else if (helas_mode==helas_mode_5cwb) then

     j12(4) = + ci * ( yc(1)*(fo(3)*fi(3)+fo(4)*fi(4)) &
          + yc(2)*(fo(5)*fi(5)+fo(6)*fi(6)))
         
     call define_gauge_dir(q,n)
     
     nq = n(0)*dble(q(0))-n(1)*dble(q(1))-n(2)*dble(q(2))-n(3)*dble(q(3))
     
     js1 = (n(0)*j12(0)-n(1)*j12(1)-n(2)*j12(2)-n(3)*j12(3)) / nq
     
     js2 = (q(0)*j12(0)-q(1)*j12(1)-q(2)*j12(2) &
          - q(3)*j12(3)-dconjg(q(4))*j12(4)) / nq     
     
     jio(3:7) = (j12(0:4) - q(0:4)*js1 -n(0:4)*js2) * d

     
  end if

  return

end subroutine jioxxx
\end{lstlisting}

\subsection{VVV vertex}

Subroutines in this section compute amplitudes (\texttt{vvvxxx}) and off-shell currents (\texttt{jvvxxx}) of the three-point vector boson vertex.
The subroutine for the amplitudes is shown explicitly in List~\ref{lst:vvvxxx}.

The coupling constants and their permutations according to the order of input particles for the three-point vector boson vertices are listed in Table~\ref{tab:VVV} and Table~\ref{tab:notation_VVV}.

\begin{table*}[]
\centering
\caption{Vertices for $VVV$: $WWZ$ and $WWA$.  All momenta are taken to be outgoing.  }
\label{tab:VVV}
\renewcommand{\arraystretch}{1.5}
\begin{tabular}{|c|c|c|c|c|}
\hline
\hline
\multicolumn{3}{|c|}{$V^{M_1M_2M_3}_{VVV}$}        & $W^-W^+Z$   &  $W^-W^+A$    \\
\hline
\hline
\multicolumn{2}{|c|}{$V^{\mu_1\mu_2\mu_3}$} &   $-ia(g^{\mu_1\mu_2}(p_1-p_2)^{\mu_3}+ g^{\mu_2\mu_3}(p_2-p_3)^{\mu_1}+g^{\mu_3\mu_1}(p_3-p_1)^{\mu_2})  $  &    $a=g\cw$  &  $a=g\sw$  \\
\hline
 \multirow{3}{4em}{$V^{4\mu_i\mu_j}$}  
 &  $V^{4\mu_2\mu_3}$& $b_1g^{\mu_2\mu_3 }$   &  $b_1=-\dfrac{g^{2}\sws v}{2\cw}$  &  $b_1=\dfrac{g^{2}\sw v}{2}$   \\  
    &  $V^{\mu_14\mu_3}$& $b_2g^{\mu_1\mu_3 }$  & $b_2=\dfrac{g^{2}\sws v}{2\cw}$ &$b_2=-\dfrac{g^{2}\sw v}{2}$  \\
   &  $V^{\mu_1\mu_2 4}$& $b_3g^{\mu_1\mu_2 }$   & $b_3=0$   
    &  $b_3=0$   \\
 \hline
 \multirow{3}{4em}{$V^{44\mu_i} $} & $V^{4 4\mu_3}$   & $ ic_1(p_1-p_2)^{\mu_3}$   &$ c_1=\dfrac{g \ctwow}{2\cw}$    &  $c_1= g\sw$    \\
 & $V^{\mu_144}$   &  $ic_2(p_2-p_3)^{\mu_1}$   & $c_2=\dfrac{g}{2}$  & $c_2=0$   \\
  &    $V^{4\mu_2 4}$   & $ ic_3(p_3-p_1)^{\mu_2}$   & $c_3=\dfrac{g}{2}$  & $c_3=0$  \\
\hline
\multicolumn{3}{|c|}{$V^{444}$}&      $0$&  $0$  \\
\hline
\end{tabular}
\end{table*}

\begin{table*}
\centering
\caption{Values of $a',\{b_1',b_2',b_3'\}, \{c_1',c_2',c_3'\}$ for $W^-W^+Z$ or $W^-W^+A$ subroutines,  depending on different charges of $1,2,3$.   $a,\{b_1,b_2,b_3\}, \{c_1,c_2,c_3\}$ are defined  in Tabel~\ref{tab:VVV}.}
\label{tab:notation_VVV}
\renewcommand{\arraystretch}{1.6}
\begin{tabular}{|c||c|c|c|c|c|c|c|}
\hline
$W_1^-W_2^+Z_3/W_1^-W_2^+A_3$   & $a'=a$  & $b_1'=b_1$  &  $b_2'=b_2$ & $b_3'=b_3$ 
& $c_1'=c_1$ & $c_2'=c_2$ &   $c_3'=c_3$ \\
\hline
$W_1^+Z_2W_3^-/W_1^+A_2W_3^-$   & $a'=a$  & $b_1'=b_2$  &  $b_2'=b_3$ & $b_3'=b_1$ 
& $c_1'=c_2$ & $c_2'=c_3$ &   $c_3'=c_1$   \\
\hline
$Z_1W_2^-W^+_3/A_1W_2^-W^+_3$   & $a'=a$  & $b_1'=b_3$  &  $b_2'=b_1$ & $b_3'=b_2$ 
& $c_1'=c_3$ & $c_2'=c_1$ &   $c_3'=c_2$   \\
\hline
\end{tabular}
\end{table*}

\begin{lstlisting}[caption=vvvxxx.f90,label=lst:vvvxxx]
!
! This subroutine computes an amplitude of the three-point coupling
! of the gauge bosons.
!
! input:
!       double complex wm(7) : vector            flow-out W-
!       double complex wp(7) : vector            flow-out W+
!       double complex w3(7) : vector            A or Z
!       double precision gc(7): coupling constant 
!
! output:
!       double complex vertex: amplitude         gamma(wm,wp,w3)
!     
subroutine vvvxxx(wm,wp,w3,gc , vertex)

  use helas_params
  use helas_const
  use helas_utils
  
  implicit none
  
  complex(kind(0.d0)), intent(in) :: wm(ndim_wv),wp(ndim_wv),w3(ndim_wv)
  complex(kind(0.d0)), intent(in) :: gc(7)
  complex(kind(0.d0)), intent(out) :: vertex

  complex(kind(0.d0)) :: xv1,xv2,xv3,v12,v23,v31,p12,p13,p21,p23,p31,p32
  double precision pwm(0:3),pwp(0:3),pw3(0:3)
  complex(kind(0.d0)) :: dwm(0:4),dwp(0:4),dw3(0:4)

  complex(kind(0.d0)) :: p12v3, p23v1, p31v2


  pwm(0) = dble( wm(1))
  pwm(1) = dble( wm(2))
  pwm(2) = dimag(wm(2))
  pwm(3) = dimag(wm(1))
      
  pwp(0) = dble( wp(1))
  pwp(1) = dble( wp(2))
  pwp(2) = dimag(wp(2))
  pwp(3) = dimag(wp(1))
      
  pw3(0) = dble( w3(1))
  pw3(1) = dble( w3(2))
  pw3(2) = dimag(w3(2))
  pw3(3) = dimag(w3(1))

  dwm(0:4) = wm(3:7)
  dwp(0:4) = wp(3:7)
  dw3(0:4) = w3(3:7)
  
  v12 = dwm(0)*dwp(0)-dwm(1)*dwp(1)-dwm(2)*dwp(2)-dwm(3)*dwp(3)
  v23 = dwp(0)*dw3(0)-dwp(1)*dw3(1)-dwp(2)*dw3(2)-dwp(3)*dw3(3)
  v31 = dw3(0)*dwm(0)-dw3(1)*dwm(1)-dw3(2)*dwm(2)-dw3(3)*dwm(3)

  xv1 = 0.d0
  xv2 = 0.d0
  xv3 = 0.d0
  if (abs(wm(3))/=0.d0) then
     if (abs(wm(3))>=max(abs(wm(4)),abs(wm(5)),abs(wm(6)))*0.1d0) &
          xv1 = pwm(0)/wm(3)
  endif
  if (abs(wp(3))/=0.d0) then
     if (abs(wp(3))>=max(abs(wp(4)),abs(wp(5)),abs(wp(6)))*0.1d0) &
          xv2 = pwp(0)/wp(3)
  endif
  if (abs(w3(3))/=0.d0) then
     if (abs(w3(3))>=max(abs(w3(4)),abs(w3(5)),abs(w3(6)))*0.1d0 ) &
          xv3 = pw3(0)/w3(3)
  endif

  p12 = (pwm(0)-xv1*wm(3))*wp(3)-(pwm(1)-xv1*wm(4))*wp(4) &
       -(pwm(2)-xv1*wm(5))*wp(5)-(pwm(3)-xv1*wm(6))*wp(6)
  p13 = (pwm(0)-xv1*wm(3))*w3(3)-(pwm(1)-xv1*wm(4))*w3(4) &
       -(pwm(2)-xv1*wm(5))*w3(5)-(pwm(3)-xv1*wm(6))*w3(6)
  p21 = (pwp(0)-xv2*wp(3))*wm(3)-(pwp(1)-xv2*wp(4))*wm(4) &
       -(pwp(2)-xv2*wp(5))*wm(5)-(pwp(3)-xv2*wp(6))*wm(6)
  p23 = (pwp(0)-xv2*wp(3))*w3(3)-(pwp(1)-xv2*wp(4))*w3(4) &
       -(pwp(2)-xv2*wp(5))*w3(5)-(pwp(3)-xv2*wp(6))*w3(6)
  p31 = (pw3(0)-xv3*w3(3))*wm(3)-(pw3(1)-xv3*w3(4))*wm(4) &
       -(pw3(2)-xv3*w3(5))*wm(5)-(pw3(3)-xv3*w3(6))*wm(6)
  p32 = (pw3(0)-xv3*w3(3))*wp(3)-(pw3(1)-xv3*w3(4))*wp(4) &
       -(pw3(2)-xv3*w3(5))*wp(5)-(pw3(3)-xv3*w3(6))*wp(6)

  vertex = -gc(1)*(v12*(p13-p23)+v23*(p21-p31)+v31*(p32-p12))

  if (helas_mode==helas_mode_5cwb) then

     vertex = vertex &
          + ci*( gc(2)*dwm(4)*v23 + gc(3)*dwp(4)*v31 + gc(4)*dw3(4)*v12)

     p12v3 = (pwm(0)-pwp(0))*dw3(0) -(pwm(1)-pwp(1))*dw3(1) &
           - (pwm(2)-pwp(2))*dw3(2) -(pwm(3)-pwp(3))*dw3(3)
     p23v1 = (pwp(0)-pw3(0))*dwm(0) -(pwp(1)-pw3(1))*dwm(1) &
           - (pwp(2)-pw3(2))*dwm(2) -(pwp(3)-pw3(3))*dwm(3)
     p31v2 = (pw3(0)-pwm(0))*dwp(0) -(pw3(1)-pwm(1))*dwp(1) &
           - (pw3(2)-pwm(2))*dwp(2) -(pw3(3)-pwm(3))*dwp(3)

     vertex = vertex &
          + (  gc(5)*dwm(4)*dwp(4)*p12v3 &
             + gc(6)*dwp(4)*dw3(4)*p23v1 &
             + gc(7)*dw3(4)*dwm(4)*p31v2 )
     
  endif
  
  return

end subroutine vvvxxx
\end{lstlisting}

%

\begin{table*}[]
\centering
\caption{Vertices for $VVH$: $WWH$ and $ZZH$. All momenta are taken to be outgoing.  }
\label{tab:VVS}
\renewcommand{\arraystretch}{1.8}
\begin{tabular}{|c|c|c|c|c|}
\hline
\hline
\multicolumn{3}{|c|}{$V^{M_1M_2}_{VVH}$}        & $W^-W^+H$   &  $ZZH$    \\
\hline
\hline
\multicolumn{2}{|c|}{$V^{\mu_1\mu_2}$} &   $ia g^{\mu_1\mu_2} $  &    $a=g m_W$  &  $a=\dfrac{g m_Z}{\cw}$  \\
\hline
\  \  \  \     \multirow{2}{4em}{$V^{4\mu_i}$} 
 &  $V^{4\mu_2}$& $b_1(p_1-p_3)^{\mu_2}$   &  $b_1=\dfrac{g}{2}$  &  $b_1=\dfrac{g}{2\cw}$   \\  
    &  $V^{\mu_1 4}$& $b_2(p_2-p_3)^{\mu_1}$  & $b_2=\dfrac{g}{2}$ &$b_2=\dfrac{g}{2\cw}$  \\
 \hline
 \multicolumn{2}{|c|}{$V^{44}$}   & $c_1$   &$c_1= -2i\lambda v $    &  $c_1= -2i\lambda v$    \\
\hline
\end{tabular}
\end{table*}

\subsection{VVS vertex}

Subroutines in this section compute amplitudes (\texttt{vvsxxx}) and off-shell currents (\texttt{jvsxxx}, \texttt{hvvxxx}) of the vector boson--vector boson--scalar vertex.
The subroutine for the amplitudes is shown explicitly in List~\ref{lst:vvsxxx}.

The coupling contants for the vector--vector--scalar vertices are listed in Table~\ref{tab:VVS}.

\begin{lstlisting}[caption=vvsxxx.f90,label=lst:vvsxxx]
!
! This subroutine computes an amplitude of the vector-vector-scalar
! coupling.
!
! input:
!       double complex v1(7) : first  vector     v1
!       double complex v2(7) : second vector     v2
!       double complex sc(3) : input  scalar     s
!       double complex gc(4) : coupling constant gvvs 
!
! output:
!       double complex vertex: amplitude         gamma(v1,v2,s)
!     
subroutine vvsxxx(v1,v2,sc,gc , vertex)

  use helas_params
  use helas_const
  use helas_utils
  
  implicit none

  complex(kind(0.d0)), intent(in) :: v1(ndim_wv),v2(ndim_wv),sc(ndim_ws)
  complex(kind(0.d0)), intent(in) :: gc(4)
  complex(kind(0.d0)), intent(out) :: vertex

  complex(kind(0.d0)) :: dv1(0:4), dv2(0:4)
  double precision :: pv1(0:3),pv2(0:3),psc(0:3)
  complex(kind(0.d0)) :: v12
  complex(kind(0.d0)) :: p13v2, p23v1


  dv1(0:4) = v1(3:7)
  dv2(0:4) = v2(3:7)

  v12 = dv1(0)*dv2(0)-dv1(1)*dv2(1)-dv1(2)*dv2(2)-dv1(3)*dv2(3)

  vertex = gc(1)*v12

  if (helas_mode==helas_mode_5cwb) then

     pv1(0) = dble( v1(1))
     pv1(1) = dble( v1(2))
     pv1(2) = dimag(v1(2))
     pv1(3) = dimag(v1(1))

     pv2(0) = dble( v2(1))
     pv2(1) = dble( v2(2))
     pv2(2) = dimag(v2(2))
     pv2(3) = dimag(v2(1))

     psc(0) = dble( sc(1))
     psc(1) = dble( sc(2))
     psc(2) = dimag(sc(2))
     psc(3) = dimag(sc(1))

     p13v2 = (pv1(0)-psc(0))*dv2(0)-(pv1(1)-psc(1))*dv2(1) &
           - (pv1(2)-psc(2))*dv2(2)-(pv1(3)-psc(3))*dv2(3)

     p23v1 = (pv2(0)-psc(0))*dv1(0)-(pv2(1)-psc(1))*dv1(1) &
           - (pv2(2)-psc(2))*dv1(2)-(pv2(3)-psc(3))*dv1(3)
     
     vertex = vertex + ci*(gc(2)*dv1(4)*p13v2 &
                         + gc(3)*dv2(4)*p23v1)
     vertex = vertex - ci*gc(4)*dv1(4)*dv2(4)

  endif

  vertex = vertex * sc(3)

  return

end subroutine vvsxxx
\end{lstlisting}

%

\begin{table*}[]
\centering
\caption{Vertices for $VVVV$. }
\label{tab:VVVV}
\renewcommand{\arraystretch}{1.9}
\begin{tabular}{|c|c|c|c|c|c|c|c|}
\hline
\hline
\multicolumn{3}{|c|}{$V^{M_1M_2M_3M_4}_{VVVV}$}        & $W^-W^+W^-W^+$   &  $W^-ZW^+Z$  & $W^-AW^+A$    &   $W^-ZW^+A$  & $ZZZZ$       \\
\hline
\hline
  \multicolumn{2}{|c|}{}     & $-ia(g^{\mu_1\mu_2}g^{\mu_3\mu_4}$  &                 &    &    &   &   \\
 \multicolumn{2}{|c|}{$V^{\mu_1\mu_2\mu_3\mu_4}$}  
 &     $+g^{\mu_1\mu_4}g^{\mu_2\mu_3}$         &      $a=g^{2}$  &  $a=-g^{2}\cws$   & $a= -g^{2}\sws$    &  $a=-g^{2}\sw\cw$     &   $a=0$\\
  \multicolumn{2}{|c|}{}     &  $-2g^{\mu_1\mu_3}g^{\mu_2\mu_4})$          &      &    &       &        & \\
\hline
\multicolumn{3}{|c|}{$V^{4\mu_i\mu_j\mu_k}$}     &  $0$ &   $0$  &  $0$   &  $0$  &  $0$   \\
\hline 
 \multirow{6}{4em}{$V^{44\mu_i\mu_j}$} 
    &  $V^{44\mu_3\mu_4}$& $ib_1g^{\mu_3\mu_4 }$   &  $b_1=\dfrac{g^{2}}{2}$  
    &  $b_1=\dfrac{g^{2}\sws}{2\cw}$    &    $b_1=0$     &    $b_1=-\dfrac{g^{2}\sw}{2}$    &      $b_1=\dfrac{g^{2}}{2\cws}$  \\  
    &  $V^{4\mu_24\mu_4}$& $ib_2g^{\mu_2\mu_4 }$     & $b_2=0$
     &$b_2=\dfrac{g^{2}\ctwows}{2\cws}$             &          $b_2=2g^{2}\sws$  & $b_2=\dfrac{g^{2}\sw \ctwow}{\cw}$         &   $b_2=\dfrac{g^{2}}{2\cws}$\\
    &  $V^{4\mu_2\mu_34}$& $ib_3g^{\mu_2\mu_3 }$  & $b_3=\dfrac{g^{2}}{2}$ 
    &$b_3=\dfrac{g^{2}\sws}{2\cw}$              &            $b_3=0$   &   $b_3=0$  &  $b_3=\dfrac{g^{2}}{2\cws}$ \\
   &  $V^{\mu_14 4\mu_4}$& $ib_4g^{\mu_1\mu_4 }$   & $b_4=\dfrac{g^{2}}{2}$ 
    &  $b_4=\dfrac{g^{2}\sws}{2\cw}$             &   $b_4=0$         &  $b_4=-\dfrac{g^{2}\sw}{2}$     &    $b_4=\dfrac{g^{2}}{2\cws}$\\
   &  $V^{\mu_14 \mu_3 4}$& $ib_5g^{\mu_1\mu_3 }$   & $b_5=0$    
   &  $b_5=\dfrac{g^{2}}{2}$              &   $b_5=0$    & $b_5=0$   &       $b_5=\dfrac{g^{2}}{2\cws}$         \\
   &  $V^{\mu_1\mu_2 4 4}$& $ib_6g^{\mu_1\mu_2 }$   & $b_6=\dfrac{g^{2}}{2}$    
   &  $b_6=\dfrac{g^{2}\sws}{2\cw}$                             &     $b_6=0$    &   $b_6=0$   &     $b_6=\dfrac{g^{2}}{2\cws}$      \\ 
 \hline
\multicolumn{3}{|c|}{$V^{444\mu_i} $}     &0    &  0   &0        &  0       &  0  \\
 \hline
\multicolumn{2}{|c|}{$V^{4444}$}&     $d $ & $d=-4i\lambda$&  $d=-2i\lambda$  
& $d=0$   &   $d=0$    &  $d=-6i\lambda$ \\
\hline
\end{tabular}
\end{table*}

\begin{table*}
\centering
\caption{Definition of $(b_1',b_2',b_3',b_4',b_5',b_6')$ in $VVVV$ by Table~\ref{tab:VVVV}. The first column is the canonical setting, the second column is obtained by $1\leftrightarrow 2$, $3\leftrightarrow 4$, the third column is obtained by $2\leftrightarrow 4$, the fourth column is obtained by $1\leftrightarrow 4$, $2\leftrightarrow 3$.  For $Z_1Z_2Z_3Z_4$, $b_1'=b_2'=b_3'=b_4'=b_5'=b_6'$, either permutation is satisfied.  We can choose the canonical permutation, i.e. the first column. }
\label{tab:notation_VVVV}
\renewcommand{\arraystretch}{1.5}
\begin{tabular}{|c||c|c|c|c|c|c|}
\hline
 $W^-_1W^+_2W^-_3W^+_4$/ $W^-_1Z_2W^+_3A_4$  & \multirow{2}{*}{$b_1'=b_1$}  &  \multirow{2}{*}{$b_2'=b_2$}     
&  \multirow{2}{*}{$b_3'=b_3'$} &  \multirow{2}{*}{$b_4'=b_4'$} &  \multirow{2}{*}{$b_5'=b_5$} &  \multirow{2}{*}{$b_6'=b_6'$}     \\
       $W^-_1Z_2W^+_3Z_4$ /   $W^-_1A_2W^+_3A_4$          &      & &  &  &   &  \\
\hline
 $W^+_1W^-_2W^+_3W^-_4$/  $Z_1W^-_2A_3W^+_4$  &  \multirow{2}{*}{$b_1'=b_1$} &  \multirow{2}{*}{$b_2'=b_5$}  &  \multirow{2}{*}{$b_3'=b_4$}  &  \multirow{2}{*}{$b_4'=b_3$}  &  \multirow{2}{*}{ $b_5'=b_2$}  &    \multirow{2}{*}{$b_6'=b_6$} \\
  $Z_1W^-_2Z_3W^+_4$/$A_1W^-_2A_3W^+_4$        &  &  &   &   &   & \\
 \hline
  $W^-_1A_2W^+_3Z_4$  & $b_1'=b_3$ & $b_2'=b_2$  & $b_3'=b_1$   & $b_4'=b_6$
 &  $b_5'=b_5$  &   $b_6'=b_4$ \\
\hline
  $A_1W^+_2Z_3W^-_4$/ $Z_1W^+_2Z_3W^-_4$  & \multirow{2}{*}{$b_1'=b_6$} & \multirow{2}{*}{$b_2'=b_5$}  & \multirow{2}{*}{$b_3'=b_3$}   & \multirow{2}{*}{$b_4'=b_4$}
 &  \multirow{2}{*}{$b_5'=b_2$}  &   \multirow{2}{*}{$b_6'=b_1$} \\
 $A_1W^+_2A_3W^-_4$               &              &  &   &   &   &  \\
\hline
\end{tabular}
\end{table*}

\subsection{VVVV vertex}

Subroutines in this section compute amplitudes (\texttt{vvvvxx}) and off-shell currents (\texttt{jvvvxx}) of the four-point vector boson vertices.
The subroutine for the amplitudes is shown explicitly in List~\ref{lst:vvvvxx}.

The coupling constants and their permutations according to the order of input particles for the four-point vector-boson vertices are listed in Table~\ref{tab:VVVV} and Table~\ref{tab:notation_VVVV}.

\begin{lstlisting}[caption=vvvvxx.f90,label=lst:vvvvxx]
!
! This subroutine computes an amplitude of the four-point coupling
! of the gauge bosons.
!
! input:
!       double complex v1(7) : first vector boson  v1
!       double complex v2(7) : second vector boson v2
!       double complex v3(7) : third vector boson  v3
!       double complex v4(7) : fourth vector boson v4
!       double precision gc(8): coupling constant  gvvvv
!                                                
! output:
!       double complex vertex: amplitude           gamma(v1,v2,v3,v4)
!     
subroutine vvvvxx(v1,v2,v3,v4,gc , vertex)

  use helas_params
  use helas_const
  use helas_utils
  
  implicit none

  complex(kind(0.d0)), intent(in) :: v1(ndim_wv),v2(ndim_wv),
  complex(kind(0.d0)), intent(in) :: v3(ndim_wv),v4(ndim_wv),
  complex(kind(0.d0)), intent(in) :: gc(8)
  complex(kind(0.d0)), intent(out) :: vertex
  
  complex(kind(0.d0)) :: dv1(0:4),dv2(0:4),dv3(0:4),dv4(0:4)
  complex(kind(0.d0)) :: v12,v13,v14,v23,v24,v34


  dv1(0:4) = v1(3:7)
  dv2(0:4) = v2(3:7)
  dv3(0:4) = v3(3:7)
  dv4(0:4) = v4(3:7)

  v12 = dv1(0)*dv2(0)-dv1(1)*dv2(1)-dv1(2)*dv2(2)-dv1(3)*dv2(3)
  v13 = dv1(0)*dv3(0)-dv1(1)*dv3(1)-dv1(2)*dv3(2)-dv1(3)*dv3(3)
  v14 = dv1(0)*dv4(0)-dv1(1)*dv4(1)-dv1(2)*dv4(2)-dv1(3)*dv4(3)
  v23 = dv2(0)*dv3(0)-dv2(1)*dv3(1)-dv2(2)*dv3(2)-dv2(3)*dv3(3)
  v24 = dv2(0)*dv4(0)-dv2(1)*dv4(1)-dv2(2)*dv4(2)-dv2(3)*dv4(3)
  v34 = dv3(0)*dv4(0)-dv3(1)*dv4(1)-dv3(2)*dv4(2)-dv3(3)*dv4(3)

  vertex = -gc(1) * (v12*v34 + v14*v23 - 2.d0*v13*v24)
  
  if (helas_mode==helas_mode_5cwb) then

     vertex = vertex &
          + gc(2)*dv1(4)*dv2(4)*v34 &
          + gc(3)*dv1(4)*dv3(4)*v24 &
          + gc(4)*dv1(4)*dv4(4)*v23 &
          + gc(5)*dv2(4)*dv3(4)*v14 &
          + gc(6)*dv2(4)*dv4(4)*v13 &
          + gc(7)*dv3(4)*dv4(4)*v12

     vertex = vertex -ci* gc(8)*dv1(4)*dv2(4)*dv3(4)*dv4(4)
     
  endif

  return

end subroutine vvvvxx
\end{lstlisting}

%

\subsection{VVVS vertex}

\begin{table*}[]
\centering
\caption{Vertices for $VVVH$: $WWZH$ and $WWAH$. }
\label{tab:VVVS}
\renewcommand{\arraystretch}{1.8}
\begin{tabular}{|c|c|c|c|c|}
\hline
\hline
\multicolumn{3}{|c|}{$V^{M_1M_2M_3}_{VVVH}$}        & $W^-W^+ZH$   &  $W^-W^+AH$    \\
\hline
\hline
\multicolumn{3}{|c|}{$V^{\mu_1\mu_2\mu_3}$}  &    $0$  &  $0$ \\
\hline
 \multirow{3}{4em}{$V^{4\mu_i\mu_j}$}  & 
  $V^{4\mu_2\mu_3}$& $b_1g^{\mu_2\mu_3 }$   &  $b_1=-\dfrac{g^{2}\sws}{2\cw}$  &  $b_1=\dfrac{g^{2}\sw}{2}$   \\  
    &  $V^{\mu_14\mu_3}$      & $b_2g^{\mu_1\mu_3 }$  & $b_2=\dfrac{g^{2}\sws}{2\cw}$ &$b_2=-\dfrac{g^{2}\sw}{2}$  \\
   &  $V^{\mu_1\mu_2 4}$& $b_3g^{\mu_1\mu_2 }$   & $b_3=0$    &  $b_3=0$   \\
 \hline
\multicolumn{3}{|c|}{$V^{44\mu_i} $}     &$0$    &  $0$    \\
\hline
\multicolumn{3}{|c|}{$V^{444}$} & $0$&  $0$  \\
\hline
\end{tabular}
\end{table*}

\begin{table*}
\centering
\caption{Definition of $b_1',b_2',b_3'$ according to the permutation of  $(b_1,b_2,b_3)$ in Table~\ref{tab:VVVS}. }
\label{tab:notation_VVVS}
\renewcommand{\arraystretch}{1.5}
\begin{tabular}{|c||c|c|c|}
\hline
$W_1^-W_2^+Z_3H_4$/$W_1^-W_2^+A_3H_4$            &    $b_1'=b_1$  &  $b_2'=b_2$  &   $b_3'=b_3$   \\
\hline
$W_1^{+}Z_2W^-_3H_4$/$W_1^+A_2W^-_3H_4$  &    $b_1'=b_2$ & $b_2'=b_3$  &  $b_3'=b_1$  \\
\hline
$W^-_1Z_2W^+_3H_4$/$W^-_1A_2W^+_3H_4$  &    $b_1'=b_1$ & $b_2'=b_3$  & $b_3'=b_2$  \\
\hline

\end{tabular}
\end{table*}

Subroutines in this section compute amplitudes (\texttt{vvvsxx}) and off-shell currents (\texttt{jvvsxx}, \texttt{hvvvxx}) of the four-point vertices among three vector bosons and a scalar.
The subroutine for the amplitudes is shown explicitly in List~\ref{lst:vvvsxx}.

The coupling constants and their permutations according to the order of input particles for the vector--vector--vector--scalar vertices are listed in Table~\ref{tab:VVVS} and Table~\ref{tab:notation_VVVS}.

\begin{table*}[]
\centering
\caption{Vertices for $VVHH$: $WWHH$ and $ZZHH$. }
\label{tab:VVSS}
\renewcommand{\arraystretch}{1.8}
\begin{tabular}{|c|c|c|c|}
\hline
\hline
\multicolumn{2}{|c|}{$V^{M_1M_2}_{VVHH}$}        & $W^-W^+HH$   &  $ZZHH$    \\
\hline
\hline
$V^{\mu_1\mu_2}$ &   $iag^{\mu_1\mu_2} $  &    $a=\dfrac{g^{2}}{2}$  &  $a=\dfrac{g^{2}}{2\cws}$ \\
\hline
\multicolumn{2}{|c|}{$V^{4\mu_i}$}  &  $0$  &  $0$  \\  
 \hline
 $V^{44}$  & $c$   &$c= -2i\lambda  $    &  $c= -2i\lambda $    \\
\hline
\end{tabular}
\end{table*}

\subsubsection{vvvsxx}

\begin{lstlisting}[caption=vvvsxx.f90,label=lst:vvvsxx]
!
! This subroutine computes an amplitude of the vector-vector-vector-
! scalar coupling.
!
! input:
!       double complex v1(7)  : first  vector     v1
!       double complex v2(7)  : second vector     v2
!       double complex v3(7)  : third vector      v3
!       double complex sc(3)  : input scalar      s
!       double complex gc(3)  : coupling constant gvvvs
!
! output:
!       double complex vertex: amplitude          gamma(v1,v2,v3,s)
!     
subroutine vvvsxx(v1,v2,v3,sc,gc , vertex)

  use helas_params
  use helas_const
  use helas_utils
  
  implicit none

  complex(kind(0.d0)), intent(in) :: v1(ndim_wv),v2(ndim_wv),v3(ndim_wv)
  complex(kind(0.d0)), intent(in) :: sc(ndim_ws)
  complex(kind(0.d0)), intent(in) :: gc(3)
  complex(kind(0.d0)), intent(out) :: vertex

  complex(kind(0.d0)) :: v12, v23, v31
  
  vertex = (0.d0, 0.d0)

  if (helas_mode==helas_mode_5cwb) then

     v12 = v1(3)*v2(3)-v1(4)*v2(4)-v1(5)*v2(5)-v1(6)*v2(6)
     v23 = v2(3)*v3(3)-v2(4)*v3(4)-v2(5)*v3(5)-v2(6)*v3(6)
     v31 = v3(3)*v1(3)-v3(4)*v1(4)-v3(5)*v1(5)-v3(6)*v1(6)
     
     vertex = ci*sc(3)*(gc(1)*v1(7)*v23 + gc(2)*v2(7)*v31 + gc(3)*v3(7)*v12)

  endif

  return

end subroutine vvvsxx
\end{lstlisting}

%

\subsection{VVSS vertex}

Subroutines in this section compute amplitudes (\texttt{vvssxx}) and off-shell currents (\texttt{jvssxx}, \texttt{hvvsxx}) of the four-point vertices among two vector bosons and two scalars.
The subroutine for the amplitudes is shown explicitly in List~\ref{lst:vvssxx}.

The coupling constants for the vector--vector--scalar--scalar vertices are listed in Table~\ref{tab:VVSS}.

\begin{lstlisting}[caption=vvssxx.f90,label=lst:vvssxx]
!
! This subroutine computes an amplitude of the vector-vector-scalar-
! scalar coupling.
!
! input:
!       double complex v1(7) : first  vector     v1
!       double complex v2(7) : second vector     v2
!       double complex s1(3) : first  scalar     s1
!       double complex s2(3) : second scalar     s2
!       double complex gc(2) : coupling constant gvvss
!
! output:
!       double complex vertex: amplitude         gamma(v1,v2,s1,s2)
!     
subroutine vvssxx(v1,v2,s1,s2,gc , vertex)

  use helas_params
  use helas_const
  use helas_utils
  
  implicit none

  complex(kind(0.d0)), intent(in) :: v1(ndim_wv),v2(ndim_wv)
  complex(kind(0.d0)), intent(in) :: s1(ndim_ws),s2(ndim_ws)
  complex(kind(0.d0)), intent(in) :: gc(2)
  complex(kind(0.d0)), intent(out) :: vertex

  complex(kind(0.d0)) :: v12

  v12 = v1(3)*v2(3)-v1(4)*v2(4)-v1(5)*v2(5)-v1(6)*v2(6)
  vertex = gc(1)*s1(3)*s2(3) * v12

  if (helas_mode==helas_mode_5cwb) then

     vertex = vertex - ci*gc(2)*v1(7)*v2(7)*s1(3)*s2(3)

  endif

  return

end subroutine vvssxx
\end{lstlisting}


 \subsection{Interface to {\tt MadGraph5}}

The new \HELAS\ library contains the subroutines to interface the \HELAS\ subroutines to the {\tt MadGraph5}~\cite{Alwall:2011uj} subroutines.
Amplitude subroutines generated by {\tt MadGraph5} can calculate helicity amplitude with 5-components vector bosons by linking with the new \HELAS\ library.

\section{Goldstone boson couplings in the SM}\label{sec:vertices}
\def\theequation{B.\arabic{equation}}
\setcounter{equation}{0}
In this appendix, we present all the Goldstone-boson couplings of the SM explicitly, 
because not only the magnitude
but also the relative signs of all the gauge-boson and the corresponding Goldstone-boson couplings should be kept exact 
in order to keep the BRST invariance of the amplitudes.
The Goldstone bosons appear in the four
sector of the SM Lagrangian,
the Higgs potential ${\cal V}_H$, the Higgs gauge interactions ${\cal K}_H$,
the gauge-fixing term ${\cal L}_{GF}$, 
and the Yukawa term ${\cal L}_Y$.
We show all of them explicitly below.

The minimal SM has just one SU(2)$_L$ doublet Higgs field $\phi(x)$.
The Lagrangian of the Higgs field is given by
\begin{align}
{\cal L}_{H} = {\cal K}_{H} - {\cal V}_{H}
\label{lag}
\end{align}
with
\begin{align}  
 {\cal K}_{H} &= (D^\mu \phi)^\dagger (D_\mu \phi), \label{KH}\\
 {\cal V}_{H} &= \frac{\lambda}{4} \Big(\phi^\dagger \phi-\frac{v^2}{2}\Big)^2,
\end{align}
where $v=(\sqrt{2}G_F)^{-1/2}\sim246$~GeV and 
\begin{align}
  \lambda=2\frac{m_H^2}{v^2}.
\end{align}

We parametrize the Higgs doublet field with hypercharge $Y=-1/2$ as
\begin{align}
 \phi = ((v+H+i \pi^0)/\sqrt{2},\ i \pi^-)^T, 
\label{phi}
\end{align}
and its $Y=1/2$ partner as
\begin{align}
\phi^c =-i\sigma^2\phi^* = (i \pi^+,\ (v+H-i \pi^0)/\sqrt{2})^T.
\label{phic}
\end{align}
The Goldstone fields $\pi^\pm(x)$ and $\pi^0(x)$ are defined
as the triplet of the custodial SU(2) transformation of the quartet field 
\begin{align}
\Phi = (\phi,\ \phi^c )\quad \to\quad  \Phi' = U \Phi U^\dagger
\end{align}
which keeps the vacuum invariant 
\begin{align}
 \langle\Phi\rangle' = U\langle\Phi\rangle U^\dagger = \langle\Phi\rangle
= \frac{v}{\sqrt{2}} 
\begin{pmatrix}1&0\\0&1\end{pmatrix}.
\end{align}
More explicitly, 
\begin{align}
\Phi(x) = \frac{1}{\sqrt{2}} \Big\{[v+H(x)]\,\mathbbm{1} +i\sum_{k=1,2,3}\sigma^k\pi^k(x)\Big\}
\end{align}
with Pauli's sigma matrices and 
\vrev{
the charge eigenstates are defined as
}
\begin{align}
 \pi^0(x)=\pi^3(x),\quad \pi^\pm(x)=\frac{\pi^1(x)\mp i\pi^2(x)}{\sqrt{2}}.
\label{piEM} 
\end{align}

By noting
\begin{align}
 \phi^\dagger \phi &=\frac{1}{2}{\rm tr}[ \Phi^\dagger \Phi] \n\\
 &=\frac{1}{2}[(v+H)^2 + (\pi^1)^2+(\pi^2)^2+(\pi^3)^2],
\label{phi2}
\end{align}
the Higgs potential term is expressed as
\begin{align} 
{\cal V}_{H}
=\frac{m_H^2}{2} \left[H+ \frac{H^2 + (\pi^1)^2+(\pi^2)^2+(\pi^3)^2}{2v}\right]^2, 
\label{VH}
\end{align}
which is manifestly invariant under the custodial SU(2) symmetry.
Note that 
\begin{align} 
 (\pi^1)^2+(\pi^2)^2+(\pi^3)^2 = 2\pi^+\pi^-+(\pi^0)^2
\end{align}
when expressed in terms of the electromagnetic charge eigenstates~\eqref{piEM}.

The covariant derivative in the kinetic term of the Higgs Lagrangian~\eqref{KH} is
\begin{align}  
 D_\mu &= \del_\mu +i\frac{g}{\sqrt{2}} (T^+W^+_\mu + T^-W^-_\mu) \n\\
       &\quad+ ig_Z (T^3-Qs_W^2)Z_\mu + ieQ A_\mu 
\end{align}
with the gauge couplings $e=gs_W=g_Zs_Wc_W$, 
the SU(2)$_L$ generators, $T^\pm=T^1\pm iT^2$ and $T^3$ with $T^k=\sigma^k/2$, 
and the electromagnetic charge operator $Q$.
For the Higgs doublet field~\eqref{phi}, the Higgs kinetic Lagrangian~\eqref{KH} reads
\begin{subequations}
\begin{align}
{\cal K}_{H} 
&=
\frac{1}{2}(\del H)^2 +(\del\pi^+)(\del\pi^-) +\frac{1}{2}(\del\pi^0)^2  \label{K1}\\
&+ 
\Big[ \frac{g^2}{4} W^+W^- +\frac{g_Z^2}{4}\frac{Z^2}{2}\Big]
 \big[(v+H)^2+(\pi^0)^2\big] \label{K2}\\
&+
\frac{g_Z}{2} Z \big[(\del\pi^0)(v+H)-(\del H)\pi^0 \big] \label{K3}\\ 
&+
\frac{g}{2} [W^+(\del\pi^-)+W^-(\del\pi^+)] (v+H) \label{K4}\\
&-
i\frac{g}{2} [W^+(\del\pi^-)-W^-(\del\pi^+)] \pi^0 \label{K5}\\
&+
i\frac{g}{2} (s_W^2g_ZZ-eA)(W^+\pi^- -W^-\pi^+)(v+H) \label{K6}\\
&+
\frac{g}{2} (s_W^2g_ZZ-eA)(W^+\pi^- +W^-\pi^+) \pi^0 \label{K7}\\
&-
\frac{g}{2}  \big[ (W^+\pi^- + W^-\pi^+)(\del H) \n\\
 &\qquad +i(W^+\pi^- - W^-\pi^+) (\del\pi^0)\big] \label{K8}\\
&-
i\Big[\Big(\frac{1}{2} -s_W^2\Big)g_ZZ+eA\Big] \big[(\del\pi^+)\pi^- -(\del\pi^-)\pi^+\big] \label{K9}\\
&+
 \Big[\frac{g^2}{2}W^+W^- + \Big(\Big(\frac{1}{2} -s_W^2\Big)g_ZZ+eA\Big)^2 \Big]\pi^+\pi^-, \label{K10}
\end{align}
\end{subequations}
where we drop repeated Lorentz indices.

Let us briefly explain each term.
The first line~\eqref{K1} gives kinetic terms of the Higgs boson and the Goldstone bosons.
The terms with $(v+H)^2$ in~\eqref{K2} give W and Z boson masses and their Higgs-boson couplings, while
the last term with $(\pi^0)^2$ contributes to $WWZZ$ and $ZZZZ$ vertices via the Goldstone (fifth) component of the Z boson.
The next terms in~\eqref{K3} and \eqref{K4} contain the bilinear Goldstone-boson--gauge-boson mixing terms 
which are removed by the $R_\xi$ gauge fixing, as given below, 
and the associate $ZHH$, $ZZH$, $WWH$ coupling terms through the Z and W Goldstone-boson components.  
All the remaining terms in~\eqref{K5}--\eqref{K10} contribute to three- or four-point vertices among the Higgs boson and the weak bosons with one or two Goldstone-boson components.

Although the Goldstone-boson interactions in the Higgs potential~\eqref{VH} take a very simple form 
due to its custodial SU(2) symmetry,
their gauge couplings look more complicated because the U(1)$_Y$ and hence U(1)$_{EM}$ gauge interactions violate  the custodial symmetry explicitly.

In the renormalizable covariant ($R_\xi$) gauge~\cite{Fujikawa:1972fe}, 
the gauge-fixing terms are
\begin{align}
 {\cal L}_{GF}&=-\frac{1}{2\xi_A}(\del^\mu A_\mu)^2
 -\frac{1}{2\xi_Z}(\del^\mu Z_\mu-\xi_Zm_Z\pi^0)^2 \n\\
&-\frac{1}{\xi_W}(\del^\mu W_\mu^+-\xi_Wm_W\pi^+) (\del^\mu W_\mu^--\xi_Wm_W\pi^-), 
\end{align}
which allow us to quantize the gauge fields, in a covariant manner, 
while removing the weak-boson--Goldstone-boson mixing terms in~\eqref{K3} and \eqref{K4}, 
and giving the Goldstone-boson masses $\xi_Z m_Z^2$ and $\xi_W m_W^2$ for $\pi^0$ and $\pi^\pm$, respectively. 
We may call the choice $\xi_A=\xi_Z=\xi_W=1$ as the Feynman--'t Hooft gauge.

The Yukawa interactions in the SM is given by
\begin{align}
{\cal L}_Y &=
-y^u_{ij} Q_i^\dagger \phi\, {u_R}_j
-y^d_{ij} Q_i^\dagger \phi^c {d_R}_j
-y^l_{ij} L_i^\dagger \phi^c {l_R}_j
+ h.c., 
\end{align}
where $i,j=1,2,3$ are flavor indices and we ignore neutrino masses.
In terms of the Higgs and Goldstone bosons in Eqs.~\eqref{phi} and \eqref{phic}, 
we find
\begin{align}
{\cal L}_Y&=
 -y^u_{ij} \Big[\frac{1}{\sqrt{2}}(v+H+i\pi^0) {u_L^\dgr}_i +i\pi^-   {d_L^\dgr}_i\Big] {u_R}_j \n\\
&\quad -y^d_{ij} \Big[\frac{1}{\sqrt{2}}(v+H-i\pi^0) {d_L^\dgr}_i +i\pi^+   {u_L^\dgr}_i\Big] {d_R}_j \n\\
&\quad -y^l_{ij} \Big[\frac{1}{\sqrt{2}}(v+H-i\pi^0) {l_L^\dgr}_i +i\pi^+ {\nu_L^\dgr}_i\Big] {l_R}_j 
+ h.c. 
\end{align}
By introducing the mass matrices
\begin{align}
M^f_{ij} = \frac{v}{\sqrt{2}} y^f_{ij} \quad 
{\rm for}\ f=u,d,l,
\end{align}
the Yukawa Lagrangian reads
\begin{align}
{\cal L}_Y&=
       -\Big[\Big(1+\frac{H+i\pi^0}{v}\Big) {u_L^\dgr}_i +i\sqrt{2}\frac{\pi^-}{v}   {d_L^\dgr}_i\Big] M^u_{ij} {u_R}_j \n\\ 
&\quad -\Big[\Big(1+\frac{H-i\pi^0}{v}\Big) {d_L^\dgr}_i +i\sqrt{2}\frac{\pi^+}{v}   {u_L^\dgr}_i\Big] M^d_{ij} {d_R}_j \n\\
&\quad -\Big[\Big(1+\frac{H-i\pi^0}{v}\Big) {l_L^\dgr}_i +i\sqrt{2}\frac{\pi^+}{v} {\nu_L^\dgr}_i\Big] M^l_{ij} {l_R}_j \n\\
&\quad+ h.c.
\end{align}
Positive mass eigenvalues are obtained by the bi-unitary transformations,
\begin{align}
M^u &= U^u_L {\rm diag}(m_u,m_c,m_t)      {U^u_R}^\dgr, \n\\
M^d &= U^d_L {\rm diag}(m_d,m_s,m_b)      {U^d_R}^\dgr, \n\\
M^l &= U^l_L {\rm diag}(m_e,m_\mu,m_\tau) {U^l_R}^\dgr,
\end{align}
that relate the flavor eigenstates with the mass eigenstates as
\begin{align}
({u_L}_1,{u_L}_2,{u_L}_3)^T &= U^u_L (u_L, c_L,  t_L)^T   = U^u_L \bm u_L, \n\\
({u_R}_1,{u_R}_2,{u_R}_3)^T &= U^u_R (u_R, c_R,  t_R)^T   = U^u_R \bm u_R, \n\\
({d_L}_1,{d_L}_2,{d_L}_3)^T &= U^d_L (d_L, s_L,  b_L)^T   = U^d_L \bm d_L, \n\\
({d_R}_1,{d_R}_2,{d_R}_3)^T &= U^d_R (d_R, s_R,  b_R)^T   = U^d_R \bm d_R, \n\\
({l_L}_1,{l_L}_2,{l_L}_3)^T &= U^d_L (e_L,\mu_L,\tau_L)^T = U^l_L \bm l_L, \n\\
({l_R}_1,{l_R}_2,{l_R}_3)^T &= U^d_R (e_R,\mu_R,\tau_R)^T = U^l_R \bm l_R,
\end{align}
where $\bm f_L$ and $\bm f_R$ ($\bm f=\bm u,\bm d,\bm l$) are column vectors of the three mass eigenstates with definite chirality.

By inserting the above relations into ${\cal L}_Y$, we obtain
\begin{align}
{\cal L}_{Y} &=
       -\Big(1+\frac{H+i\pi^0}{v}\Big) \bm u_L^\dgr       {\rm diag}(m_u,m_c,m_t)   \bm u_R \n\\
&\quad -i\sqrt{2}\frac{\pi^-}{v} \bm d_L^\dgr V^\dgr {\rm diag}(m_u,m_c,m_t) \bm u_R                           \n\\
&\quad -\Big(1+\frac{H-i\pi^0}{v}\Big) \bm d_L^\dgr       {\rm diag}(m_d,m_s,m_b)   \bm d_R                         \n\\
&\quad -i\sqrt{2}\frac{\pi^+}{v} \bm u_L^\dgr V     {\rm diag}(m_d,m_s,m_b)    \bm d_R                         \n\\
&\quad -\Big(1+\frac{H-i\pi^0}{v}\Big) \bm l_L^\dgr       {\rm diag}(m_e,m_\mu,m_\tau)  \bm l_R                       \n\\
&\quad -i\sqrt{2}\frac{\pi^+}{v} \bm \nu_L^\dgr       {\rm diag}(m_e,m_\mu,m_\tau)  \bm l_R                      
+ h.c.,
\end{align}
where $\bm \nu_L=({\nu_{e}}_L,{\nu_\mu}_L,{\nu_\tau}_L)^T$.
The Cabibbo--Kobayashi--Maskawa (CKM) matrix by
\begin{align}
V = {U^u_L}^\dgr U^d_L
\end{align}
appears for the $\pi^\pm$ couplings of quarks.
This is the final expression, which determines all the Goldstone boson couplings with fermions in the SM.
We note that, 
because we do not introduce neutrino mass terms,
$\bm \nu_L$ gives the massless flavor eigenstates in the diagonal charged-lepton mass basis.
We also note that
we set $V = {\rm diag}(1,1,1)$ as a default for collider physics.

\section{Polarization vectors}\label{sec:polvector}
\def\theequation{C.\arabic{equation}}
\setcounter{equation}{0}
Since the polarization vectors of the vector boson field play important roles 
in deriving the parton shower (PS)~\cite{Hagiwara:2020tbx} or the Feynman diagram (FD) gauge representation of the gauge boson propagators,
we summarize our definitions and their basic properties in this appendix.

We define the polarization four vectors as 
\begin{subequations}
\begin{align}
  \eps^\mu(q,\lam=\pm1)&=\frac{1}{\sqrt{2}}(0,\mp1,-i,0)^T, \label{polT}\\
  \eps^\mu(q,\lam=0)  &=(0,0,0,1)^T,
\end{align}\label{polvec}%
\end{subequations}
in the rest frame of the time-like momenta,
\begin{align}
  q^\mu=(Q,0,0,0)^T
\label{q_time}
\end{align}
with $Q=\sqrt{q^2}$.
The phase convention in Eq.~\ref{polvec} follows the quantum mechanics
where the application of the raising/lowering operator, $J_\pm=J_1\pm iJ_2$, 
obtain the three states without sign change.  
They are eigen vectors of the angular momentum operator $J_3$
\begin{align}
  J_3\,\eps^\mu(q,\lam)=\lam\,\eps^\mu(q,\lam).
\end{align}
For the space-like momenta, we define the polarization vectors in the Breit frame,
\begin{align}
  q^\mu=(0,0,0,Q)^T
\label{q_space}
\end{align}
with $Q=\sqrt{-q^2}$,
where the transverse ($\lam=\pm1$) polarization vectors remain the same as in~\eqref{polT},
while the longitudinal ($\lam=0$) polarization vector is 
\begin{align}
  \eps^\mu(q,\lam=0)  &=(1,0,0,0)^T.
\end{align} 
All the polarization four-vectors satisfy
\begin{align}
  q_\mu\eps^\mu(q,\lam)=0,
\end{align} 
and the normalization
\begin{align}
  \eps_\mu(q,\lam)^*\eps^\mu(q,\lam')=-\eta_\lam\,\delta_{\lam\lam'}, 
\end{align} 
where $\eta_\pm=1$ and $\eta_0=\sgn(q^2)$.
They also satisfy the completeness condition
\begin{align}
  &\sum_{\lam=\pm1}\eps^\mu(q,\lam)^*\eps^\nu(q,\lam)
  +\sgn(q^2)\eps^\mu(q,0)\eps^\nu(q,0) \n\\
  &=-g^{\mu\nu}+\frac{q^\mu q^\nu}{q^2} \n\\
  &=-g^{\mu\nu}_{j=1},
\label{complete}
\end{align} 
giving the spin-1 polarization tensor.

We now introduce the reduced longitudinal polarization vector as
\begin{align}
  \epstld^\mu(q,0) = \eps^\mu(q,0) - \frac{q^\mu}{Q} 
\end{align}
for both time-like and space-like four momenta.
It is instructive to note its explicit form in the rest frame~\eqref{q_time}
of the time-like momentum
\begin{align}
  \epstld^\mu(q,0) = (-1,0,0,1)^T,
\end{align}
and in the Breit frame~\eqref{q_space} of the space-like momentum
\begin{align}
  \epstld^\mu(q,0) = (1,0,0,-1)^T.
\end{align}
They are light-like vectors
\begin{align}
  \epstld^\mu(q,0)\, \epstld_\mu(q,0) =0, 
\end{align}
and transform under the boost along the positive $z$-axis as
\begin{align}
  B_3(y)\,\epstld^\mu(q,0) =e^{-iK_3y}\epstld_\mu(q,0) =e^{-y}\epstld^\mu(q,0). 
\end{align}
All four components of $\epstld^\mu(q,0)$ scale as $e^{-y}$, 
which vanishes at high energies ($y\gg1$), where
\begin{align}
  q^\mu=\left\{ 
   \begin{array}{ll}
   Q (\cosh{y}, 0, 0, \sinh{y})^T  &\quad{\rm for}\ q^2=Q^2>0, \\
   Q (\sinh{y}, 0, 0, \cosh{y})^T  &\quad{\rm for}\ q^2=-Q^2<0.
  \end{array}\right.
\label{q_rapidity}  
\end{align}
The reduced longitudinal polarization vector is orthogonal to the transverse polarization vectors
\begin{align}
  \epstld^\mu(q,0)\, \eps_\mu(q,\pm)=0,
\end{align}
while
\begin{align}
  \epstld^\mu(q,0)\, q_\mu=-\frac{q^2}{Q}=-\sgn(q^2)\, Q.
\end{align}
It is because of this property, the $\epstld^\mu(q,0)$ polarization state of the massive vector bosons
can mix with the spin-0 (Goldstone boson) state.

Now, by introducing the light-cone vector
\begin{align}
  n^\mu = ( \sgn(q^0) , -\vect{q}/|\vect{q}| )^T
\label{nmu_app}
\end{align}
and the corresponding propagator tensor~\cite{Hagiwara:2020tbx}
\begin{align}
   \tilde{g}^{\mu\nu}
  =g^{\mu\nu}-\frac{n^\mu q^\nu + q^\mu n^\nu}{n\cdot q}, 
\end{align}
we find
\begin{align}
  \epstld^\mu(q,0) 
  =\tilde g^\mu_{\ \nu}\eps^\nu(q,0)
  = -\sgn(q^2) \frac{Q\,n^\mu}{n\cdot q}.
\label{epstld0_app}
\end{align}
It shows that our reduced longitudinal polarization vector is proportional
to the light-cone gauge vector~\eqref{nmu_app}, 
and its magnitude scale as 
\begin{align}
	\frac{Q}{n\cdot q}=\frac{Q}{|q^0|+|\qvec|}=e^{-y}
\end{align}
\vrev{
where the rapidity $y$ is given by \eqref{q_rapidity}.
}
The `completeness' condition~\eqref{complete} reduces to
\begin{align}
  &\sum_{\lam=\pm1}\eps^\mu(q,\lam)^*\eps^\nu(q,\lam)
  +\sgn(q^2)\tilde\eps^\mu(q,0)\tilde\eps^\nu(q,0) \n\\
  &=-g^{\mu\nu}+\frac{n^\mu q^\nu + q^\mu n^\nu}{n\cdot q} \n\\
  &=-\tilde{g}^{\mu\nu},
\end{align}
giving the physical meaning of the special (FD) gauge propagator.

We note 
\begin{subequations}
\begin{align}
  \tilde{g}^{\mu}_{\ \nu} \eps^\nu(q,\pm)&=\epstld^\mu(q,\pm)=\eps^\mu(q,\pm), \\
  \tilde{g}^{\mu}_{\ \nu} \eps^\nu(q,0)&=\epstld^\mu(q,0)=\eps^\mu(q,0) - \frac{q^\mu}{Q}.
\end{align}
\end{subequations}
It should also be noted that 
\begin{subequations}
\begin{align}
  \tilde{g}^{\mu}_{\ \nu} n^\nu&=0, \\
  \tilde{g}^{\mu}_{\ \nu} q^\nu&=-q^2 \frac{n^\mu}{n\cdot q}=Q\,\epstld^\mu(q,0),
\end{align}
\end{subequations}
and hence
\begin{align}
  \tilde{g}^{\mu}_{\ \nu}\tilde{g}^{\nu}_{\ \rho}
  &=\tilde{g}^{\mu}_{\ \rho}+q^2 \frac{n^\mu n_\rho}{(n\cdot q)^2} 
   ={P_T}^\mu_{\ \rho},
\end{align}
where ${P_T}^\mu_{\ \rho}$ denotes the transverse polarization tensor
\begin{align}
  q_\mu{P_T}^\mu_{\ \rho} =  n_\mu{P_T}^\mu_{\ \rho} = 0
\end{align}
in the light-cone gauge.
It is worth noting that the tensor $\tilde{g}^{\mu\nu}$ is not projective,
and hence cannot be used to perform Schwinger summation of 1PI (truncated) propagators.
This is in contrast to the spin projector for $j=1$~\eqref{complete} and for $j=0$
\begin{align}
  g^{\mu\nu}_{j=0}=\frac{q^\mu q^\nu}{q^2},
\end{align} 
which satisfy
\begin{subequations}
\begin{align}
  &{g_j}^{\mu}_{\ \nu}{g_{j'}}^{\nu}_{\ \rho} =\delta_{jj'}{g}^{\mu}_{\ \rho}, \\
  &g^{\mu\nu}_{j=1}+g^{\mu\nu}_{j=0}=g^{\mu\nu}.
\end{align}
\end{subequations}
%

%
%
%
%
\bibliographystyle{JHEP}
\bibliography{bibnewhelas}

\end{document}